\documentclass[11pt]{article}
\usepackage{amsmath}
\usepackage{amsthm}
\newtheorem{definition}{Definition}
\newtheorem{theorem}{Theorem}
\newtheorem{corollary}[theorem]{Corollary}

\theoremstyle{plain}
      \newtheorem{assumption}{Assumption}

\newtheorem{assumptionalt}{Assumption}[assumption]
\newenvironment{assumptionp}[1]{
  \renewcommand\theassumptionalt{#1}
  \assumptionalt
}{\endassumptionalt}

\usepackage{hyperref}
\usepackage{wasysym}
\usepackage[font=footnotesize]{caption}
\usepackage[T1]{fontenc}
\usepackage{latexsym}
\usepackage[english]{babel}
\usepackage{cite}
\usepackage{enumerate}
\usepackage[shortlabels]{enumitem}
\pdfoutput=1
\usepackage[utf8]{inputenc}
\usepackage{dsfont}
\usepackage{diagbox}
\usepackage{amsfonts}
\usepackage{mathtools}
\allowdisplaybreaks[4]         
\usepackage{amssymb}
\usepackage{euscript}     
\usepackage{braket}
\usepackage{amssymb}
\usepackage{starfont}
\usepackage{color,soul}         
\usepackage{braket}
\usepackage{tensor}        
\usepackage{amsthm}
\usepackage{graphicx}
\usepackage{slashed}
\usepackage{leftidx}
\usepackage{subfigure}
\usepackage{bbm}
\usepackage{empheq}
\usepackage[dvipsnames]{xcolor}
\definecolor{outerspace}{rgb}{0.25, 0.29, 0.3}
\definecolor{scarlet}{rgb}{1.0, 0.13, 0.0}
\usepackage[header,title,page,titletoc]{appendix}  
\definecolor{princetonorange}{rgb}{1.0, 0.56, 0.0}
\definecolor{WildStrawberry}{rgb}{1.0, 0.26, 0.64}
\definecolor{rossocorsa}{rgb}{0.83, 0.0, 0.0}
\definecolor{navyblue}{rgb}{0.0, 0.0, 0.5}
\usepackage[numbers,sort&compress]{natbib}  
\usepackage{float}
\usepackage[paper=a4paper,margin=1in]{geometry}
\newtheorem{prop}{Proposition}

\newcommand{\norm}[1]{\left\lVert#1\right\rVert}
\usepackage{titlesec}

\pdfoutput=1
\usepackage{amsmath}
\usepackage{color}
\usepackage{amsfonts}
\usepackage{amssymb}
\usepackage{graphicx}
\usepackage{geometry}
\usepackage{amssymb,epsfig,subfigure}
\usepackage{amssymb}
\usepackage{hyperref}
\usepackage{comment}
\usepackage[font=footnotesize]{caption}
\usepackage[T1]{fontenc}
\usepackage[utf8]{inputenc}
\usepackage{latexsym}

\usepackage{cancel}

\usepackage[numbers,sort&compress]{natbib}  
\usepackage{float}

\usepackage{dsfont}
\usepackage{diagbox}

\usepackage{mathtools}
\allowdisplaybreaks[4]

\usepackage{euscript}     
\usepackage{braket}
\usepackage{starfont}
\usepackage{color,soul}         
\usepackage{tensor}        
\usepackage{amsthm}

\usepackage{slashed}
\usepackage{leftidx}
\usepackage{bbm}

\makeatletter
\renewcommand\section{\@startsection {section}{1}{\z@}%
                                 {-3.5ex \@plus -1ex \@minus -.2ex}
                                   {2.3ex \@plus.2ex}%
                                   {\normalfont\large\bfseries}}
\renewcommand\subsection{\@startsection{subsection}{2}{\z@}%
                                   {-3.25ex\@plus -1ex \@minus -.2ex}%
                                     {1.5ex \@plus .2ex}%
                                     {\normalfont\bfseries}}
\renewcommand\subsubsection{\@startsection{subsubsection}{3}{\z@}%
                                   {-3.25ex\@plus -1ex \@minus -.2ex}%
                                     {1.5ex \@plus .2ex}%
                                     {\normalfont\itshape}}
\makeatother

\def\pplogo{\vbox{\kern-\headheight\kern -29pt
\halign{##&##\hfil\cr&{\ppnumber}\cr\rule{0pt}{2.5ex}&\ppdate\cr}}}
\makeatletter
\def\ps@firstpage{\ps@empty \def\@oddhead{\hss\pplogo}%
  \let\@evenhead\@oddhead 
}
\def\maketitle{\par
 \begingroup
 \def\thefootnote{\fnsymbol{footnote}}
 \def\@makefnmark{\hbox{$^{\@thefnmark}$\hss}}
 \if@twocolumn
 \twocolumn[\@maketitle]
 \else \newpage
 \global\@topnum\z@ \@maketitle \fi\thispagestyle{firstpage}\@thanks
 \endgroup
 \setcounter{footnote}{0}
 \let\maketitle\relax
 \let\@maketitle\relax
 \gdef\@thanks{}\gdef\@author{}\gdef\@title{}\let\thanks\relax}
\makeatother

\numberwithin{equation}{section}

\newcommand\eea{\end{eqnarray}}
\newcommand\bea{\begin{eqnarray}}

\def\beq{\begin{equation}}
\def\eeq{\end{equation}}

\newcommand{\be}{\begin{equation}}
\newcommand{\ee}{\end{equation}}
\newcommand{\ba}{\begin{align}}
\newcommand{\ea}{\end{align}}
\newcommand{\bg}{\begin{gather}}
\newcommand{\eg}{\end{gather}}
\newcommand{\bseq}{\begin{subequations}}
\newcommand{\eseq}{\end{subequations}}

\usepackage{hyperref}
\hypersetup{
    colorlinks,
    citecolor=rossocorsa,
    filecolor=navyblue,
    linkcolor=navyblue,
    urlcolor=navyblue
}

\begin{document} 

\begin{titlepage}

\begin{center}

\phantom{ }
\vspace{3cm}

{\bf \Large{A generalization of the DHR theorem for higher form symmetries}}
\vskip 0.5cm
Horacio Casini${}^{*}$, Javier M. Mag\'an${}^{\dagger}$
\vskip 0.05in
\small{ ${}^{*\,\dagger}$\textit{Instituto Balseiro, Centro At\'omico Bariloche}}
\vskip -.4cm
\small{\textit{ 8400-S.C. de Bariloche, R\'io Negro, Argentina}}

\small{${}^{\dagger}$ \textit{Departament de F\'isica Qu\`antica i Astrof\'isica, Institut de Ci\`encies del Cosmos, }}
\vskip -.4cm
\small{\textit{Universitat de Barcelona, Mart\'i i Franqu\`es 1, E-08028 Barcelona, Spain}}

\begin{abstract}

The Doplicher-Haag-Roberts (DHR) reconstruction theorem shows that standard ($0$-form) internal symmetries are associated to groups in relativistic quantum field theory in spacetime dimension $D>2$.  In particular, non-invertible symmetry structures in $D>2$ correspond to the choice of a subtheory of a unique parent one, where the symmetry is a compact group. We present a theorem that generalizes this result to higher form symmetries. We first re-formulate the DHR theorem in terms of Haag duality violations (HDV) for regions with non-trivial homotopy group $\pi_0$ in the finite index case. In this light, the theorem states that the category associated with such HDV is the dual of a group, and it can be extended to spontaneous symmetry breaking scenarios. Then, after eliminating $\pi_0$ sectors via DHR reconstruction, we show that the HDV corresponding to regions with non-trivial $\pi_i$, $0<i<D-2$, are associated with abelian groups. Physically, the result shows that generalized order/disorder parameters in $D>2$ are labeled by such groups, in agreement with the case of confinement order parameters in Yang-Mills theories (Wilson and 't Hooft loops). For the special case of $D=4n$ and loops of dimension $k=2 n-1$, the group is further required to have a Hermitian character table. This does not rule out the possibility of an extra $\mathbb{Z}_2$ factor that is not achievable by Lagrangian gauge models.  In the way we find a new proof of the group-like origin of internal symmetries, and analyze the sectors for more general regions, e.g., direct sums, knots, and links. In particular, we find that generalized knot order parameters are classified by the unknot order parameters, and the commutator of knot non-local operators is determined by the linking number.
\end{abstract}

\vspace{-.5cm}

\end{center}

\small{\vspace{5.0 cm}\noindent 
${}^{\text{\text{*}}}$horaciocasini@gmail.com\\
${}^{\dagger}$magan@fqa.ub.edu
}

\end{titlepage}

\setcounter{tocdepth}{2}

{\parskip = .4\baselineskip \tableofcontents}
\newpage


\section{Introduction}

The main objective of this article concerns a first principles classification of confinement order parameters \cite{PhysRevD.10.2445,tHooft:1977nqb}. In perturbative approaches, QFTs exhibiting confinement are formulated in terms of Yang-Mills gauge theories, i.e. QFT's based on a gauge connection associated with a local gauge invariance principle. In these cases, confinement order parameters are given by holonomies of the gauge potential and their duals, the Wilson and 't Hooft loops. They are labeled by elements or representations of the center of the gauge group, they fuse following the appropriate abelian rules, and their commutator is controlled by the linking number. Is this scenario mandatory in QFT? or else, are there loop order parameters labeled by more general fusion categories? The main result of this article is the demonstration that this type of order parameters are necessarily labeled by abelian groups with Hermitian character tables, with trivial self-statistics and whose commutator is computable from the linking number of the loops.

Although the previous questions contain the main physical justification for the present work, it can also be motivated from various other perspectives, connecting with various different research communities. We now expand on these motivations. 

\noindent \textbf{A generalization of the DHR theorem:} The DHR (Doplicher-Haag-Roberts) theorem  \cite{Doplicher:1971wk,Doplicher:1973at,doplicher1990there} marks a milestone in the conceptual development of QFT. For the first time, the reason behind the existence of the fermion-boson alternative and the spin-statistics connection was addressed with full generality and shown to follow from the basic postulates of relativity, quantum mechanics and locality. In this investigation, the question about statistics becomes entangled and inseparable to the one about the internal symmetries of the theory. Hence, the DHR theorem is at the same time a proof of the relation between spin and statistics, and a classification of possible global internal symmetries in terms of  compact groups for spacetime dimensions $D>2$. 

The starting point of the DHR theorem is the analysis of the Superselection Sectors (SS) of the theory that satisfy a particular locality criterion. This DHR-criterion states that the SS is indistinguishable from the vacuum outside a sufficiently large ball-shaped region. Intuitively, this criterion is just specializing the type of ``charge'' considered, i.e. the DHR-criterion concentrates on charges $\psi$ which can be localized in a ball. In appropriate scenarios/limits, this means such charges can be generated by point-like Wightman fields.  These charge carrying operators $\psi$ naturally act on an enlarged Hilbert space, and this leads to the DHR ``reconstruction'' of a bigger theory ${\cal B}\supset {\cal A}$, incorporating the charged fields. 

Given this starting point, the two main results of the DHR analysis are the following. There is a precise notion of category associated with DHR sectors. This category has irreducible (simple) objects, conjugates, direct sums,  and crucially, in $D>2$ it is \emph{symmetric}, which means that the statistics of the fusion category furnish a representation of the permutation group. In modern language, the DHR category is a symmetric monoidal $C^*$ category. This turns out to imply that the quantum dimensions of the sectors are integers and the category is isomorphic to the representation category of a compact group \cite{Deligne2007,Doplicher1989}. This first result ultimately led to the second main result. The relation between the two theories ${\cal A}$ and ${\cal B}$ gives a global compact symmetry group $G$, such that ${\cal A}={\cal B}/G$ is the fix point algebra under the action of the group. The charged operators are labeled by representations of the group, $\psi_r^i$, $i=1\cdots,d_r$, where $r$ label irreducible representations of dimension $d_r$. It is worth remarking that this enlarged structure is completely determined by the original theory of chargeless operators ${\cal A}$. The charged operators still retain locality properties, but in general can be fermionic.

From the perspective of the Landau paradigm, the DHR analysis would imply that standard local Landau order parameters are labeled by irreducible representations of a group and are indeed charged under the action of such group of internal symmetries of the theory. But ``charges'' are not necessarily attached to balls or local Wightman fields. This observation leads to the concept of generalized charges and symmetries (next item), motivating a generalization of the DHR analysis and theorem to such charges.
 
\noindent \textbf{Classification of SSB and Generalized symmetries:} Though the perspective of global SS is very fruitful, there are important scenarios going beyond the DHR setup. The first one, well known in the literature, is that it does not capture the structure of charged operators in spontaneously broken symmetry (SSB) scenarios. In this case, the charged states belong to the original Hilbert space and do not give place to SS \cite{roberts1974spontaneously}.  
The second is a description of generalized symmetries \cite{tHooft:1977nqb,Gaiotto:2014kfa}, see recent reviews on the subject \cite{McGreevy:2022oyu,Schafer-Nameki:2023jdn,Brennan:2023mmt,Bhardwaj:2023kri,Shao:2023gho}. This new powerful generalization of the concept of symmetry in QFT refers to symmetries that act on multidimensional objects. An example is the existence of Wilson or 't Hooft loops (generalized magnetic and electric fluxes) in gauge theories. These operators are acted upon by automorphisms generated by the dual loop operators, and these automorphisms are the generalized symmetries. In this sense, the loop operators can be understood as ``generalized charges'', and are indeed charged under the notion of generalized symmetries \cite{Gaiotto:2014kfa}. However, it is worth remarking that these charges do not create SS out of the vacuum state (in flat space) simply because they do not have vanishing vacuum expectation values. In this  precise sense, generalized symmetries are always in a SSB scenario, forcing us to understand such generalization of DHR first.

From this perspective, it is interesting to inquire for a classification of generalized symmetries. Indeed, as mentioned above and using the language of \cite{Gaiotto:2014kfa}, part of the DHR theorem is a classification of $0$-form symmetries in terms of compact groups for $D>2$. Notice that this classification is dual to the one in the previous item, which first seeks to classify charges. But, intuitively, it seems natural that a classification of the possible categories associated to generalized charges (the category associated with the order parameters) already contains a classification of the possible categories associated with the generalized symmetries (the category of disorder parameters), and viceversa. This will be expanded upon rigorously elsewhere. In this vein, the present article can be seen and motivated as a classification of generalized symmetries in $D>2$. In particular we will obtain a complete classification in $D=4$ ($D=3$ is fully covered by DHR).

There are a couple of previous results in this subject worth mentioning. In \cite{Gaiotto:2014kfa} it was shown that group-like higher-form symmetries display abelian fusion rules. More precisely, the categorical fusion rules associated with higher form symmetries are necessarily abelian. Therefore, if they correspond to a group, the group is necessarily abelian. This leaves open the possibility that the category, while abelian, is not related to a group, and could be non invertible. A heuristic argument supporting the abelian group nature of higher-form was given in \cite{Casini:2020rgj}. Part of the motivation for this article came from the intention of turning the argument of \cite{Casini:2020rgj} into a precise theorem.

\noindent \textbf{Classification of Haag duality violations:} A more intrinsic and local algebraic approach to symmetries in QFT can be identified from the following well-known observation. Consider the case in which there are DHR superselection sectors. The charged operators $\psi_r^i$ do not belong to the observable algebra (the neutral operators under the action of the symmetry). However, the observable algebra  ${\cal A}(B_1\cup B_2)$ on two disjoint balls $B_1,B_2$, will contain the neutral combinations $\sum_i \psi_r^{(1) i}(\psi_r^{(2) i})^\dagger$ where each charged operator is localized on one ball. These operators clearly commute with the algebra of the complementary region though they cannot be produced by the algebras of neutral operators on each of the two balls. This leads to a violation of Haag duality for two balls:
\be
{\cal A}(B_1\cup B_2)\subsetneq({\cal A}((B_1\cup B_2)'))'\,,
\ee
where we are calling $A'$ to the causal complement of the region $A$, and the prime on an algebra corresponds to the commutant algebra. These operators $\sum_i \psi_r^{(1) i}(\psi_r^{(2) i})^\dagger$ are known in the DHR framework since they are the ``charge transporters'', or intertwiners, of the DHR category,  moving the localization region of the DHR sector.

Now, if instead on focusing on the DHR category \cite{Doplicher:1971wk,Doplicher:1973at,doplicher1990there} or the symmetry category \cite{Gaiotto:2014kfa}, we focus on this type of operators, a straightforward generalization of the idea of symmetries in QFT follows \cite{Casini:2020rgj,Casini:2021zgr}. In a QFT there are von Neumann algebras $\mathcal{A}(B)$ for each ball (or more precisely its causal closure: a double cone in the space-time). A minimal self-consistent assignation of algebras for general regions is the {\sl additive} net, which we will call ${\cal A}$. This assigns to every region $R$ the  algebra
\be
{\cal A}(R)\equiv{\cal A}_{\textrm{add}}(R)\equiv \bigvee_{B \, \textrm{ball},\,  B\subset R} {\cal A}(B)\,, 
\ee 
where the symbol ${\cal A}_1\vee {\cal A}_2=({\cal A}_1\cup {\cal A}_2)''$ means the minimal von Neumann algebra generated by the two. This is the mathematical counterpart of the expectation that the operator content of the theory is ultimately generated by operators localized in arbitrarily small regions of space. This net naturally satisfies causality
\be
{\cal A}(R) \subseteq  ({\cal A}(R'))'\,.
\ee
We thus have two canonical algebras associated with any given region, namely, the additive algebra ${\cal A}(R)$ and the maximal algebra $\hat{{\cal A}}(R)\equiv {\cal A}(R')' $. The latter is the maximal assignation for $R$ that is  compatible with causality. These two algebras mathematically correspond to the two notions of ``locality'' that appear in QFT. The first algebra ${\cal A}(R)$ conforms with the idea that observables localized in a certain region $R$ are generated by observables localized in smaller regions also contained in $R$. The second algebra $\hat{{\cal A}}(R)$ conforms with the idea that the algebra of $R$ should be the algebra of observables that commute with the ones that can be additively generated in $R'$. 
The key question is whether these two notions coincide or not. In the first case it is said that there is Haag duality for $R$. In the second case there are Haag Duality Violations (HDV) for $R$. In this latter case there are operators (non-local or HDV operators) in the region $R$ that commute with locally generated operators outside $R$, but that cannot be created additively in the region $R$. These HDV operators naturally form order parameters for the phases of the theory \cite{Casini:2019kex,Casini:2020rgj,Casini:2021zgr}, providing an intrinsic and local starting point for a generalized Landau paradigm.

The fact that HDV constitute an algebraic description of generalized symmetries has been extensively developed in \cite{Casini:2019kex,Casini:2020rgj,Magan:2020ake,Pedro,Casini:2021zgr,casini2021generalized,Magan:2021myk,Benedetti:2022zbb,Casini:2022rlv,Benedetti:2023owa,Benedetti:2023ipt,benedetti2023charges,Casini:2023vrb,Benedetti:2024dku,Benedetti:2024utz,Shao:2025mfj,Jia:2025bui,Martinek:2025xik,Evans:2025msy,Abate:2025ywp} (see also \cite{tHooft:1977nqb,inproceedings}). They also lead to generalized DHR endomorphisms. These HDV are associated to regions of certain topology. For example, the case of SSB of a global symmetry leads to HDV for a single ball, and gauge theories may give HDV for regions with the topology of a donut at a spatial slice. 
A key consequence of the violation of duality in a region $R$ is the violation of duality in the complementary region $R'$. This simply follows from von Neumann's double commutant theorem  ${\cal A}''={\cal A}$.  Finally, the maximal algebras for $R$ and $R'$ cannot commute with each other. Such commutation would imply $\hat{{\cal A}}(R)\subseteq (\hat{{\cal A}}(R'))'={\cal A}(R)$, which is not possible if the inclusions are strict. Hence $\hat{\cal A}(R)$ is not a causal net. The connection of HDV and generalized symmetries arises from the fact that an inclusion of algebras is associated to a structure of endomorphisms and viceversa. For a unitary and causal QFT we have the bonus that there will be dual generalized symmetries that greatly constrain the possible categories. Arguably, the HDV structure is more primary and natural from the point of view of the local structure of the theory than the SS. It also provides a deeper notion of symmetry in QFT. Indeed, the categories of topological operators, as well as the generalized DHR categories can be derived from the HDV structure, something we will describe elsewhere. 

From this perspective, it is natural to inquire for a general classification of HDV in QFT, for different region topologies and spacetime dimensions, and this is yet another motivation for this work. We note there are significant previous results in the subject in the context of $D=2$ CFT's. These works provided much of the inspiration for the present analysis. In particular, building on the generalization of the reconstruction theorem to $D=2$ in \cite{Longo:1994xe}, and previous particular analysis in loop models \cite{Xu1997JONESWASSERMANNSF}, Ref. \cite{Kawahigashi:1999jz} analyzed the structure of HDV in conformal nets in $D=2$. This ultimately led to the classification of associated DHR categories for minimal models \cite{Kawahigashi:2002px,Kawahigashi:2003gi} (i.e. those unitary CFTs with $c<1$). Such classifications results have been re-derived recently in \cite{Benedetti:2024utz} using the standard CFT  formalism \cite{BELAVIN1984333,francesco2012conformal}, i.e. standard OPE methods in minimal models. Finally, it is interesting to note that some of the basic assumptions that we will use below, such as the notion of transportability, can be proven in $D=2$ CFTs \cite{Kawahigashi:1999jz}.

\noindent \textbf{On the existence of non-invertible symmetries in $D>2$:} In a couple of interesting and influential articles \cite{Choi:2022jqy,Cordova:2022ieu}, it was claimed that there exist genuine non-invertible $0$-form symmetries in $D>2$ dimensions, in particular in standard QED and pion electrodynamics in $D=4$, see \cite{Shao:2023gho} for a review. This claim is in direct tension with the DHR theorem mentioned above, which states that $0$-form symmetries ultimately descend from groups. This tension has received almost no attention in the literature, neither from the algebraic nor from the high-energy communities. In \cite{Benedetti:2023owa}, we investigated some of the particular scenarios where non-invertible symmetries were claimed to exist, such as QED, pion electrodynamics and electromagnetic duality. In all these examples we explicitly found the symmetry was a $U(1)$ group. Still, it was unclear whether non-invertible symmetries could potentially exist on basic grounds in $D>2$, for example in SSB scenarios or for higher-form symmetries. This interesting debate also motivates a first principles analysis. Below we will find that the result of this analysis clarifies that more general symmetry categorical structures ultimately descend from groups (through quotients) in $D>2$, and this differentiates them from the $D=2$ scenario. In more precise terms, after DHR reconstruction all symmetries are in fact groups.

\noindent \textbf{On generalized link/knots order parameters:} As mentioned above, in e.g. four spacetime dimensions, there are generalized order parameters associated to ring-like regions. In Yang-Mills theories, these are precisely the confinement order parameters, the Wilson and 't Hooft loops. From the perturbative gauge theory analysis, it is clear that, say, Wilson loops can be defined for more general ring-like regions, such as knots, since these are defined by the holonomies of the gauge potential. Similar statement holds for the dual 't Hooft loop. But from a first principles perspective, it is not clear whether QFT's exist where the amount and categorical nature of generalized order parameters depend on the precise knot chosen (up to homotopy transformations in the ambient spacetime). In principle, these regions are different from the unknot one, and could display new generalized symmetries (associated to the complementary region of the knot).  Experience from topological field theories in three dimensions \cite{Witten:1988} suggests this is not the case, and that the sectors on those general regions are isomorphic to the sectors of the unknot. Also, for links, one expects that the sectors are the $n$-tensor product of the category of the unknot, where $n$ is the number of disconnected components. Still, this is of course obscure from an abstract and local QFT perspective. In this vein, the last motivation for this work concerns the analysis of sectors and symmetries in these more general knotted and linked regions.

Summarizing, we seek a classification of confinement order parameters, and/or generalized DHR endomorphisms, and/or generalized symmetries, and/or HDV in local relativistic QFT.

\noindent \textbf{Outline of the article:}  The article is organized as follows. In the next section \ref{SecII}, we introduce the basic assumptions and the simplest consequences. Much of this material is contained in the paper by Longo and Rehren \cite{Longo:1994xe},  but we introduce a novel and purely algebraic description of transportability that is naturally self dual. We also introduce a new property (weak modularity) about distributivity of operations $\vee,\cap$ among the algebras of the net. Then, the first important task is to translate the DHR theorem into the language of HDV. This is accomplished in section \ref{SecIII}  closely following, when possible, the path of the work \cite{Kawahigashi:1999jz} by Kawahigashi, Longo and Muger on $D=2$ CFT's.  This actually constitutes a key bedrock for the subsequent analysis. On one hand, the DHR theorem is reflected in a first classification result about the structure of HDV for disconnected regions or their complement, which contains $D-2$ dimensional shells. These HDV are classified in terms of groups or their duals for $D>2$. On the other hand, the DHR reconstruction allows to eliminate this particular type of HDV. We call such a QFT $\pi_0$-complete. Once the theory is $\pi_0$-completed, much of the structure simplifies and we are able to prove the invertibility of $0$-form symmetries and the statistics of associated sectors in a novel fashion, providing in this regard a complementary approach to the DHR theorem. Then, using again the $\pi_0$-complete QFT, in section \ref{SecIV} we demonstrate that the general structure of HDV for higher dimensional loops is given by abelian groups, our main new classification result. This is accomplished by an algebraic dimensional reduction.  In section \ref{SecV} we show other consequences of $\pi_0$-completeness, in particular the transportability to connected sums of regions. This allows us to classify in section \ref{SecVI} the case of arbitrarily linked and knotted regions for $D=4$. In this section we also discuss further restrictions to the group for $D=4$ (and more generally the self dual case in $D=4n$) that imply it has the form of a group square $G=H\times H^*$, with the possibility of an extra $\mathbb{Z}_2$ factor. We will end in section \ref{SecVII} with some  discussion and open problems.

\section{Assumptions and direct consequences}\label{SecII}

 In this section we introduce the basic assumptions, concepts, and notation necessary to describe HDV in a general way. We also develop the most immediate consequences. The majority of the assumptions are well known in the context of algebraic QFT, but less familiar in other communities working in QFT. The list of assumptions is not intended to be optimal. Our intention here is rather to abstract from concrete physical examples the circle of ideas from which the results follow in a natural way. In other words, we expect some of the assumptions to follow from other ones and more basic principles, and it is probable that some of them can be significantly weakened. Also, not all of them are needed for all the results in the following sections and we will be explicit in what assumptions are needed in each case. The main innovations contained in this section are the concept of transportability for HDV sectors and a circle of related ideas, as well as a new property we call ``weak modularity''.  

\bigskip

\begin{assumption}\label{as:1}
We take the space $E$ to be topologically $\mathbb{R}^d$. A {\sl region} is any regular open (equal to the interior of the closure) subset $R\subseteq E$. For any region $R$ there is a von Neumann algebra ${\cal A}(R)\subseteq {\cal B}({\cal H})$, acting on a separable Hilbert space ${\cal H}$. The algebras are non trivial for non trivial regions. We also assume irreducibility, ${\cal A}(E)={\cal B}(\cal {H})\label{00}$.
\end{assumption}

The set $E$ is ``the space'' (of dimension $d=D-1$) rather than spacetime (of dimension $D$). This indicates we are interested in properties or features that are not directly related to relativity or space-time symmetries. These features will be topological in nature and can be associated to regions in the $t=0$ surface in the relativistic context. The underlying picture is that for any region in this surface there is one causally complete region in spacetime, and the assigned algebra corresponds to this spacetime region. However, the set-up could in principle be applied to more general scenarios than relativistic QFT. We also note that although this assumption is not satisfied by strictly topological field theories since these have trivial algebras for bounded regions in $E$, it can apply to theories that have a topological infrared limit.

 The choice of regular open regions for the assignment of algebras is the natural match to the strong additivity property below. Regular open sets form a complete Boolean algebra with the complement $X'=\textrm{int}(E-X)$, the meet $\vee_{i\in I} X_i=\textrm{int}(\overline{\cup_{i\in I} X_i})$ and the intersection $\wedge_{i\in I} X_i=\textrm{int}(\cap_{i\in I} X_i)$. Consequently, for a region $R$, we take the complement $R'=\textrm{int}(E-R)$, and write $R_1-R_2=R_1\cap R_2'$. Finally, it is worth remarking that we will generally consider a much smaller and simpler set of regions in this paper. In general we will have in mind regions having boundaries that are $(d-1)$-dimensional piecewise smooth embedded submanifolds of $E$. We will also restrict attention exclusively to bounded regions or their unbounded complements. Finally, we emphasize that the properties we will be discussing do not depend on the specific shape but only on its topology. For example, we will use the term ``ball'' to refer to bounded regions in $E$ with the topology of a ball.

 The following is the mathematical formulation of the principle of Einstein causality.
\bigskip

\begin{assumption}\label{as:2}
  Locality: algebras for disjoint regions commute, 
\be
R_1, \,R_2\subseteq E\,, \qquad R_1\cap R_2=\emptyset\, \implies {\cal A}(R_1)\subseteq ({\cal A}(R_2))'\,.
\ee
\end{assumption}
This allows us to define the {\sl maximal algebra} $\hat{\cal A}(R)$ of a region $R$ as 
\be
\hat{\cal A}(R)=({\cal A}(R'))'\,,
\ee
such that for any $R$ we have the inclusion ${\cal A}(R)\subseteq \hat{\cal A}(R)$. When ${\cal A}(R)= \hat{\cal A}(R)$ we say there is {\sl Haag's duality} for $R$. In the other scenario, i.e. when the inclusion is strict ${\cal A}(R)\subset \hat{\cal A}(R)$, we say  there is a Haag Duality Violation (HDV) for $R$, and the operators in $\hat{\cal A}(R)$ that are not in ${\cal A}(R)$ are HDV operators (for $R$).  Notice that if there is HDV for $R$ there is also HDV for $R'$. This is a direct consequence of von Neumann's double commutant theorem $({\cal A}(R))''={\cal A}(R)$.  
It also follows that while ${\cal A}(R)$ and ${\cal A}(R')$ commute, the maximal algebras for complementary regions $\hat{\cal A}(R)$ and $\hat{\cal A}(R')$ cannot commute with each other if there are HDV.

\bigskip

\begin{assumption}\label{as:7}
  Strong additivity: for any collection of regions $R_i, i\in I$,  
\be
\vee_{i\in I} {\cal A}(R_i)=  {\cal A}({\rm int}(\overline{\cup_{i\in I} R_i}))\;. 
\ee
\end{assumption}

In the relativistic setting, this assumption encapsulates two ideas. The first is that operators can be formed by other operators with a smaller support. The second is a form of causal evolution law in spacetime. For example, it indicates that the algebra of a double cone can be produced by a set of smaller double cones covering the spatial ball at the base of the double cone. In this sense, it extends the time slice axiom to bounded regions; see \cite{haag1962postulates} and \cite{araki1964neumann,araki1963lattice} for the case of the free scalar field.

This form of additivity is termed ``strong'' because it implies that the algebras of disjoint open regions with a common piece of their $d-1$-dimensional boundaries generate the algebra of a region without any gap in the common boundary (see similar use in \cite{Kawahigashi:1999jz} for the case $d=1$). In the framework of Wightmann fields, this property is suggested because smearing a field operator in a $d-1$ dimensional spatial surface is not enough to form an operator in Hilbert space.  
Given strong additivity, we often call an operator in ${\cal A}(R)$ an {\sl additive operator} of $R$. In contrast, HDV operators which belong to $\hat{\cal A}(R)$, but are not additive in $R$, will also be called {\sl non-local operators} of $R$.

Note that from this assumption it follows the property of isotony, namely $R_1\subseteq R_2\rightarrow {\cal A}(R_1)\subseteq {\cal A}(R_2)$. In turn,  this also implies isotony for the maximal algebras $R_1\subseteq R_2\rightarrow \hat{{\cal A}}(R_1)\subseteq \hat{{\cal A}}(R_2)$. In conjunction with assumption~(\ref{as:1}) it also follows that all ${\cal A}(R)$ are factors, namely, ${\cal A}(R)\cap ({\cal A}(R))'=\mathbb{C}$. This is because the algebras for complementary regions commute and, by strong additivity, we have ${\cal A}(R)\vee {\cal A}(R')={\cal A}(E)={\cal B}(\cal {H})$. Taking commutants in this relation it follows ${\cal A}(R)\cap ({\cal A}(R))'\subseteq \hat{\cal A}(R)\cap \hat{\cal A}(R')=({\cal A}(R'))'\cap ({\cal A}(R))'=\mathbb{C}$, what gives the factoriality. Then, it also follows that algebras of disjoint regions have trivial intersection.

The dual (by taking commutants) of this strong additivity relation gives that arbitrary intersections of maximal algebras is the maximal algebra of the intersection
\be
 \cap_{i\in I} \hat{{\cal A}}(R_i)=  \hat{{\cal A}}(\textrm{int}(\cap_{i\in I} R_i))\,. 
\ee
In particular, we have $\hat{{\cal A}}(R')\cap \hat{{\cal A}}(R)=({\cal A}(R))'\cap \hat{{\cal A}}(R)=\mathbb{C}$.
Then ${\cal A}(R)$ is an irreducible subfactor of $\hat{{\cal A}}(R)$.

If Haag duality holds for any region $R$ we say the theory is {\sl complete} \cite{Casini:2021zgr}. In this case, under the assumption of strong additivity, arbitrary use of $\cap, \vee, '$, for the algebras of regions gives place to the algebra of the region formed by the corresponding operations on the regions. This gives an isomorphism between the Boolean algebra of regions into a Boolean algebra of subalgebras of ${\cal B}({\cal H})$. However, we will be interested in analyzing failures of Haag duality for non-complete theories. In these scenarios, the operations $\cap, \vee$ are in general not distributive.

  \bigskip 

\begin{assumption}\label{as:5} Factor type. 
${\cal A}(R)$ for non trivial $R\subset E$ is a type III  factor. 
 \end{assumption}
Evidence for this property comes from explicit models as well as some general derivations. These give generically the unique type III$_1$ hyperfinite factor \cite{buchholz1987universal,gabbiani1993operator,brunetti1993}, e.g. for the Rindler wedge in QFTs satisfying the Wightman axioms; see also \cite{Witten:2018lha} for a recent review. Also, a general proof that the wedge von Neumann algebras are type III$_1$ factors can be found in \cite{Longo:1979dw}. Such proof does not assume the Bisognano-Wichmmann property or Wightman fields (but only the uniqueness of the vacuum vector). 

From a physical and practical perspective, this assumption reflects some regularity conditions at short distances. This is an important  assumption since in particular it implies Borchers property \cite{cmp/1103839939}, commonly called ``property B''. This is stated as follows. The net of algebras satisfies property B if for any two double cones $K_1$ and $K_2$, where the closure of one is included in the other $\bar{K}_1\subseteq K_2$, then any non-zero projector $P\in K_1$ is equivalent to the identity with respect to $K_2$, namely there is a partial isometry $V\in K_2$ such that $V^{\dagger}V=\mathds{1}$ and $VV^{\dagger}=P$. For type III algebras this property is immediate since every projection is the range projection of a partial isometry inside the algebra. The fact that any projector is equivalent to the identity within the algebra is useful for decomposing endomorphisms into direct sums.

Before stating the next assumption we introduce the notion of conditional expectation and finite index, see \cite{Longo:1994xe} for a more detailed review. A conditional expectation $\varepsilon_{\cal M}$ from an algebra ${\cal M}$ to a subalgebra ${\cal N}\subseteq {\cal M}$ is a linear positive map that is the identity on the subalgebra, and satisfies the bimodular property $\varepsilon_{\cal M}(n_1 m n_2)=n_1 \varepsilon_{\cal M}(m) n_2$, for any $m \in {\cal M}$ and $n_1,n_2 \in {\cal N}$. For irreducible subfactors, namely those factors  $\mathcal{N}\subset \mathcal{M}$ for which $\mathcal{N}'\cap \mathcal{M}=\mathds{1}$, if a conditional expectation exists, it is unique, and it is faithful (the kernel do not contain positive elements).  
Conditional expectations are useful for extending states from the subalgebra to the algebra as $\omega_{\cal M}=\omega_{\cal N}\circ \varepsilon_{\cal M}$. This state is invariant under $\varepsilon_{\cal M}$, that is, $\omega_{\cal M}(m)=\omega_{\cal M}(\varepsilon(m))$. Starting from a faithful state $\omega_{\cal N}$ and taking the GNS representation for $\omega_{\cal M}$ one gets a cyclic and separating vector for ${\cal M}$ in ${\cal H}$, invariant under the conditional expectation, and which is cyclic for ${\cal N}$ in a subspace ${\cal H}_0$. The projection $e$ into this subspace is called the Jones projection.\footnote{See \cite{Longo:1994xe} for a review of the Jones ladder in type III algebras.} It belongs to ${\cal N}'$ and it implements the conditional expectation as
\be
e \,m\, e=\varepsilon_{\cal M}(m)\, e\,.
\ee
We have ${\cal N}'={\cal M}'\vee e$.  If a dual conditional expectation $\varepsilon_{{\cal N}'}:{\cal N}'\rightarrow {\cal M}'$ also exists, the subfactor is said to be of finite index. Calling $e'$ to the corresponding Jones projection we now have ${\cal M}={\cal N}\vee e'$. 
In this case, there is a notion of relative size for the algebras ${\cal M}$ and ${\cal N}$. This is the Jones index \cite{Jones1983,KOSAKI1986123,Longo:1989tt}. There are various equivalent definitions. For finite index $\lambda$, see \cite{Longo:1994xe}, these definitions boil down to
\be
\varepsilon_{\cal M}(e')= \lambda^{-1}\, \mathbf{1}\,.
\ee
The index is the same for the two dual subfactor inclusions, i.e. we have $\varepsilon_{{\cal N}'}(e)= \lambda^{-1}\, \mathbf{1}$, where $\varepsilon_{{\cal N}'}$ is the dual conditional expectation.

\bigskip 

\begin{assumption}\label{as:7}
The inclusion ${\cal A}(R)\subseteq \hat{\cal A}(R)$ is of finite index.
 \end{assumption}

This assumption implies the larger algebra is generated by the smaller one and a finite set of non-local operators with different ``charges'', as we review shortly.  An important example concerns subalgebras invariant under an internal symmetry group of finite order. The index for this example is the order of the group. A finite index is also expected for sectors associated with regions with non contractible loops in non abelian gauge theories \cite{Casini:2020rgj}. When we have a discrete but infinite number of charges for the non local operators, a conditional expectation might still exist, but the index will not be finite. In this case the dual conditional expectation does not exist, though a dual weight exists \cite{CONNES1980153,Longo:1994xe}. This is the case e.g. of sectors coming from a global compact but continuous symmetry group, and, presumably, the case of loop sectors in QED \cite{Benedetti:2023owa}. Some other cases where the dual non commuting non local operators are both labeled by continuous parameters, such as the case of ring sectors for  free Maxwell field, may be called ``non-compact''. On general grounds non-compact cases are expected to correspond to free massless models \cite{benedetti2023charges}. 

We will use the finite index assumption mainly for technical reasons. It will allow us to cover an important case (e.g.  confinement order parameters) without introducing further technical difficulties that are presumably alien to the problem at hand. This will also allow us to profit from previous developments in the literature. The assumption means we will restrict attention to regions $R$ with finite index, that by the transportability assumption below will include all regions with the same topology as $R$. Notice that when some bounded region has a finite index, it will follow there are some other regions with infinite index, e.g., the ones formed by an infinite set of disjoint regions with the same topology. Then, the assumption also means we are restricting attention to simpler regions that exclude these other cases.

In the present QFT context, we have subfactors associated with each region $R$, namely the subfactor $\mathcal{A}(R)\subseteq\hat{\mathcal{A}}(R)$. We will call $\varepsilon_{R}$, $e_R$, $\lambda_R$ to the quantities corresponding to the subfactor associated to the region $R$, e.g. $\varepsilon_R: \hat{{\cal A}}(R)\rightarrow {\cal A}(R)$ is the unique conditional expectation mapping the maximal algebra to the additive algebra in $R$, and the corresponding Jones projection and Jones index are $e_R$ and $\lambda_R$ respectively. We use similar nomenclature for the complementary region $R'$.
 The dual structure of inclusions can be pictured via the quantum complementary diagram \cite{Casini:2020rgj,Magan:2020ake}
\begin{eqnarray}
\hat{\mathcal{A}}(R) & \overset{\varepsilon_R}{\longrightarrow} & \mathcal{A}(R)\nonumber \\
\updownarrow' &  & \updownarrow'\\
\mathcal{A}(R') & \overset{\varepsilon_{R'}}{\longleftarrow} & \hat{\mathcal{A}}(R')\,.\nonumber 
\end{eqnarray}
The conditional expectations $\varepsilon_R$ and $\varepsilon_{R'}$ are dual to each other  \cite{CONNES1980153,Longo:1994xe}.

Let us then further expand on the finite index scenario, following \cite{Longo:1994xe}. We first introduce some further concepts. An endomorphism of an algebra ${\cal A}$ is a linear mapping $\rho:\mathcal{A}\rightarrow \rho(\mathcal{A})\subseteq \mathcal{A}$ respecting the product operation $\rho(ab)=\rho(a)\rho(b)$ and the involution $\rho(a^{\dagger})=\rho(a)^{\dagger}$.  In particular $\rho(\mathds{1})=\mathds{1}$. The inner equivalence class $\rho\sim\rho_{u}$, with $\rho_{u}=u\rho u^{\dagger}$ and $u\in \mathcal{A}$ unitary, is called a sector. Notice that $\rho(\mathcal{A})$ can be strictly contained in $\mathcal{A}$, i.e. $\rho(\mathcal{A})\subset \mathcal{A}$ can be an strict inclusion, with associated Jones index greater than one. The {\sl dimension} $d_\rho$ of the endomorphism equals the square root of the Jones index associated with this inclusion 
 \cite{Longo:1989tt}. 
A sector is called ``irreducible'' if the previous subfactor is irreducible, namely $\rho(\mathcal{A})'\cap \mathcal{A}=\mathds{1}$. This nomenclature parallels that of the standard notion of dimensions and irreducible representations in group theory as will be transparent in a moment. Let us note that this notion of dimension is sometimes called ``quantum dimension'', the rational being it is not necessarily an integer number, as follows through its relation to the Jones index, whose values were classified in \cite{Jones1983}.

Given a subfactor ${\cal A}\subset {\cal B}$ an operator $\psi\in {\cal B}$ satisfying the commutation relation
\be
\psi a=\rho(a)\psi\;,\quad a\in {\cal A}\,,
\ee
is called a charged intertwiner for the sector $\rho$ of ${\cal A}$. It takes the identity sector and gives the endomorphism $\rho$. These types of endomorphisms of ${\cal A}$ are associated to the particular subfactor.  Then $\psi^{\dagger}$ goes in the opposite direction and $\psi^{\dagger}\psi$ commutes with $\mathcal{A}$. If the subfactor is irreducible it follows that $\psi^{\dagger}\psi\propto\mathds{1}$.  This is still the case if there are further independent intertwiners for the same sector. This defines an inner product in such space of intertwiners as $\psi^{\dagger}\psi '\equiv \langle \psi^{\dagger}\psi '\rangle\,\mathds{1}$. 
Choosing an orthonormal basis $\psi^i$ for the particular irreducible endomorphism we have $\psi^{i\,\dagger} \psi^j=\delta_{ij}$. Then these operators are isometries, and the multiplication in the reverse order $\psi^i \psi^{i\, \dagger}$ is a projector that belongs to $\rho({\cal A})'\cap {\cal B}$.

Endomorphisms can be ``added up'' to form new endomorphisms. This works as follows. Since we are working with type III algebras, we can always take a partition of unity into a set of projectors $\mathds{1}=\sum_r p_r$. The number of projectors of this partition is arbitrary. Each projector is equivalent to the identity within the algebra so that partial isometries $\omega_r$ can be chosen to satisfy
\be 
\omega_r \omega_r^\dagger=p_r\,,\quad\mathds{1}=\sum_r \omega_r\omega_r^{\dagger}\,,\quad \omega_r^{\dagger}\omega_{r'}=\delta_{r r'} \mathds{1}\,,\quad \omega_r\in\mathcal{A}\;.
\ee
 This allows to define direct sums as
\be 
\rho\simeq \oplus_r \rho_r \,\,\,\,\,\,\,\,\Rightarrow \,\,\,\,\,\,\,\,\rho(a)\equiv \sum_r \omega_r\,\rho_r(a)\,\omega_r^{\dagger},\,\,\,\,\,\,\,\,\,\,\omega_r,a\in\mathcal{A}\;. 
\ee
Then, by construction $\omega_r$ intertwines $\rho_r$ with $\rho$ within $\mathcal{A}$ since $\omega_r\,\rho_r=\rho\,\omega_r $.

Going in the opposite direction, a reducible endomorphism $\rho$ can be written as a sum of irreducible endomorphisms. By definition, for a reducible endomorphism $\rho(\mathcal{A})'\cap \mathcal{A}\neq\mathds{1}$. If we choose a maximal partition of unity by projectors $p_r$ inside $\rho(\mathcal{A})'\cap \mathcal{A}$, then again $p_r=\omega_r\,\omega_r^{\dagger}$, and we can define irreducible endomorphisms as $\rho_r\equiv \omega_r^{\dagger}\,\rho\,\omega_r$. It is then simple to verify that $\rho$ is the direct sum, as defined above, of such irreducible endomorphisms.

Further important structure arises by considering a state for the algebras. In particular, given an inclusion $\mathcal{A}\subset \mathcal{B}$ and a cyclic and separating vector for both algebras $\vert \Omega\rangle$ (that always exists for type III factors), the {\sl canonical endomorphism} \cite{Longo:1989tt} is defined by
\be  \label{canon}
\gamma (\mathcal{B})\equiv j_{\mathcal{A}}j_{\mathcal{B}} (\mathcal{B}) \subset \mathcal{A}\;.
\ee
This uses the modular conjugations, e.g. $j_{\mathcal{B}}(b)\equiv J_{\mathcal{B}} \,b \, J_{\mathcal{B}}$ and similarly for $j_{\mathcal{A}}(a)\equiv J_{\mathcal{A}} \,a \,J_{\mathcal{A}}$, associated with each algebra and the vector $|\Omega\rangle$. This is an endomorphism for $\mathcal{B}$. We can use it to define a (dual) canonical endomorphism in the subalgebra by restricting the action $\rho (\mathcal{A})\equiv \gamma\vert_{\mathcal{A}}\subset \mathcal{A}$. The canonical endomorphism plays a role analogous to the regular representation in group theory. In the symmetry group scenario it exactly decomposes as the regular representation. All canonical endomorphisms (for different vectors) are unitarily equivalent to each other \cite{Longo:1994xe}. If $\mathcal{A}\subset \mathcal{B}$ has  finite index $\lambda$, then the canonical endomorphism can be expressed as a direct sum of all irreducible endomorphisms $\rho_r$ induced by the subfactor, as
\be \label{candec}
\rho \simeq \oplus_r N_r\, \rho_{r} \;,
\ee
where $N_r$ are some integer multiplicities, the identity endomorphism appears exactly once, and where the sum only includes a finite number of terms. The reason is that the index $\lambda$ can be proven to be the dimension of the canonical endomorphism, which is additive in the expected manner, namely $\lambda=d_{\rho}=\sum_r\, N_r\,d_r$, where $d_r$ are the dimensions of the irreducible endomorphisms. For the case of conditional expectations with infinite index the discussion stays  the same but the canonical endomorphism is written as a sum of an infinite number of irreducible sectors, as it is the case for compact Lie symmetry groups.

The previous expression (\ref{candec}) means we can find a set of partial isometries $\omega_r^i\in \mathcal{A}$, $i=1,\cdots, N_r$, satisfying
\be \label{omegpart}
\omega_r^{i\dagger}\omega_{r'}^{i'}=\delta_{ii'}\delta_{rr'}\,, \quad\sum_{r,i}\, \omega_r^i \,\omega_r^{i\dagger}=\mathds{1} \,,\quad \omega_r^i\, \rho_{r}(a) =\rho(a)\,\omega_r^i\,,\quad a\in \mathcal{A}\;,
\ee
and the canonical endomorphism explicitly reads
\be \label{endoexpl}
\rho(a)=\sum_{r,i}\, \omega_{r}^i\,\rho_r(a)\,\omega_{r}^{i\dagger}\;.
\ee
The isometry $\omega$ intertwining the identity representation with the canonical endomorphism is unique (for irreducible subfactors). In the finite index scenario, Ref. \cite{Longo:1994xe} then shows that this structure allows to reconstruct $\mathcal{B}$  from $\mathcal{A}$ and a further isometry $v\in \mathcal{B}$. This isometry $v$ intertwines the identity representation and the canonical endomorphism $\gamma$ of $\mathcal{B}$, namely $vb=\gamma(b)v$. The isometry $v$ creates or destroys the canonical endomorphism. The triple $\lbrace\gamma, v,\omega\rbrace$ is called a ``Q-system'', see \cite{bischoff2015tensor} for a complete introduction and references. In such triple, $\omega$ is seen as an intertwiner between $\gamma$ and $\gamma^2$, and the following relations are satisfied
\bea\label{Qsystem} 
\omega^*\,v &=& \lambda^{-1/2}\mathds{1}=\omega^*\,\gamma (v)\,, \\
\omega\,\omega^* &=&\gamma(\omega^*)\omega\,, \\ \omega\,\omega &=&\gamma(\omega)\omega\;.
\eea
Such data $\lbrace\gamma, v,\omega\rbrace$ completely characterizes the subfactor $\mathcal{A}\subset\mathcal{B}$ and associated Jones ladder \cite{Longo:1994xe}. 

In particular, with such canonical isometry $v$ one can find charged isometries $\psi_r^i\in \mathcal{B}$ for all the irreducible representations $\rho_r$ appearing in the decomposition~(\ref{candec}). More concretely 
\be \label{chargephi}
\psi_r^i\equiv \sqrt{\frac{\lambda}{d_r}}\,\omega_r^{i\,\dagger}\,v\;,
\ee
and it can shown that $N_r\le d_r$ \cite{Rehren:1993yu}. The previous relation implies $\psi_r^i \, a=\rho_r(a)\,\psi_r^i$, namely the isometries intertwine the identity endomorphism with the irreducible ones. Inversely one can also write the canonical isometry as a linear combination of all the others
\be 
v=\sum_{r,i}\,\sqrt{\frac{d_r}{\lambda}}\,\omega_r^i\,\psi_r^i\;,
\ee
which shows $v$ to be a ``master charge'' creating excitations in all irreducible representations. This is an isometry in the sense that $v^\dagger v=\mathds{1}$ and the range is the Jones projection of the dual inclusion, namely $vv^\dagger=e'$, so that $\varepsilon (vv^\dagger)=\mathds{1}/\lambda$. Finally, this algebraic structure can be used to write any operator $b\in\mathcal{B}$ using only operators from the subalgebra $\mathcal{A}$ and the charged isometries:
\be 
b=\sum_{r,i}\,d_r\,\varepsilon(b\,\psi_r^{i\dagger})\,\psi_r^i=\sum_{r,i} a_{i,r}\, \psi_r^i\;, \label{expa}
\ee
which is the desired expression since $a_{i,r}=d_r\,\varepsilon(b\,\psi_r^{i\dagger})\in\mathcal{A}$ gives the unique (operator)-coefficient in the expansion of $b$. For infinite index, although the isometries $\omega_r^i$ and charged isometries $\psi_r^i$ might exist, the master charge operator $v$ will not.

Returning to the inclusion ${\cal A}(R)\subset \hat{\cal A}(R)$, in the finite index scenario, eq. (\ref{expa}) gives the maximal algebra in terms a finite number of non-local (charged) operators with coefficients in the additive algebra. As the operators in (\ref{expa}) close an algebra (namely the larger algebra), the charged operators $\psi_r^i$ close an algebra between themselves with ``OPE coefficients'' in the subalgebra ${\cal A}$. This construction can be seen as a cross product construction. The larger algebra is the cross product of the smaller by the charged isometries. In the case of groups, this is the cross product with the dual of the group. What is interesting is that everything is controlled by the data $\lbrace\gamma, v,\omega\rbrace$ defining the Q-system, since $\gamma(v)$ can be computed from this data and contains the precise OPE of the charged operators \cite{Longo:1994xe}.

\begin{definition} \label{def1} Two regions $R_1, R_2$, $R_1\subset R_2$, are {\sl transportable} to each other when, defining $R=R_2-R_1=R_1'-R_2'$,  we have
 $\hat{{\cal A}}(R_1)\vee {\cal A}(R)=\hat{\cal A}(R_2)$ and, dualy,  $\hat{{\cal A}}(R_2')\vee {\cal A}(R)=\hat{\cal A}(R_1')$.\end{definition}

Transportability is then a self dual notion. It is also transitive, if $R_1\subset R_2$, and $R_2\subset R_3$ are transportable, then $R_1\subset R_3$ is transportable. This follows from strong additivity. The first condition,  $\hat{{\cal A}}(R_1)\vee {\cal A}(R)=\hat{\cal A}(R_2)$, implies all non-local operators of the larger region can be constructed with the ones of the smaller one and additive operators. The second condition, $\hat{{\cal A}}(R_2')\vee {\cal A}(R)=\hat{\cal A}(R_1')$, implies that all non-local operators of the smaller region $R_1$ (that commute with ${\cal A}(R)$ and do not commute with $\hat{\cal A}(R_1')$) are also non local operators of the larger one (because they could not commute with $\hat{{\cal A}}(R_2')$).

The following is essentially  a combination of some propositions in \cite{Longo:1994xe} and \cite{Kawahigashi:1999jz}.

\begin{prop} Under assumptions 1,2,3,4,5  when $R_1\subset R_2$ we have that the following three properties are equivalent to each other and to transportability: 

\label{prop1}

\begin{enumerate}

\item 
The conditional expectation $E_{R_2}$ extends $E_{R_1}$ and $E_{R_1'}$ extends $E_{R_2'}$ (this implies the index is the same for all these regions $\lambda_{R_1}=\lambda_{R_1'}=\lambda_{R_2}=\lambda_{R_2'}$).  

\item The conditional expectation $E_{R_2}$ extends $E_{R_1}$ and there is a vector  cyclic and separating for both $\hat{{\cal A}}(R_1)$ and $\hat{{\cal A}}(R_2)$ and invariant under $E_{R_2}$.   

\item A canonical endomorphism $\gamma_{R_1}$ of $R_1$ extends to a canonical endomorphism  $\gamma_{R_2}$ of $R_2$, where the extension is the identity in the relative commutant $(\hat{\cal A}(R_1))'\cap {\cal A}(R_2)$.

\end{enumerate}

\end{prop}

\noindent {\bf Proof:}  Let us call $R=R_2-R_1=R_1'-R_2'$. By definition $E_{R_2}$ maps $\hat{{\cal A}}(R_2)$ to ${\cal A}(R_2)$ leaving ${\cal A}(R_2)$ invariant. We then have $E_{R_2}({\cal A}(R))={\cal A}(R)$. Then, by the bimodule property of conditional expectations and isotony of maximal algebras, we have $E_{R_2}(\hat{{\cal A}}(R_1))\subseteq ({\cal A}(R))'\cap {\cal A}(R_2)$. By the  assumption $\hat{{\cal A}}(R_2')\vee {\cal A}(R)=\hat{{\cal A}}(R_1')$ we have, taking commutants, $({\cal A}(R))'\cap {\cal A}(R_2)={\cal A}(R_1)$. Then the conditional expectation restricts to the unique one for $R_1$. The same happens dually because we assume $\hat{{\cal A}}(R_1)\vee {\cal A}(R)=\hat{{\cal A}}(R_2)$.  The index is the same for dual conditional expectations and do not increase under restriction. Then it is the same for $R_1,R_2,R_1',R_2'$ (see proposition 5-a of \cite{Kawahigashi:1999jz}). This shows transportability implies (1).

On the other hand, starting from (1), let $e_{R_2}$ be a Jones projection that implements $E_{R_2}$. Then it also implements $E_{R_1}$. It follows that $\hat{\cal A}(R_2')=({\cal A}(R_2))'=e_{R_2}\vee {\cal A}(R_2')$ and  $\hat{\cal A}(R_1')=({\cal A}(R_1))'=e_{R_2}\vee {\cal A}(R_1')$. Since by strong additivity we have ${\cal A}(R_1')={\cal A}(R_2')\vee {\cal A}(R)$ we get $\hat{\cal A}(R_1')={\cal A}(R)\vee\hat{{\cal A}}(R_2')$. Using in the same way the extension of the dual conditional expectations we conclude that (1) implies transportability.

In order to prove (1) implies (2) consider a faithful state $\phi$ on ${\cal A}(R_2)$ whose GNS representation is cyclic (and separating) for both ${\cal A}(R_2)$ and ${\cal A}(R_1)$. This is always possible for type III factors. Then the state $\omega=\phi\circ E_{R_2}$ is faithful for $\hat{{\cal A}}(R_2)$ and its GNS representation is a vector $|\omega\rangle$ cyclic and separating for $\hat{{\cal A}}(R_2)$ and invariant under the conditional expectation, in a Hilbert space ${\cal H}$. It is then also separating for $\hat{{\cal A}}(R_1)$. We have to show that is cyclic for $\hat{{\cal A}}(R_1)$. This follows from the fact that the dual conditional expectation also extends. This implies it is implemented by the same Jones projection $e_{R_1'}$, giving $\hat{{\cal A}}(R_2)={\cal A}(R_2)\vee e_{R_1'}$,  $\hat{{\cal A}}(R_1)={\cal A}(R_1)\vee e_{R_1'}$. But by construction the GNS representation gives that $[{\cal A}(R_2)|\omega\rangle]$ and $[{\cal A}(R_1)|\omega\rangle]$ are the same Hilbert subspaces ${\cal H}_0\subset {\cal H}$. Therefore, if $|\omega\rangle $ is cyclic for $\hat{{\cal A}}(R_2)$ it is cyclic for  $\hat{{\cal A}}(R_1)$. This concludes the proof of (1) implies (2). 

The condition (2) was called a standard net of subfactors in \cite{Longo:1994xe}. That (2) implies (3) is theorem 3.2 of \cite{Longo:1994xe} (does not need finite index nor additivity). 

We finally show that (3) implies (1) under the finite index assumption. First we notice that (3) directly implies that $\rho_{R_1}\equiv \gamma_{R_1}\vert_{\mathcal{A}(R_1)}$ extends to $\rho_{R_2}\equiv \gamma_{R_2}\vert_{\mathcal{A}(R_2)}$ with the constraint that $\rho_{R_2}\vert_{\mathcal{A}(R)}=\mathcal{A}(R)$. This implies that the isometry $\omega_{R_1}\in {\cal A}(R_1)$ intertwining the identity with $\rho_{R_1}$, namely satisfying $\omega_{R_1} a_{R_1}=\rho_{R_1}(a_{R_1})\omega_{R_1}$, is also an intertwiner from the identity sector to  $\rho_{R_2}$. This follows from strong additivity of the additive algebra $\mathcal{A}(R_2)=\mathcal{A}(R)\vee \mathcal{A}(R_1)$ and that
\be 
\omega_{R_1} a_{R_1}a_{R}=\rho_{R_1}(a_{R_1})\omega_{R_1}a_{R}=\rho_{R_1}(a_{R_1})a_{R}\omega_{R_1}=\rho_{R_2}(a_{R_1} a_{R})\omega_{R_1}\,,
\ee
where in the first equality we have used the fact that $\omega_{R_1}$ intertwines the identity with $\rho_{R_1}$, in the second that $\omega_{R_1}\in \mathcal{A}(R_1)$ and therefore commutes with any $a_{R}\in \mathcal{A}(R)$, and in the third the fact that $\rho_{R_2}$ extends $\rho_{R_1}$ being trivial in $ \mathcal{A}(R)$. The fact $\omega_{R_2}=\omega_{R_1}$, and that these isometries induce the conditional expectations as \cite{Longo:1994xe}
\be 
E_{R_1}(\hat{a}(R_1))=\omega_{R_1}^*\,\gamma_{R_1}(\hat{a}(R_1))\omega_{R_1}\,,\,\,\,\,\,\,\,\,E_{R_2}(\hat{a}(R_2))=\omega_{R_2}^*\,\gamma_{R_2}(\hat{a}(R_2))\omega_{R_2}=\omega_{R_1}^*\,\gamma_{R_2}(\hat{a}(R_2))\omega_{R_1}\,,\nonumber
\ee
implies that $E_{R_2}\vert_{\hat{\mathcal{A}}(R_1)}=E_{R_1}$. This is the first condition in (1). To show the dual condition we observe that (3) gives  that $\gamma_{R_2}\vert_{\hat{\mathcal{A}}(R_1)}=\gamma_{R_1}$. This implies that $v_{R_2}$, the isometry intertwining the identity with $\gamma_{R_2}$, also intertwines the identity with $\gamma_{R_1}$, namely $v_{R_2}=v_{R_1}$. This follows since 
\be
v_{R_2} \hat{a}(R_2)=\gamma_{R_2}(\hat{a}(R_2))v_{R_2}\,\,\rightarrow\,\,v_{R_2}\hat{a}(R_1)=\gamma_{R_2}(\hat{a}(R_1))v_{R_2}\,\,\rightarrow\,\,v_{R_2}\hat{a}(R_1)=\gamma_{R_1}(\hat{a}(R_1))v_{R_2}\;,\nonumber
\ee
where in the first arrow we used isotony of maximal algebras. We thus only need to further show that $v_{R_2}\in \hat{\mathcal{A}}(R_1)$. But this follows from the triviality of the extension $\gamma_{R_2}$ in $\mathcal{A}(R)$, which implies that $v_{R_2} a_R=a_R v_{R_2}$ for any $a_R\in \mathcal{A}(R)$. Then, by strong additivity, $v_{R_2}\in ({\cal A}(R))'\cap \hat{\cal A}(R_2)\subset ({\cal A}(R)\cup {\cal A}(R_2'))'=\hat{\cal A}(R_1)$.\footnote{The fact that these two isometries are equal in $R_2$ and $R_1$ means that the full Q-system is transported from one region to the other.} Since $v_{R_2}=v_{R_1}$ we have that 
\be 
\hat{\cal A}(R_2)={\cal A}(R)\vee\hat{{\cal A}}(R_1)\;.
\ee
We now follow the same lines as before when we first derive (1) from this notion of transportability. In particular we know that $E_{R_1'}\vert_{\mathcal{A}(R)}=\mathcal{A}(R)$ and that taking commutants in the previous equation we see ${\cal A}(R_2')=(\mathcal{A}(R))' \cap \mathcal{A}(R_1')$. This implies $E_{R_1'}\vert_{\hat{\cal A}(R_2')}\subset (\mathcal{A}(R))' \cap \mathcal{A}(R_1')={\cal A}(R_2')$. $\square$

The condition (2) is the starting point of \cite{Longo:1994xe}, where this property is called a {\sl standard net of subfactors} for the net (here of two elements) of subfactor inclusions corresponding to $R_1$ and $R_2$.    Strong additivity, which was not assumed in \cite{Longo:1994xe}, allows to prove the equivalence between the three conditions and with our definition of transportability, that emphasizes it is a self-dual notion. Note the definition of transportability does not require finite index and could in principle be applied more generally. The conditions (2) and (3) are not stated dually here, but as the proposition shows, they also imply, under the finite index assumption, the dual statements.  The last condition (3), derived from (2) \cite{Longo:1994xe},  gives the deepest meaning to transportability. It implies that the sector theory coming from the subfactor inclusion is isomorphic for $R_1$ and $R_2$. It also implies that non local operators for $R_1$ are non local for $R_2$, and the non local operator expansion of the algebra $R_2$ (in the form of eq. \ref{expa}) can be chosen with exactly the same non local operators as an expansion for $R_1$. Equivalently, the full Q-system $\lbrace\gamma ,v,\omega\rbrace$ transports. The same isomorphism between sectors occur for the dual regions $R_1',R_2'$, with the same index, but of course the dual theories of sectors on the additive algebras  (say for $R_1$ and $R_1'$) will in general not be isomorphic to each other. 

\begin{definition} \label{def2} We say that two regions $R_1\subset R_2$ are {\sl topological deformations} of each other if $R_2-R_1$ is a  (topological) ball and $R_1$ and $R_2$ have the same topology. In this case the same happens for the complements $R_2'\subset R_1'$.\end{definition} 

\bigskip

\begin{assumption}\label{as:9}
Topological transportability. Two regions $R_1\subset R_2$ which are topological deformations are transportable.  
 \end{assumption}
 
In the context of the DHR theorem, see below, a similar form of transportability (restricted to deformations of ball sectors) is usually assumed. In two dimensions, for multi-interval sectors, transportability was shown from the other assumptions (including Haag duality for intervals and the split property) in \cite{Kawahigashi:1999jz}. In $d=2$ CFT's it also follows from the diffeomorphism transformations induced by the stress tensor. It is physically natural to conjecture that the existence of a stress tensor should allow to show topological transportability in higher dimensions. 

Inspired by the idea of local deformation operators, presumably produced by the stress tensor, we can abstract the following two ideas that imply topological transportability.  We write $A\Subset B$ when $A\subset B$ and the boundaries are separated by a finite distance.

\begin{definition} A QFT has {\sl local transporters} if for any two (topological) balls $B_1, B_2$, such that $B_1 \cap B_2$ and $B_2-B_1$ are topological balls, and any  region $R$ such $B_1\Subset R$, there is a unitary $u\in {\cal A}(R)$ such that $u\, {\cal A}(B_2)\, u^*\subset {\cal A}(B_2-B_1)$. \end{definition} 

 Notice that this local transporter cannot be produced directly by standard localized translations constructed with the split property \cite{Doplicher:1984zz,buchholz1986noether}. The latter can be defined to operate as translations in $B_1$ and do nothing outside $R$. But their action is non local in $R-B_1$. By strong additivity  this property extends to arbitrary regions $R_1,R_2$, instead of topological balls $B_1,B_2$, provided their intersection is a topological ball, and that  $R_2$ has the same topology as $R_2-R_1$.

 Another natural assumption is the following continuity property.

\begin{definition} A QFT satisfies {\sl continuous isotony}   if for any nested sequence $R_i$, $i \in \mathbb{N}$, $R_{i+1}\subset R_{i}$, where all $R_i$ and  $R=\wedge_i R_i$ have the same topology, we have $\cap_i {\cal A}(R_i)={\cal A}(R)$.\end{definition}

See  \cite{horuzhy2012introduction} for related properties. The topological condition is necessary to warrant there is no change of topology from the sequence to the limit. Otherwise there are counterexamples. If there are topological changes in general we have rather that $\wedge_i {\cal A}(R_i)\subseteq \hat{{\cal A}}(R)$.  

\begin{prop} Continuous isotony, strong additivity, and the existence of local transporters imply topological transportability. \end{prop}

\noindent {\bf Proof: } Let $R_1\subset R_2$ be two regions with the same topology and $R=R_2-R_1$ be a topological ball. We will show $\hat{\cal A}(R_2')\vee {\cal A}(R)=\hat{\cal A}(R_1')$. Define $R_1^\epsilon$ as the set of points at a distance at most $\epsilon$ for $R_1$. Using a local transporter $u_\epsilon \in {\cal A}(R_1')$ we obtain
$
u_\epsilon \, {\cal A}(R_2)\, u_{\epsilon}^\dagger\subseteq {\cal A}(R_2)\cap {\cal A}(R_1^\epsilon)
$. 
By continuous isotony, taking a decreasing sequence $\epsilon_i\rightarrow 0$, we get $\cap_i u_{\epsilon_i} \, {\cal A}(R_2)\, u_{\epsilon_i}^\dagger= {\cal A}(R_1)$. 
Taking commutants $\vee_i u_{\epsilon_i} \, \hat{\cal A}(R_2')\, u_{\epsilon_i}^\dagger= \hat{\cal A}(R_1')$. Since $u_{\epsilon_i}\in {\cal A}(R_1')={\cal A}(R)\vee {\cal A}(R_2')$ the result follows. The other equation $\hat{\cal A}(R_1)\vee {\cal A}(R)=\hat{\cal A}(R_2)$ follows in the same  way. $\square$  

As a partial converse, it is not difficult to show that topological transportability plus strong additivity imply continuous isotony.

\bigskip 

\begin{assumption}\label{as:8}
  The split property. For two disjoint  regions $R_1, R_2$, separated a finite distance, where one of them is bounded,  there exist a type I factor ${\cal N}$ such that  ${\cal A}(R_1)\subset {\cal N}$, $\hat{\cal A}(R_2)\subset {\cal N}'$.
\end{assumption}

This is a UV regularity condition. It is related to the existence of a partition function (absence of Hagedorn temperature), or that the number of degrees of freedom in the theory is not increasing too fast with energy. The usual proof of the split property relies on a nuclearity assumption for the set of vectors $e^{-\beta H} A|0\rangle$, where $H$ is the Hamiltonian, $|0\rangle$ the vacuum state, and $A\in {\cal A}(R_1)$, $\norm{A}\le 1$ 
\cite{buchholz1986causal,buchholz1990nuclear}. The spatial separation between $R_1$ and $R_2$ implies a finite time in which the time evolution of ${\cal A}(R_1)$ still commutes with ${\cal A}(R_2)$ and this implies the split property under the nuclearity condition. The usual presentation of the split property is for arbitrary double cones, but the nuclearity condition for arbitrary double cones implies the one for arbitrary bounded regions, and extends the split property to this more general case, including regions with general topologies.

As ${\cal B}({\cal H})\simeq {\cal N}\otimes {\cal N}'$ the split property implies an isomorphism  ${\cal A}(R_1)\vee \hat{\cal A}(R_2)\simeq {\cal A}(R_1)\otimes \hat{\cal A}(R_2)$. This tensor product structure for the commuting algebras of separated regions will be useful to establish several results. In particular, the distributivity of the operations $\vee\,, \cap$ for certain algebras. 

  A consequence of the split property is that for bounded regions or their complements, without using topological transportability, it is possible to show that non local operators can be localized arbitrarily near the boundary of the region: 

\begin{prop} The split property and strong additivity imply that if  $R_1\Subset R$, where both $R,R_1$ are bounded or the complement of bounded regions, then $\hat{\cal A}(R)={\cal A}(R_1)\vee \hat{{\cal A}}(R-R_1)$. 
\end{prop}

\noindent {\bf Proof: }  We can take a split ${\cal B}({\cal H})={\cal N}\otimes {\cal N}'$, where ${\cal N}$ contains ${\cal A}(R_1)$ and is contained in ${\cal A}(R)$. It is ${\cal N}\subset \hat{{\cal A}}(R)$. By the property of the distributivity of the intersection of algebras when there is a tensor product of factors, and one of the factors is included in the algebra \cite{ge1996tensor,zacharias2001splitting},  
$\hat{{\cal A}}(R)=\hat{{\cal A}}(R)\cap ({\cal N}\vee {\cal N}')= {\cal N}\vee (\hat{{\cal A}}(R)\cap {\cal N}')$. By strong additivity this cannot be larger than ${\cal A}(R_1)\vee {\cal A}(R-R_1)\vee (\hat{{\cal A}}(R)\cap {\cal N}')$, but also cannot be smaller, because all operators added in this second expression belong to $\hat{{\cal A}}(R)$. Finally, by the same reason we can replace this by $\hat{{\cal A}}(R)={\cal A}(R_1)\vee \hat{{\cal A}}(R-R_1)$ $\square$

This proposition indicates that non local operators in $\hat{{\cal A}}(R)$ have representatives in $\hat{\cal A}(R-R_1)$, but the converse will in general not be true, that is, some of the non local operators of $\hat{\cal A}(R-R_1)$ may be additive in $R$.

\begin{assumption}\label{as:3} Essential duality. For  two disjoint (topological) balls $B_1,B_2$ 
we have  
$\hat{\mathcal{A}}(B_1)\subset (\hat{\mathcal{A}}(B_2))'$.
\end{assumption}

This establishes the mutual locality of ``charged'' operators that are in $\hat{\cal A}(B)$ but not in ${\cal A}(B)$, when their support is disjoint. 
Because of the assumption of topological transportability, essential duality follows if Haag duality holds for any single region. The Bisognano-Wichmann result gives Haag duality for Wightman theories for the Rindler wedge (the region on one side of a plane in $E$ in our current representation) \cite{Bisognano:1975ih}. This  establishes essential duality in this case. 

Essential duality is a weakening of the quite generally expected property of Haag duality for balls, or more generally, for topological balls, as appropriate under the assumption of topological transportability:
\begin{assumptionp}{\ref*{as:3}$'$}\label{as:3'} Haag duality for topological balls, 
$\mathcal{A}(B)=(\mathcal{A}(B'))'\equiv\hat{\mathcal{A}}(B)$.
\end{assumptionp}

 Haag duality for balls  holds for example for conformal field theories \cite{brunetti1993modular}. Failure of Haag duality for balls  is in general associated to spontaneously broken global symmetries \cite{doplicher1990there,roberts1974spontaneously}.  This case will also turn out to be relevant for the discussion of sectors associated to non trivial topologies in dimensions $d\ge 3$. In any of these cases, however, the relevant assumption of essential duality is expected to hold. This allows to cure the failure of Haag duality for balls by the construction of the dual net. We will review this construction in section \ref{SecIII.I}. 

The physical content of Haag duality for balls is that any operator that by causality can be considered localized in a bounded region $R$ ultimately belongs to the additive algebra generated by local degrees of freedom in a ball containing $R$. Then, in such QFTs there are no genuinely non-local operators that cannot be produced by local degrees of freedom. The notion of non local operator is necessarily relative to the chosen localization region.

 Eventually (especially in sections \ref{SecV} and \ref{SecVI}) we are going to use a property that holds in concrete examples, and is tightly related to transportability, the split property and some continuity conditions, though we were not able to prove it from the previous assumptions.  This concerns the distributivity of $\cap$ and $\vee$ on algebras. In general there is no distributivity of these operations among algebras, as would be the case for a Boolean lattice. Strong additivity implies the distributivity ${\cal A}(A)\cap ({\cal A} (B)\vee {\cal A}(C))=({\cal A}(A)\cap {\cal A}(B))\vee ({\cal A}(A)\cap {\cal A}(C))$  when $A\subseteq (B\vee C)$.  
We also have distributivity when $B\Subset A$, and $B,C$ are separated by a non zero distance, because of the split property. The next assumption is a generalization of this idea.

\begin{assumption}  Weak modularity. $
B\Subset A\implies {\cal A}(A)\cap ({\cal A}(B)\vee {\cal A}(C))= {\cal A}(B)\vee ({\cal A}(A) \cap {\cal A}(C))$. \label{weak} 
\end{assumption}

 The name comes from the modular property of lattice theory, in which it is only required that $B\subseteq A$ rather than the condition $B\Subset A$ used here. There are counterexamples to the modular property without this stronger condition. Notice that taking commutants the relation is equivalent to the same relation for the maximal algebras. 

Weak modularity can also be thought as a generalization of topological transportability, and we will use it to prove transportability in some special cases. 

\begin{prop} Weak modularity, strong additivity, and topological transportability for topological balls imply topological transportability. \label{pp22}\end{prop}

{\sl Proof:} Let $R_1\subset R_2$ be two regions with the same topology and $R=R_2-R_1$ be a topological ball. Then there is a region $B\subset R_1$ with the topology of a ball, and such that $R\cup B$ is also a topological ball and $R_1-B$ is separated by a non zero distance from $R$. Then, by strong additivity,  ${\cal A}(R')\cap {\cal A}(R_2)={\cal A}(R')\cap ({\cal A}(R\cup B)\vee {\cal A}(R_1-B))$, and by weak modularity this equals $({\cal A}(R')\cap {\cal A}(R\cup B))\vee {\cal A}(R_1-B)$.
 The intersection ${\cal A}(R')\cap {\cal A}(R\cup B)\subseteq \hat{\cal A}(R')\cap {\cal A}(R\cup B)={\cal A}(B)$ by transportability of topological balls. Then ${\cal A}(R')\cap {\cal A}(R\cup B)={\cal A}(B)$. Then by strong additivity ${\cal A}(R')\cap {\cal A}(R_2)= {\cal A}(R_1)$, what is equivalent to $\hat{{\cal A}}(R_2')\vee \hat{\cal A}(R)=\hat{{\cal A}}(R_1')$. Here $\hat{\cal A}(R)$ can be replaced by ${\cal A}(R)$ by transportability of topological balls, giving $\hat{{\cal A}}(R_2')\vee {\cal A}(R)=\hat{{\cal A}}(R_1')$. The dual relation holds in the same way. $\square $

\section{The DHR reconstruction theorem and $\pi_0$-completeness}\label{SecIII}

In this section we revisit the DHR theorem  \cite{Doplicher:1971wk,Doplicher:1973at,doplicher1990there} from various different perspectives. The main objective is to relate the theorem to, or even formulate it as, the ``cure'' of certain HDV sectors in regions with non-trivial homotopy group $\pi_0$ (disjoint regions) or their dual. Curing such HDV leads to a parent QFT which is ``$\pi_0$-complete'', i.e. a QFT with no HDV specifically due to disjoint regions \cite{Casini:2021zgr}. For the time being, we will focus on HDV for regions with non trivial $\pi_0$ group. The rational is to focus on one type of sectors at a time. In this vein, a region composed of two rings has non-trivial $\pi_0$, but the $\pi_0$-complete QFT might display HDV in the two ring region due to the existence of Wilson loops \cite{Casini:2020rgj}. So more precisely, a $\pi_0$-complete QFT is one satisfying Haag duality for regions composed on an arbitrary number of disconnected (topological)-balls, or their duals. The notion of $\pi_0$-completeness will prove to be of key importance throughout the article.

In the way, in this section we describe a new derivation of the group-like nature of $0$-form symmetries in $\pi_0$-complete QFTs. Such QFTs are the standard starting point in the high energy/string theory community, where symmetries act faithfully on the local algebra, which in turn requires local charged operators in all representations, see ref. \cite{Harlow:2025cqc} for a recent discussion.\footnote{We note that $\pi_0$-complete QFTs satisfy the notion of ``disjoint additivity'' in \cite{Harlow:2025cqc}.} The new derivation we present below shows more directly that all selection rules associated with $0$-form symmetries in $D>2$ can be ultimately derived from a group-like symmetry. The main assumption to derive this result is that there is a non-trivial subalgebra invariant under the symmetry, and this is ensured e.g. when the theory has a local stress energy tensor.

Before curing HDV for regions formed by an arbitrary number of disconnected (topological)-balls it is important to cure HDV for one ball itself. We turn to this now.

\subsection{Completing HDV for single balls. Compactification of the space.}\label{SecIII.I}

If there is essential duality but not Haag duality for balls, Haag duality for balls can be recovered by extending the net to the {\sl dual net} \cite{roberts1974spontaneously,buchholz1992new}. For a ball, the algebra of the dual net is defined as ${\cal A}^d(B)=\hat{\cal A}(B)$. For more general regions we define it by
\be
\label{add}
{\cal A}^d(R)=\vee_{B\subseteq R} \,\hat{\cal A}(B)\,,
\ee
where the index runs over all balls $B$ included in $R$. Isotony makes this definition self-consistent. 

For a bounded ball $B$ we have $\hat{{\cal A}}^d(B')=({\cal A}^d(B))'=(\hat{\cal A}(B))'={\cal A}(B')$. Since ${\cal A}^d(B')\subset \hat{{\cal A}}^d(B')$ and ${\cal A}(B')\subset{\cal A}^d(B')$, then ${\cal A}(B')={\cal A}^d(B')$. Therefore $\hat{\cal A}^d(B')={\cal A}^d(B')$, and ${\cal A}^d(B)=\hat{\cal A}^d(B)$. Hence there is Haag duality for balls. Note this implies, by topological transportability, that for unbounded single component regions $R_0$ the algebra ${\cal A}(R_0)$ already contains ${\cal A}^d(B)=\hat{\cal A}(B)$ for any $B\subset R_0$. In this sense, the infinity can be thought as a ``charged point'' that transfers its charge to all regions that contain it.  
More generally, decomposing a general region in connected components as $R= \cup_{i=1}^k R_i\cup R_0$, where $R_0$ is a (possible) single component unbounded region, and $R_i$ are disjoint single component bounded regions, the algebra ${\cal A}^d(R)=\vee_{i=1}^k {\cal A}^d(B_i)\vee {\cal A}(R)$, where $B_i$ are any set of balls such that $B_i\subset R_i$, $i=1,\cdots,k$. This again follows by topological transportability. Dually, for the maximal algebras we have $\hat{\cal A}^d(R)=\hat{\cal A}(R)\cap_i {\cal A}^d(B_i)'$, where $B_i$ are balls contained in each of the bounded connected components of $R'$.

In fact, we have the following:

\begin{prop} \label{dualnet} For a net ${\cal A}(R)$ satisfying assumptions 1-9 the dual net ${\cal A}^d(R)$ satisfies assumptions 1-8',9.
\end{prop}
\noindent 
{\sl Proof:} Causality for the dual net ${\cal A}^d$ follows directly from essential duality for ${\cal A}$. Strong additivity for ${\cal A}^d$ follows from strong additivity and topological transportability for ${\cal A}$. Then, as happens for the original net, the algebras ${\cal A}^d$ are factors, and ${\cal A}(R)\subseteq {\cal A}^d(R)\subseteq \hat{\cal A}^d(R)\subseteq \hat{\cal A}(R)$ form irreducible subfactors. The conditional expectation $E_R$ restricts to the unique one from  $\hat{\cal A}^d(R)$ to ${\cal A}(R)$, and it is of finite index. The dual conditional expectation from $({\cal A}(R))'$ to $(\hat{\cal A}^d(R))'$ restricts to the unique finite index conditional expectation from  $({\cal A}^d(R))'$  to $(\hat{\cal A}^d(R))'$. Hence all inclusions are finite index. 
Finite index implies all these algebras are of type III \cite{loi1992theory}.

To prove topological transportability of the net ${\cal A}^d$ consider a topological deformation $R_1\subset R_2$. The algebra $\hat{\cal A}^d(R_2)$ is generated by a finite number of operators  $\tilde{\psi}^{(2)}_i$ of $\hat{\cal A}(R_2)$, combined with ${\cal A}(R_2)$. These operators  commute with ${\cal A}^d(B_j)$ for balls $B_j$ in the connected components of $R_2'$. 
By transportability of the net ${\cal A}$ there are unitaries $u_i$  in ${\cal A}(R_2)$ such that $u_i\, \tilde{\psi}_i^{(2)}= \tilde{\psi}^{(1)}_i \in \hat{\cal A}(R_1)$. These still commute with ${\cal A}^d(B_j)$, so they belong to $\hat{\cal A}^d(R_1)$. Therefore $\hat{\cal A}^d(R_2)\subseteq \hat{\cal A}^d(R_1)\vee {\cal A}(R_2)$ and then $\hat{\cal A}^d(R_2)= \hat{\cal A}^d(R_1)\vee {\cal A}^d(R)$. The dual equation is proved similarly.    

Under the finite index assumption, the split property between separated regions $R_1,R_2$ in the original net extends to a split between ${\cal A}^d(R_1)$ and ${\cal A}^d(R_2)$. It will be enough to proof the split between two arbitrary separated single component regions $R_1, R_2$, where $R_2$ is unbounded, since for multicomponent regions we can use nested splits constructed using these two component splits. Indeed, $R_1,R_2$ can be taken to be the regions lying on each side of an arbitrary closed bounded surface, separated from this surface an arbitrary small distance, and we can use this type of splits to partition algebras of arbitrary regions into the tensor product of their connected components.  
The proof essentially copies   
  \cite{longo2003conformal}, lemma 22. 
We have to show that the conditional expectation  $E^{(12)}:{\cal A}^d(R_1)\vee {\cal A}^d(R_2)\rightarrow {\cal A}(R_1)\vee {\cal A}(R_2)$ 
  satisfies $E^{(12)}(a_1 a_2)=E^{(12)}(a_1)E^{(12)}(a_2)$ for $a_1\in {\cal A}^d(R_1), a_2 \in {\cal A}^d(R_2)$. 
   Then, we can compose a normal product state of ${\cal A}(R_1)\vee {\cal A}(R_2)$ with $E^{(12)}$ to obtain a normal product state of ${\cal A}^d(R_1)\vee {\cal A}^d(R_2)$. The existence of these normal product states implies the split property \cite{d1983interpolation}. 
  To show the required property of $E^{(12)}$ is sufficient to note that 
   ${\cal A}^d(R_2)={\cal A}(R_2)$, because $R_2$ is unbounded and single component. Then by the properties of a conditional expectation $E^{(12)}(a_1 a_2)=E^{(12)}(a_1) a_2=E^{(12)}(a_1)E^{(12)}(a_2)$. The same argument shows the split between ${\cal A}^d(R_1)$ and $\hat{\cal A}^d(R_2)$.
   
  To prove the weak modularity (\ref{weak}) we note the right hand side of this equation is always trivially included in the left hand side. Then we have to show that ${\cal A}^d(A)\cap {\cal A}^d(B\cup C)$ is included in ${\cal A}^d(B)\vee ({\cal A}^d(A)\cap {\cal A}^d(C))$. For a ball ${\cal A}^d(B)={\cal A}(B)\vee\{\psi_i\}$ for some finite number of ``charged'' operators $\psi_i$. These are the ones that do not commute with $\hat{\cal A}(B_1')$ for any $B_1\supset B$. 
  Now for any two regions ${\cal A}^d(R_1)\cap {\cal A}^d(R_2)$ is included in $\hat{\cal A}^d(R_1\cap R_2)$ and this last algebra is formed by ${\cal A}^d(R_1\cap R_2)$ plus some non local operators $\tilde{\psi}_i$. We can choose these non local operators uncharged by adequately combining them with the operators $\psi_i$ of balls in the region. An uncharged operator in ${\cal A}^d(R_1)\cap {\cal A}^d(R_2)$ is uncharged in both algebras, and then belongs to the intersection ${\cal A}(R_1)\cap {\cal A}(R_2)$. Therefore  ${\cal A}^d(R_1)\cap {\cal A}^d(R_2)=({\cal A}(R_1)\cap {\cal A}(R_2))\vee {\cal A}^d(B_i)$, for balls $B_i$ in the connected components of $R_1\cap R_2$. Applying this to ${\cal A}^d(A)\cap {\cal A}^d(B\cup C)$, and using weak modularity for the net ${\cal A}$ (that implies ${\cal A}(A)\cap {\cal A}(B\cup C)\subseteq {\cal A}(B)\vee ({\cal A}(A)\cap {\cal A}(C))$) it follows that this algebra is included in ${\cal A}^d(B)\vee ({\cal A}^d(A)\cap {\cal A}^d(C))$.
 $\square$

In conclusion, the dual net construction allows to canonically extend the additive net in such a way to cure the HDV for single balls. 

Let us define an unbounded (topological) ball $B'$ as the complement of a bounded one $B$. Compactifying $\mathbb{R}^d$ to a sphere $\mathbb{S}^d$ with a marked point $i$ at infinity, an unbounded ball is one in the sphere that contains the point $i$ in the interior. Similarly, for any region $R$ we say it is unbounded or bounded if it contains or not the point $i$ in the sphere representation of the space. In this paper we will restrict attention to regions that are bounded, or their unbounded complements. That is, we will not be interested in regions with boundary containing the point $i$ on the sphere. Although we will not consider such regions (and associated HDV sectors), we remark that in the present framework, the so-called Buchholz-Fredehangen (BF) selection criteria \cite{Buchholz:1981fj}, dealing with representations of the observable algebra that are unitarily equivalent to the vacuum outside a cone that extends to infinity, give place to HDV associated to unbounded conical regions.  Upon compactification of the space $E$, a cone extending to infinity becomes region with a corner representing the point at infinity. Then, BF sectors can be analized with similar tools, but applied to these singular regions. Most of the concepts, definitions and results developed above and below then apply as well to these cases, although we will not consider them here.

We note the assumptions, except for topological transportability and essential duality, make no difference between bounded and unbounded regions. When there is no Haag duality for balls, it is immediate that there is no transportability between bounded and unbounded balls (a bounded ball contains ``charges'' while an unbounded one contain ``twists''). 
For the net ${\cal A}^d$ that obeys Haag duality for balls, on the contrary, there is transportability between balls, bounded and unbounded. This follows directly from Haag duality and strong additivity. For this net, 
the property of topological transportability gets enhanced to deformations that are unbounded while preserving the topology in $\mathbb{S}^d$. That is, we can use deformation regions $R=R_2-R_1$ in definition (\ref{def2}) and assumption (\ref{as:9}) which have the topology of the complement of a bounded ball:
\begin{prop}
Haag duality for balls, strong additivity, weak modularity, and topological transportability imply topological transportability for deformations by unbounded balls (balls that contain the infinity in $\mathbb{S}^d$). 
\end{prop}

\noindent {\sl Proof:} The proof follows from weak modularity, and is essentially the same as the one of proposition (\ref{pp22}), where we have to use Haag duality $\hat{\cal A}(R)={\cal A}(R)$ for the unbounded ball deformation $R$.  $\square$

Notice this is the only place where we will use weak modularity until section \ref{SecV}. Of course we could  avoid the use of weak modularity here by simply postulating transportability for unbounded regions.  Then we have

\begin{prop} 
A net (such as ${\cal A}^d$) that satisfies assumptions 1-8',9 can be considered a net with the same assumptions on $E=\mathbb{S}^d$, where no distinction is made between bounded and unbounded regions. 
\end{prop}

\subsection{HDV for disjoint balls and DHR endomorphisms}

There exists a further net extension that is able to cure HDV for several disjoint balls. Equivalently, following the nomenclature of \cite{Casini:2021zgr}, this achieves a $\pi_0$-complete net of algebras. In the present setup, this is the result of the DHR theorem \cite{Doplicher:1971wk,Doplicher:1973at,doplicher1990there}. Roughly, the output of this theorem is that such a completion always exists for $D>2$, it is unique, the statistics of charged operators are necessarily bosonic or fermionic, and in such completion the symmetry originates in a compact group. This will be the subject of the following section. 
The objective of this section is to connect a key element in the theorem, the DHR endomorphisms, to the violation of Haag duality in multiple disjoint balls. Much of the following steams from several theorems and lemmas of a seminal article by Kawahigashi, Longo and Muger \cite{Kawahigashi:1999jz}. The setup of this last article is for $D=2$ theories but the relevant results, that we briefly review below, are  generalizable to higher dimensions given our previous assumptions.

Given a net $\mathcal{A}$ that has HDV in topological balls, we showed in the previous section that we can cure it by going to the dual net  ${\cal A}^d$. Then, without loss of generality, we assume in this section that our net satisfies Haag duality for balls. Now consider the situation in which there is HDV in a region $R=B_1\cup B_2$ which is the disjoint union of two balls, i.e. we have an inclusion
\be 
\mathcal{A}(R)\subset \hat{\mathcal{A}}(R)\;.
\ee
The fact this is an strict inclusion means the canonical endomorphism $\gamma_R$ of the maximal algebra $\hat{\mathcal{A}}(R)$ and its restriction to the additive part $\rho_R\equiv \gamma_R\vert_{\mathcal{A}(R)}$, see definitions \eqref{canon} and \eqref{candec}, are non-trivial. Equivalently, these endomorphisms are reducible as sectors of their respective algebras. In particular we have
\be 
\rho_R\simeq \oplus_s N_s\, \rho_{s}\;,
\ee
for some irreducible sectors $\rho_{s}$ of $\mathcal{A}(R)$ that appear with multiplicities $N_s$. For conformal nets in one dimension, the main result of Ref. \cite{Kawahigashi:1999jz} was to unravel the nature of such direct sum. Under the previous assumptions, the proof generalizes to higher dimensions in a quite direct way.

We start by reviewing some standard definitions. Given the net ${\cal A}$ in $E=\mathbb{R}^d$, the {\sl quasilocal algebra} is the $C^*$ algebra ${\cal U}=\overline{\cup_{B} {\cal A}(B)}$, where the union is over all (bounded) balls $B$ and the closure is in the norm topology. In the same way a $C^*$ algebra ${\cal U}(R)$ can be defined for any unbounded region $R$ as the inductive limit of the union of ball algebras included in $R$. 
A {\sl localizable representation} of the net is a representation $\pi$ of ${\cal U}$ in a separable Hilbert space such that $\pi\vert_{{\cal U}(E-B)}$ is unitarily equivalent to $id\vert_{{\cal U}(E-B)}$ for some ball $B$.  
Physically, localized representations are representations induced by charged states that can be localized in a ball. Inequivalent non trivial irreducible representations represent different superselection sectors. This defines a class of representations/superselection sectors of direct interest. 

An endomorphism $\sigma$ of ${\cal U}$ is {\sl localized} if there is a ball $B$ such that 
\be
\sigma (\mathcal{A}(B)) \subseteq \mathcal{A}(B)\;,
\ee
and
\be
\sigma (x) = x\,,\,\,\, x\in {\cal A}(B_1)\,,\quad \forall\,B_1\subset B',
\ee
i.e., it acts as the identity in the complement of the ball. 
 Two localized endomorphisms $\sigma_1,\sigma_2$ are equivalent, $\sigma_1\simeq \sigma_2$, if 
\begin{equation}
\sigma_1 = \sigma_u\,\circ \sigma_2\,,\quad \sigma_u(x)=u\,x\,u^* \,,\quad u\,u^*=1\,, \,\,u\in {\cal U}\,. \label{sec}
\end{equation}
From Haag duality for balls, the previous equation tells us that $u$ belongs to a ball containing the localization of both $\sigma_1$ and $\sigma_2$. The equivalence classes of unitarily equivalent  localized  endomorphisms as in (\ref{sec}) are called sectors.
A simple argument \cite{Doplicher:1971wk}, see Sec 5 of \cite{haag2012local} for a review, shows that localizable  representations are unitarily equivalent to localized  endomorphisms
\be 
\pi\simeq  \sigma\;, 
\ee
where in the right hand side the endomorphism is assumed to be applied within the vacuum representation. Under the assumption of Haag duality for balls, there is in fact a one to one correspondence between localizable  representations and sectors.

Given two localized endomorphisms $\sigma_1,\sigma_2$ the composition $\sigma_1 \sigma_2$ and the direct sum $\sigma_1 \oplus \sigma_2$ are localized endomorphisms (we are assuming the local algebras are type III so direct sums can be locally defined as explained in the previous section).  A localized endomorphism is irreducible when $\sigma ({\cal A}(B))'\cap {\cal A}(B)=\mathbb{C}$. In this case it cannot be decomposed into non trivial direct sums, otherwise it can always be decomposed.  
An irreducible endomorphism $\sigma$ localized in $B$ is of finite dimension if $\sigma ({\cal A}(B))\subseteq {\cal A}(B)$ is of finite index $d_\sigma^2$, where $d_\sigma$ is called the dimension of $\sigma$ \cite{Longo:1989tt}. Only finite dimension localized endomorphisms will play a role in what follows and we will restrict attention to this case. Indeed, very general considerations for relativistic theories imply finite dimensions \cite{haag2012local,guido1992relativistic}. 
For an irreducible localized endomorphism (of finite dimension) $\sigma$ there exist a {\sl conjugate} localized  irreducible endomorphism  $\bar{\sigma}$ for which
\be 
\sigma\bar\sigma \simeq \imath \oplus \cdots\;,
\ee
where $\imath$ stands for the identity representation. The composition of an irreducible  sector and its conjugate contains the identity exactly once. 

The localized endomorphism $\sigma$ is 
{\sl transportable} if for any ball $\tilde{B}$ there is a  equivalent endomorphism $\tilde{\sigma}$ localized in $\tilde{B}$. A localized and transportable endomorphism of finite dimension is called a {\sl DHR endomorphism}. We note that if there is translation symmetry, transportability of localized endomorphisms between translated balls follow from the split property. The transporter can be taken as a local implementation of translations \cite{buchholz1992new}. We will not need this result here.

Given these preliminaries, consider $R=B_1\cup B_2$ a region composed by two disjoint separated balls, and two irreducible localized  endomorphisms $\nu$ and $\sigma$ localized respectively in $B_1$ and $B_2$. The composition $\nu\sigma$ naturally restricts to  an endomorphism of $\mathcal{A}(R)$, that is irreducible by the split property.  From theorem $9$ in Ref. \cite{Kawahigashi:1999jz} we have (using Haag duality for balls, finite index and strong additivity)

\begin{prop}\label{nunubar}
The endomorphism $\nu\sigma$ belongs to the dual canonical endomorphism $\rho$, i.e. $\nu\sigma\prec \rho$, if and only if $\nu\simeq \bar{\sigma}$. In this case $\nu\sigma\prec \rho$ with multiplicity one.
\end{prop}

{\sl Proof:} See theorem $9$ in \cite{Kawahigashi:1999jz}. We sketch the proof here. By definition of the dual canonical endomorphism, an irreducible endomorphism of $\mathcal{A}(R)$  appears in the decomposition of $\rho$ if and only if it is intertwined to the identity with an intertwiner in the maximal algebra $\hat{\mathcal{A}}(R)$ \cite{Longo:1994xe}. So if $\nu\sigma\prec \rho$ then we have an isometry $v$ in $\hat{\mathcal{A}}(R)$ intertwining the identity and $\nu\sigma$. The extension from the endomorphisms of the localized subfactor in $R$ to the ones in the net can be accomplished via strong additivity.
Since $\hat{\mathcal{A}}(R)\subset \mathcal{A}(B)$ for any ball $B\supset R$, this is only possible if $\nu\simeq \bar{\sigma}$. Reversely, if $\nu\simeq\bar{\sigma}$, we have an isometry $v$ in a ball containing $R$ intertwining the identity and $\nu\sigma$. But from the intertwining definition of this isometry, it necessarily belongs to $\hat{\mathcal{A}}(R)$. Furthermore, given that $\nu$ is irreducible, there is only one such isometry, so $\sigma\bar\sigma$ appears with multiplicity one. $\square$

Finite dimension of $\sigma$ is also immediate from the finite index assumption. This proposition shows that if $\sigma_r$ is a system of DHR irreducible  endomorphisms, then $\bigoplus \sigma_r\bar{\sigma}_r\prec \rho$. Therefore, if there are non trivial DHR sectors there is a HDV for $R=B_1\cup B_2$.  

Indeed, all irreducible sectors appearing in $\rho$ are in fact of the form $\sigma_r\bar{\sigma}_r$ for irreducible DHR endomorphism $\sigma_r$. This goes in two steps. The first is a slight reformulation of corollary 14 and Lemma $27$ in Ref. \cite{Kawahigashi:1999jz}, which are general already and we write as
\begin{prop} \label{prodirr}
Let $\mathcal{A}_1$ and $\mathcal{A}_2$ be nets satisfying assumptions $1-5,7,8'$.  
If $\sigma$ is an irreducible localized endomorphism of $\mathcal{A}_1\otimes \mathcal{A}_2$ transportable between a ball $B$ and a disjoint ball $\tilde{B}$ then
\be
\sigma\simeq \sigma_1 \otimes \sigma_2 \;, \label{39}
\ee
with $\sigma_i$ irreducible localized endomorphisms of $\mathcal{A}_i$, transportable between $B$ and $\tilde{B}$. 
\end{prop}

\noindent {\sl Proof:} 
   Consider the localizable representation $\pi$ of ${\cal U}$ corresponding to $\sigma$ localized in $B$. $\pi$ restricted to the quasilocal $C^*$-algebra ${\cal U}_1$ corresponding to ${\cal A}_1$ gives a representation $\pi_1$. Identifying ${\cal A}_1\otimes 1$ with ${\cal A}_1$ this is a localizable representation of ${\cal U}_1$. Therefore there is a unitary $V:{\cal H}\otimes {\cal H}\rightarrow {\cal H}$ such that $\sigma_1(x)= \sigma_V\,\pi_1(x)$ is an endomorphism of ${\cal U}_1$ localized in $B$. Let $\sigma_u$ be an inner automorphism in ${\cal U}$ such that $\tilde{\sigma}=\sigma_u \,\sigma$ is localized in $\tilde{B}$.  Correspondingly, we have another representation $\tilde{\pi}$ that restricts to a representation $\tilde{\pi}_1$ of ${\cal A}_1$, and  an endomorphism $\tilde{\sigma}_1(x)= \sigma_{\tilde{V}}\,\sigma_u \,\pi_1(x)$ of ${\cal U}_1$ localized in $\tilde{B}$. Then $\tilde{\sigma}_1(x)=\sigma_{\tilde{V}}\,\sigma_u\,\sigma_{V^*}  \, \sigma_1(x)$.  The endomorphism $\sigma_{\tilde{V}}\,\sigma_u\,\sigma_{V^*}$ has to be inner in ${\cal U}_1$ by Haag duality and this shows that $\sigma_1$ and $\tilde{\sigma}_1$ are transported to each other.  

    A localized representation $\pi$ is a factor representation if $\pi({\cal U})''$ is a factor and is type I if $\pi({\cal U})''$ is type I. Lemma 13 and Corollary 14 of \cite{Kawahigashi:1999jz} show that for a net ${\cal A}_1$ (from the assumptions $1-5,7,8'$), if a transportable sector between the two balls gives a factorial representation it is of type I.  
   The proof of this corollary essentially shows a factor representation not of type I entails the existence of an infinite set of inequivalent localized endomorphisms and this clashes with finite index of the two ball regions. 
   
   Then Lemma 27 on \cite{Kawahigashi:1999jz} gives the stated result (\ref{39}). Essentially $\pi_1({\cal A}_1)''$ and $\pi_2({\cal A}_2)''$ ($\pi_2$ the restriction of $\pi$ to ${\cal U}_2$) commute and span ${\cal B}({\cal H}\otimes {\cal H})$ because $\pi$ is irreducible, so they are factors, and of type I from the preceding discussion. Therefore $\pi\simeq \pi_1\otimes \pi_2$.    
$\square $

We have assumed transportability of the subfactors $\mathcal{A}(R)\subset\hat{\mathcal{A}}(R)$, the structure of the subfactor does not depend on the relative position of the two balls conforming $R$, nor on their sizes. Then, following Lemma 31 in \cite{Kawahigashi:1999jz}, one constructs the net for $B\subset E$, $\tilde{\cal A}(B)={\cal A}(B)\otimes {\cal A}(\bar{B})$, where $\bar{B}$ is the reflection of $B$ in a fixed plane. Given an irreducible endomorphism in the decomposition of the dual canonical endomorphism $\rho$ of two disjoint balls $B, \bar{B}$ the split property allows to see this as an endomorphism of the net $\tilde{\cal A}$. This is clearly transportable to balls further away from the plane of reflection. Transportability of the two ball subfactor also allows to show this endomorphism can be extended and is a localized endomorphism of the net $\tilde{\cal A}$. The previous proposition then allows to decompose it into irreducibles $\sigma_i\sigma_j$ of the original net, that are transportable, and have to be conjugate to each other by proposition \ref{nunubar}.    
 This geometric scenario allows a straightforward generalization of Lemma $31$ in Ref. \cite{Kawahigashi:1999jz}, which we state here for completeness

\begin{prop} \label{propcangen}
Let $\sigma_r$ be the set of irreducible DHR endomorphisms for the net $\mathcal{A}$ satisfying assumptions $1-7,8'$. Then the dual canonical endomorphism of the two ball region $R=B_1\cup B_2$ is unitarily equivalent to
\be \label{cangen}
\rho\simeq \bigoplus_r \sigma_r\bar{\sigma}_r\;,
\ee
where the $\sigma_r$ are localized in $B_1$ and the $\bar{\sigma}_r$ are localized in $B_2$.
\end{prop}
{\sl Proof:} The proof copies the one in Lemma $31$ in Ref. \cite{Kawahigashi:1999jz}. $\square$

From this and the fact that the index of an inclusion equals the dimension of the canonical endomorphism \cite{Longo:1994xe}, it follows that 
the global index $\mu$ associated with the violation of Haag duality in two balls $\mathcal{A}(R)\subset\hat{\mathcal{A}}(R)$ is 
    \be 
\mu= \sum_r\,d_r^2\;.
    \ee

The next question is whether regions $E_n=B_1\cup B_2\cup\cdots \cup B_n$ composed by $n$ disjoint balls bring conceptually new information to the problem. The answer is negative. Everything is controlled by the two-ball inclusion. To show this in detail we now generalize Lemma $42$  in \cite{Kawahigashi:1999jz}. In this case, the proof is slightly different in higher dimensions because the topology of the complement of $n$ disjoint balls changes.

\begin{prop}\label{mun}
The Jones index $\lambda_n$ associated with the violation of Haag duality in a region $E_n=B_1\cup B_2\cup\cdots \cup B_n$ composed of $n$ disjoint balls is
    \be 
\lambda_n= \mu^{n-1}\;,
    \ee
    where $\mu\equiv E_2=\sum_r\,d_r^2$.
\end{prop}
{\sl Proof:}  We proceed by induction. For $n=1$, the relation is trivially true due to Haag duality. For $n=2$ it is also true by definition. Now we consider $E_{n+1}=B_1\cup\cdots \cup B_{n+1}$. The inclusion $\mathcal{A}(E_{n+1})\subset\hat{\mathcal{A}}(E_{n+1})$ is the composition of two inclusions
\be 
\mathcal{A}(E_{n+1})=\mathcal{A}(E_{n})\vee \mathcal{A}(B_{n+1})\subset \hat{\mathcal{A}}(E_{n})\vee \mathcal{A}(B_{n+1})\subset\hat{\mathcal{A}}(E_{n+1})\;.
\ee
The index is multiplicative under compositions of inclusions \cite{Longo:1994xe}. The first inclusion, due to the split property, is isomorphic to the inclusion $\mathcal{A}(E_{n})\otimes \mathcal{A}(B_{n+1})\subset\hat{\mathcal{A}}(E_{n})\otimes \mathcal{A}(B_{n+1})$ so that the previous relation implies $\lambda_{n+1}=\lambda_n \tilde{\lambda}$, where $\tilde{\lambda}$ is the index of the second inclusion $\hat{\mathcal{A}}(E_{n})\vee \mathcal{A}(B_{n+1})\subset\hat{\mathcal{A}}(E_{n+1})$. We need to prove $\tilde{\lambda}=\lambda_2$. It is easy to see this last inclusion is an irreducible subfactor of finite index,  hence with unique conditional expectation $\varepsilon$. This restricts to the conditional expectation of the inclusion ${\cal A}(B_n\cup B_{n+1})\subset \hat{\cal A}(B_n\cup B_{n+1})$, that has index $\mu$. The argument is similar to the one in the proof of proposition \ref{prop1}. $\varepsilon$ takes elements in $\hat{\cal A}(B_n\cup B_{n+1})$ and send them to $\hat{\cal A}(E_{n-1})'\cap (\hat{\mathcal{A}}(E_{n})\vee \mathcal{A}(B_{n+1}))={\cal A}(B_n\cup B_{n+1})$.
Now, taking a ball $B$ including $E_n$ and separated from $B_{n+1}$ we have that the conditional expectation of the inclusion ${\cal A}(B\cup B_{n+1})\subset \hat{\cal A}(B\cup B_{n+1})$, having index $\mu$, has conditional expectation that restricts to $\varepsilon$. This uses again the same argument. By strong additivity this conditional expectation maps $\hat{\mathcal{A}}(E_{n})\vee \mathcal{A}(B_{n+1})$ to ${\cal A}(B-E_n)'\cap {\cal A}(B\cup B_{n+1})=\hat{\cal A}(E_n)\vee {\cal A}(B_{n+1})$, where in the last step we use the split property. The index do not increase under restriction, and this proofs the statement. $\square$

Using this proposition we arrive at a generalization of Theorem $43$ in \cite{Kawahigashi:1999jz} to higher dimensions:
\begin{theorem}\label{theonball}
    Let $\sigma_r$ be the set of irreducible DHR endomorphisms for a net $\mathcal{A}$. Then the dual canonical endomorphism of the the $n$-ball region $E_n=B_1\cup\cdots\cup B_n$ is equal to
\be \label{cangen}
\rho\simeq \bigoplus_{r_1,\cdots,r_n} N_{r_1\cdots r_n}^{0}\sigma_{r_1}\sigma_{r_2}\cdots\sigma_{r_n}\;,
\ee
where $\sigma_{r_i}$ is localized in $B_i$ and where $N_{r_1\cdots r_n}^{0}$ is number of times the identity representation appears in the composition $\sigma_{r_1}\sigma_{r_2}\cdots\sigma_{r_n}$.
\end{theorem}
{\sl Proof:} The fact that $\sigma_{r_1}\sigma_{r_2}\cdots\sigma_{r_n}$ appears with multiplicity $N_{r_1\cdots r_n}^{0}$ is obvious and follows the same logic as proposition \eqref{nunubar}. There are such number of independent isometries in $\hat{A}$ destroying the sector $\sigma_{r_1}\sigma_{r_2}\cdots\sigma_{r_n}$ by definition of such number, but by their interwining definition all of them belong to $\hat{A}(E_n)$. Therefore $\bigoplus_{r_1,\cdots,r_n} N_{r_1\cdots r_n}^{0}\sigma_{r_1}\sigma_{r_2}\cdots\sigma_{r_n}\prec\rho$. To show there are no more in this case, one proceeds by direct computation of the index. Using Frobenious reciprocity one computes
\bea
&\sum\limits_{r_1\cdots r_n}&N_{r_1\cdots r_n}^{0}d_{r_1}d_{r_2}\cdots d_{r_n}=\sum\limits_{r_1\cdots r_{n-1}}\left(\sum\limits_{r_n} N_{r_1\cdots r_{n-1}}^{\bar{r}_n}d_{r_n}\right)\,d_{r_1}d_{r_2}\cdots d_{r_{n-1}}=\nonumber\\ &=&\sum\limits_{r_1\cdots r_{n-1}} \left(d_{r_1}d_{r_2}\cdots d_{r_{n-1}}\right)^2=\left(\sum\limits_{r} d_r^2\right)^{n-1}=\mu^{n-1}\;.
\eea
Since this already gives the $n$-interval index by the previous proposition \eqref{mun}, there cannot be further irreducible sectors in the decomposition of the canonical endomorphism. $\square$

The result of proposition (\ref{mun})  can be extended further. Suppose we have $R=\cup_i R_i$, $i=1\,\cdots,n$,  where $R_i\subseteq B_i$, and $B_i$ separated balls, and $B^1_i\supseteq R_i$ is another set of smaller balls included in the disconnected components $R_i$. Then, we have the nested subfactors $\hat{\cal A}(\cup_i B_i)\supseteq {\cal A}(\cup_i B_i)$, $\hat{\cal A}(\cup_i R_i)\supseteq \vee_i \hat{\cal A}(R_i)$, and $\hat{\cal A}(\cup_i B^1_i)\supseteq {\cal A}(\cup_i B^1_i)$. The conditional expectation of the first subfactor restricts to the other two. The first and last subfactors have the same index $\mu^{n-1}$, and so the second subfactor has the same index. Now, composing the conditional expectations of the subfactor $\hat{\cal A}(\cup_i R_i)\supseteq \vee_i \hat{\cal A}(R_i)$ and $\vee_i \hat{\cal A}(R_i)\supseteq {\cal A}(\cup_i R_i)$, and using the split property in this last subfactor,  we get that the index 
$\lambda_{\cup_i R_i}= \mu^{n-1}\, \prod_i \lambda_{R_i}$. 
We can proceed in the same way with each $R_i$ decomposing into regions included in disjoint (topological) balls, until we get the same expression where the $R_i$ are either single component regions, or {\sl linked} regions, which are multicomponent but for which it is not possible to take two disjoint balls separating the region in two. Let us consider only unlinked regions for simplicity. We get then $n=c(R)$, the number of connected components in $R$. For each single component $R_i$ we can compute the index for the region $R_i'$. This may have several connected components according to the existence of non trivial ``shells'' in $R_i$, that is, non contractible $d-1$ dimensional spheres. We again assume they are unknotted.  The $R_i'$ cannot contain shells because this would contradict $R_i$ being single component. Adding $c(R_i')-1$ for each $R_i$ we get $c(R')-1$. In doing this decomposition we arrive at uniquely defined regions $\tilde{R}_s$ that are single components of the complement of the single components of $R$. We can call them blocks for $R$. The blocks are {\sl $\pi_0$-trivial} in the sense that are single component regions whose complement is also single component. Note the blocks are in general not included in $R$ nor in $R'$.  The blocks for $R'$ are just the complements of the blocks for $R$. 
If all the blocks for $R$ are topological balls we say $R$ is  {\sl $\pi_0$-reducible}. In this case no linking is possible.  
Then we arrive at
\begin{prop}\label{10}
For unlinked regions, the index $\lambda_R=\mu^{c(R)+c(R')-2}\, \prod_s \lambda_{\tilde{R}_s} $, where $c(R)$ and $c(R')$ are the number of connected components in $R$ and $R'$ respectively, $\tilde{R}_s$ are blocks for $R$,  which are regions having $ c(\tilde{R}_s)=c(\tilde{R}_s')=1$. For $\pi_0$-reducible regions we get $\lambda_R=\mu^{c(R)+c(R')-2}$.
\end{prop}
Note that for $d=1,2$ all regions are $\pi_0$-reducible, and this gives the index of an arbitrary region.

This result neatly shows that if we extend the theory and cure the two-ball HDV, i.e. achieve $\mu=1$, then we cure all HDV for $\pi_0$-reducible regions, in particular,  $n$-ball HDV and the ones for regions containing also shells. In turn, to cure the two ball inclusion, since the irreducible endomorphism appearing there are all of the form $\rho_r\bar{\rho}_r$, we just need to extend the net $\mathcal{A}$ to a new net $\mathcal{B}$ which contains partial isometries $\psi_r\in\mathcal{B}$ satisfying
\be 
\psi_r a=\sigma_r(a)\,\psi_r\,\,\,\,\,\,\,\textrm{for all} \,r\;,
\ee
i.e. intertwining the identity sector with the charged sector. This way, the sector $\sigma_r\bar{\sigma}_r$ will be destructible in the extended net $\mathcal{B}$ and we will get $\mu=1$. By the theorem above, such  a net would be free of HDV in regions composed of disjoint balls, i.e. it would be a $\pi_0$-complete net.

\subsection{ DHR reconstruction}

That a net extension curing HDV for regions composed of several balls is possible is part of the content of the DHR reconstruction theorem \cite{Doplicher:1971wk,Doplicher:1973at,doplicher1990there}. The key starting observation in this approach is that the DHR endomorphisms for $d>1$ form a symmetric $C^*$ category with conjugates and subobjects. This implies that such category is isomorphic to the category of representations of a compact group \cite{Deligne2007,Doplicher1989}, and the statistics furnishes a representation of the permutation group (no braiding). In particular, the DHR irreducible endomorphisms $\sigma_r$, are associated to the irreducible representations $r$ of a compact group $G$, have integer dimensions, and their fusion is the one of the representations of $G$. This allows, given a net ${\cal A}$ with DHR endomorphisms defined in a Hilbert space ${\cal H}_{\cal A}$, to construct a new net ${\cal B}$ in a bigger Hilbert space ${\cal H}_{\cal B}$, extending ${\cal A}$, ${\cal A}\subset {\cal B}$, where these endomorphisms are trivialized.  
 For a nice review of the derivation of such result see \cite{haag2012local}.  

Let us recall the main results of this construction. Given a ball $B$ and a complete set of irreducible DHR endomorphisms of ${\cal A}$ localized in $B$, the fact they trivialize in ${\cal B}$ is expressed by the existence of isometries $\psi_r^i\in {\cal B}(B)$, $i=1,\cdots,d_r$,  such that
\bea
\psi_r^i\, a=\sigma_r(a)\,\psi_r^i\,,\quad a\in\mathcal{A}(B)\;.\label{sigma}
\eea  
For the trivial representation $\psi_1=\mathds{1}$. 
Furthermore, these isometries can be chosen to satisfy the basic relations of 
 Cuntz algebras \cite{Cuntz:1977ut}
\be \label{cuntz}
\psi_r^{i\dagger}\psi_r^j=\delta_{ij}\,,\quad\sum\limits_{i}\psi_r^i\psi_r^{i\dagger}=\mathds{1}\;.
\ee
The endomorphisms can then be implemented by the isometries
\be
\sigma_r(a)=\sum_i \psi^i_r\, a \, \psi^{i\,\dagger}_r\,.
\ee
In the finite index case, $G$ is a finite group. Then the two-ball index is $\mu=\sum_r d_r^2=|G|$, and the $n$-ball index is $|G|^{n-1}$.  As in (\ref{expa}), an arbitrary element $b\in {\cal B}(B)$ can be written as
\be
b=\sum_{r,i} a_{r,i}\, \psi_r^i\,,\quad a_{r,i}\in {\cal A}(B)\,.\label{expa1}
\ee
Further, there is a group of unitary operators $g\in G$ acting on ${\cal H}_{\cal B}$, such that ${\cal H}_{\cal A}$  is the subspace of ${\cal H}_{\cal B}$ invariant under the action of this group.  The isometries $\psi_r^i$ furnish a complete set of charged operators under the group:
\be
g\, \psi_r^i\,g^*= R^{(r)\,i}_{\,\,\,\,\,\,\,j}(g) \, \psi_r^j\,, \label{tyty}
\ee
where $R^{(r)\,i}_{\,\,\,\,\,\,\,j}(g)$ are the $d_r$ dimensional matrices of representation $r$. The group symmetry gives place to a local conditional expectation $E: {\cal B}(B)\rightarrow {\cal A}(B)$ by $E_G(b)=a_1$, or, equivalently, 
\be
E_G(b)=|G|^{-1}\, \sum_{g\in G} g \, b\, g^*\,,\quad b\in {\cal B}(B)\,. 
\ee
In terms of the projector $e_0$ to the space ${\cal H}_{\cal A}$, i.e. in terms of the common Jones projection for all ball algebra inclusions $\mathcal{A}(B)\subset\mathcal{B}(B)$, we have
\be
e_0 \, b\, e_0=E_G(b)\, e_0\,.
\ee
The formulas  (\ref{sigma}), (\ref{cuntz}), and (\ref{expa1}) imply the product of operators in ${\cal B}(B)$  has the explicit form
\be
b_1 \, b_2=\sum_{r_1,i_1,r_2,i_2,r_3,i_3} a_{r_1,i_2}^1\, \sigma_{r_2}(a_{r_1,i_2}^2)\, E_G(\psi_{r_1}^{i_1}\, \psi_{r_2}^{i_2}\, \psi_{r_3}^{i_3\, *})\, \psi_{r_3}^{i_3}\,.
\ee

Within the net, the definition of the algebra is completed without adding any further operator. Naively, this suggests there are no charged operators outside the ball $B$, but this is not the case. By transportability, the Cuntz algebra is now transported to every other ball. More explicitly, in the DHR category we have intertwiners (sometimes called charged transporters) moving the localization region from ball $B$ to ball $\tilde{B}$, i.e. there are unitary operators $u_{B\tilde{B}}\in\mathcal{A}$ satisfying
\be 
u_{B\tilde{B}}\,\sigma_{B}=\sigma_{\tilde{B}}\,u_{B\tilde{B}}\,,\quad u_{B\tilde{B}}\in\mathcal{A}\;.
\ee
The operators $u_{B\tilde{B}}$ belong to the maximal algebra of the two balls, i.e. they are the HDV operators. One can think of this operator as transporting a charge from one place to another, or as creating a charge in one ball and destroying another in the other ball. But the previous relation implies
\be 
\psi_{\tilde{B}}^i \,  a=\sigma_{\tilde{B}}(a)\, \psi_{\tilde{B}}^i\,,\quad a\in {\cal A}\,,\quad \psi_{\tilde{B}}^i\equiv u_{B\tilde{B}}\,\psi^i\;.
\ee
Then, by the localization property of $\sigma_{\tilde{B}}$, the new charged isometry $\psi_{\tilde{B}}^i$ commutes with local local operators $x\in\mathcal{A}(\tilde{B}')$, with $\tilde{B}'$ the spatial complement to $\tilde{B}$, and therefore it is naturally associated to $\mathcal{B}(\tilde{B})$. It is easy to see the conditional expectation $E_G$ extends to any ball. 

The analysis of the statistics of the DHR endomorphisms shows that the new net cannot (always) be chosen local. The price to pay for trivializing all DHR sectors is that the construction gives in the general case a {\sl graded-local} net. This means there is a unitary $V=(-1)^F$ (the fermion parity, where $F$ is the fermion number) in the center of $G$, with $V^2=1, V^*=V=V^{-1}$, such that 
\be
V \, \psi^i_r  =\pm \, \psi_r^i\,V\,,\quad  (\textrm{same sign for } i=1\,,\cdots, d_r)\,.   
\ee
Therefore ${\cal B}(B)$ is left globally invariant by the action of $V$, and is generated by even and odd parity operators ${\cal B}_{\pm}(B)$, $b=b_++b_-$, $V \, b_\pm\, V=\pm b_\pm$. An algebra left globally invariant by $V$ is called a graded algebra.  
 For elements of two separated balls $b_\pm^i\in {\cal B}_\pm(B_i)$, $i=1,2$, we have {\sl normal commutation relations}:
\be
b^1_{+}\, b^2_{+}=b^2_{+}\, b^1_{+}\,,\quad b^1_{+}\, b^2_{-}=b^2_{-}\, b^1_{+}\,,\quad b^1_{-}\, b^2_{-}=-b^2_{-}\, b^1_{-}\,. 
\ee

To summarize, the statement of DHR reconstruction theorem, that goes beyond the finite index/ finite group case, can be phrased as \cite{Doplicher:1971wk,Doplicher:1973at,doplicher1990there} (see also the books \cite{haag2012local,kastler1990algebraic,Halvorson:2006wj}):

\begin{prop}
For an irreducible local net ${\cal A}$, where algebras are defined for balls in $\mathbb{R}^d$, $d>1$, acting on a separable Hilbert space ${\cal H}_{\cal A}$, satisfying Haag duality for balls, and property B, there is a unique (up to unitary equivalence) irreducible   graded-local extension ${\cal B}$ in a separable Hilbert space ${\cal H}_{\cal B}\supseteq {\cal H}_{\cal A}$ such that the system of DHR endomorphisms of ${\cal A}$ is trivialized in ${\cal B}$ (eq. (\ref{sigma})), there is a compact group $G$ of unitaries in ${\cal H}_{\cal B}$ (the grading operator $V$ is in the center of $G$) leaving ${\cal B}(B)$ globally invariant for any ball $B$, and where the algebra ${\cal A}(B)\subseteq {\cal B}(B)$ is formed by the elements invariant under $G$.    
\end{prop}

It is worth to mention that for relativistic theories with Lorentz invariance, the usual connection between spin and statistics holds. Concretely, the SS created from the vacuum by the charged operators $\psi$ have half integer angular momentum whenever the statistics of the sector is fermionic. See \cite{guido1995algebraic} for these results and further references. It is also possible to prove the SS are Lorentz and translation covariant of positive energy \cite{guido1992relativistic,Buchholz:1981fj}.  See also the analysis of \cite{carpi2005classification} with slightly different focus and  assumptions, but not requiring finite index. 

\subsection{The $\pi_0$-complete net}

The assumptions of the DHR reconstruction are a small subset of our assumptions for the net ${\cal A}$: it is only needed an irreducible net defined on balls (part of assumption 1), locality (assumption 2), the property B (contained in  assumption 4) and Haag duality for balls (assumption 8'). However, the rest of our assumptions were important to reduce the structure of HDV for two balls to the DHR endomorphisms.  

Given the other properties assumed for ${\cal A}$ we now seek to understand which ones are transferred to ${\cal B}$. First we define ${\cal B}$ for an arbitrary region in an additive way as in (\ref{add}):
\be
{\cal B}(R)=\vee_{B\subseteq R, B\,\textrm{ball}} \,{\cal B}(B)\,.
\ee
 This is self consistent by isotony.   
 It is immediate by transportability of the DHR sectors and strong additivity of ${\cal A}$ that this definition gives a strong additive ${\cal B}$. We call this net the {\sl $\pi_0$-completion} of ${\cal A}$ (if ${\cal A}$ satisfies only essential duality its $\pi_0$-completion is the one for ${\cal A}^d$). We remark the net ${\cal A}$ as a net of subalgebras of ${\cal B}$ in the bigger Hilbert space ${\cal H}_{\cal B}$ is not irreducible. We have ${\cal A}'=G$, ${\cal A}''= G'$. The center of ${\cal A}''$,   $G\cap G'$, is the abelian algebra formed by the global projectors $e_r$ on the different irreducible charges.

Next we have to define what replaces the commutant for graded algebras. A graded algebra is one that is left globally stable by the action of $V$. The following is standard. Define the unitary 
\be
Z=\frac{1+i\, V}{1+ i}\,,\quad Z^*=Z^{-1}\,,\quad Z^2=V\,.
\ee
Define $b^t=Z\, b\, Z^*$, that gives $b_+^t=b_+$, $b_-^t=i\, V\, b_-$. It follows that the fermion parity decomposition of two spatially separated elements $b_1,b_2$ satisfy graded commutativity can be equivalently written as  $[b_1,b_2^t]=0$. Then we can define the {\sl graded-commutant} of a generic graded algebra ${\cal B}$ as
\be
{\cal B}^\triangledown=({\cal B}^t)'=({\cal B}')^t\,, \quad {\cal B}^{\triangledown\triangledown}={\cal B}.
\ee
The graded commutant is the maximal graded algebra satisfying graded-commutativity with ${\cal B}$. On a bosonic algebra it coincides with the ordinary commutant. As it is the usual commutant followed by a unitary, many properties of the usual commutant remain. For example, the graded commutant of a factor is a factor, it keeps the type of the factor, and the index of subfactors.   

Locality is of course replaced by graded-locality for the net ${\cal B}$, that we write as:
\begin{assumptionp}{\ref*{as:2}$'$}
Graded locality. There is a unitary $V$, $V^2=1$, leaving local algebras globally invariant. For any $R$, we have ${\cal B}(R)\subseteq {\cal B}(R')^\triangledown$.   
\end{assumptionp}
 The definition of maximal algebras is also generalized accordingly 
\be
\hat{\cal B}(R)={\cal B}(R')^\triangledown\,.
\ee
With this definition for the maximal algebras, the finite index assumption and the definition of transportability are unchanged. 
Graded Haag duality is naturally the case ${\cal B}(R)= {\cal B}(R')^\triangledown$, or $\hat{\cal B}(R)={\cal B}(R)$.

The analysis of the properties of the net ${\cal B}$ follows a similar path as the proof of proposition \ref{dualnet}. 
It follows  by strong additivity and irreducibility that the ${\cal B}(R)$ are factors and ${\cal B}(R)\subseteq \hat{\cal B}(R)$ form irreducible subfactors. The subfactors ${\cal A}(R)\subseteq {\cal B}(R)$ and ${\cal A}(R)\subseteq \hat{\cal B}(R)$ are also irreducible because ${\cal A}(R)\vee {\cal B}(R')\supseteq {\cal A}(R)\vee {\cal A}(R')\vee {\cal B}(B)={\cal B}({\cal H})$ for any ball $B\subset R'$. Both of these  subfactors are of finite index because the restriction of the conditional expectation $E^{\cal A}_R \circ E_G$ is the unique one in each case, and is of finite index. Considering the dual conditional expectations and its restriction it follows that ${\cal B}(R)\subseteq \hat{\cal B}(R)$ is of finite index. All these factors are then of type III \cite{loi1992theory}.

For a generic region ${\cal B}(R)={\cal A}(R)\vee_i {\cal B}(B_i)$, where $B_i$ are balls included in each connected component of $R$. This follows from transportability of the DHR sectors. 
From a faithful state in ${\cal A}(R)$, composing with the conditional expectations we get a faithful neutral state in ${\cal B}(R)$ and $\hat{\cal B}(R)$. As ${\cal B}(R)$ is globally invariant under $G$, the corresponding Jones projection of this last inclusion is invariant under $G$. The same holds for the  
  dual Jones projection. Therefore   in the equation $e_{R'}\vee {\cal B}(R)= \hat{\cal B}(R)$ the operator $e_{R'}$ can be taken as a neutral element, in particular, bosonic, and then the same holds for a complete set of non local isometries associated to the irreducible sectors of the inclusion. 
  The conditional expectation of this inclusion commutes with the group operations, in particular with $V$. These considerations lead to a straightforward generalization of proposition \ref{prop1} for the case of the graded net. 

 Given $R_1\subset R_2$, since the non local operators in $\hat{\cal B}(R_2)$ and $\hat{\cal B}(R_1)$ can be chosen uncharged, transportability in the net ${\cal B}$ follows directly if we have transportability in ${\cal A}$. In particular we have topological transportability in ${\cal B}$.

The appropriate split property for a graded net writes ${\cal B}(R_1)\subset {\cal N}$, ${\cal B}(R_2)^t\subset {\cal N}'$, or ${\cal B}(R_2)\subset {\cal N}^\triangledown$, for a graded type I factor ${\cal N}$:
\begin{assumptionp}{\ref*{as:8}$'$} \label{gradsplit}
  The graded split property. For two disjoint regions $R_1, R_2$, separated a finite distance, where one of them is bounded,  there exist a type I factor ${\cal N}$ such that  ${\cal B}(R_1)\subset {\cal N}$, ${\cal B}(R_2)\subset {\cal N}^\triangledown$.
\end{assumptionp}
As in the proof of proposition \ref{dualnet}, under the finite index assumption, the split property between separated regions $R_1,R_2$ in the original net extends to a split between ${\cal B}(R_1)$ and ${\cal B}(R_2)$. It is enough to proof the split between two arbitrary separated single component regions $R_1, R_2$. 
Following  
  \cite{longo2003conformal}, lemma 22, the proof just requires that the conditional expectation  $E^{(12)}:{\cal B}(R_1)\vee {\cal B}(R_2)\rightarrow {\cal A}(R_1)\vee {\cal A}(R_2)$ 
  satisfies $E^{(12)}(b_1 b_2)=E^{(12)}(b_1)E^{(12)}(b_2)$ for $b_1\in {\cal B}(R_1), b_2 \in {\cal B}^d(R_2)$. Charged elements for each algebra cannot produce an element of the additive algebra ${\cal A}(R_1)\vee {\cal A}(R_2)$, but rather a non local operator in $\hat{\cal A}(R_1 \cup R_2)$. Therefore, the property holds, and using a split state in ${\cal A}(R_1)\vee {\cal A}(R_2)$ composed with the conditional expectation we produce a split state in ${\cal B}(R_1)\vee {\cal B}^t(R_2)$, given that fermion operators have zero expectation value. It is clear from the construction that the split factors will be globally invariant under $G$, in particular under $V$. 

Weak modularity of ${\cal B}$ can also be proved from the one of ${\cal A}$ as in proposition \ref{dualnet}.
 We have to show that ${\cal B}(A)\cap {\cal B}(B\cup C)$ is included in ${\cal B}(B)\vee ({\cal B}(A)\cap {\cal B}(C))$, starting from the same inclusion for the net ${\cal A}$. As non local operators for the net ${\cal B}$ can be chosen neutral, it follows that ${\cal B}(A)\cap {\cal B}(B\cup C)\subseteq ({\cal A}(A)\cap {\cal A}(B\cup C))\cup {\cal B}(A\cap (B\cup C))$. Then the sought inclusion follows.

Let us consider the following inclusions. Using the split property, the inclusion ${\cal A}(R)\subseteq {\cal B}(R)$ has index $|G|^{c(R)}$, with $c(R)$ the number of connected components. The inclusion ${\cal B}(R)\subseteq \hat{\cal B}(R)$ has index $\lambda^{\cal B}(R)$. The inclusion $\hat{\cal B}(R)\subset {\cal A}(R')'$ has the same index as the inclusion ${\cal A}(R')\subseteq {\cal B}(R')$, that is, $|G|^{c(R')}$. Finally consider the inclusion ${\cal A}(R)\subseteq {\cal A}(R')'$. Since ${\cal A}(R')'=\hat{\cal A}(R)\vee {\cal B}(B)\vee G$, where $B$ is any ball in $R$, it follows that the index of this inclusion is $\lambda^{\cal A}(R)\, |G|^2$. Then, composing the indices of nested inclusions we have
\begin{prop}
The following identity between the indices corresponding to the region $R$ in the nets ${\cal A}$ and ${\cal B}$ holds
\be
\lambda^{\cal A}(R)= |G|^{c(R)+c(R')-2}\,  \lambda^{\cal B}(R)\,.
\ee
In particular, from proposition \ref{10},  for $\pi_0$-reducible regions $\lambda^{\cal B}(R)=1$, and for $\pi_0$-trivial regions we have $\lambda^{\cal B}(R)=\lambda^{\cal A}(R)$.   
\end{prop}

The trivial index for $\pi_0$-reducible regions is equivalent to (graded) Haag duality for these regions.   

We have shown the following 
\begin{prop} For a net ${\cal A}$ satisfying assumptions 1-8',9, the $\pi_0$-completion ${\cal B}$ of ${\cal A}$ is a graded net satisfying assumptions 1-2',3-6,7',9. In addition, all $\pi_0$-reducible regions, in particular regions formed by disjoint balls or their complements, are graded-Haag dual in ${\cal B}$. 
\end{prop}
\noindent 

In particular we have

\begin{corollary}
A $\pi_0$-complete theory in $d=1$ or $d=2$ is complete (it does not have HDV for bounded regions or their complements). 
\end{corollary}

 That is, under our assumptions, all HDV for $d=2$ (for bounded regions and their complement) are cured in a unique way by extending the theory to the field algebra ${\cal B}$ as a result of DHR theorem, and all theories with HDV (for bounded regions and their complements) are the fixed point of a complete theory under the action of a group. For $d=1$ CFT's, our assumptions give completely rational models in the sense of \cite{Kawahigashi:1999jz}. These models can always be completed \cite{Longo:1994xe}, though in several inequivalent ways. 
 The relation between the complete model and the submodel in $d=1$ in general is not given by a group action. See for example \cite{cmp/1104200513}.

\subsection{A new derivation of invertibility of $0$-form symmetries from $\pi_0$-completeness}

 In this section we provide a new derivation of the invertibility of $0$-form symmetries, starting from a $\pi_0$-complete net. The present starting point is then more in line with the high energy literature \cite{Gaiotto:2014kfa}, where charged operators are always assumed to be part of the algebra. Therefore, what we seek is a certain converse of the above reconstruction: starting from a $\pi_0$-complete net, we show the possible subnets are determined by the invariant algebras under symmetry groups acting locally on the net. This will show the above reconstruction in a different, somewhat simpler, light. This will be relevant for the understanding of the relation between the net ${\cal B}$ and ${\cal A}$ when ${\cal A}$ is not Haag dual for balls, i.e. the case of spontaneous symmetry breaking. This will be important for the subsequent analysis of HDV corresponding to loops in higher dimensions.

We start with two nets ${\cal A}$, ${\cal B}$ defined on balls, forming a net of (irreducible) subfactors $\mathcal{A}\subset \mathcal{B}$,  i.e. we have an inclusion
\be 
\mathcal{A}(B)\subset\mathcal{B}(B)\;,
\ee
for any ball $B$. The intuition is that $\mathcal{B}(B)$ is the complete theory, where we have all charges and symmetries acting on them. Then $\mathcal{A}(B)$ is the subnet invariant under the symmetry. One might wonder if such invariant $\mathcal{A}(B)$ exists. By definition, if the QFT has a stress tensor, this is invariant under the symmetry, so in those cases $\mathcal{A}(B)$ can be thought as the subnet generated by the stress tensor.

In this section and in the one that follows, both nets in their own Hilbert spaces satisfy assumptions 1,2',3-6,7', in particular, they are graded-local. ${\cal B}$  satisfies Haag duality for  balls. We assume the inclusion is transportable in the sense that the canonical endomorphism extends for each pair of included balls $B\subset B_1$, being the identity in the complement ${\cal A}(B_1)\cap {\cal B}(B)^\triangledown$. Equivalently, the conditional expectation $E_B$ for the subfactor extends in any pair of included balls and there is a common cyclic and separating vector $\Omega$, invariant under the conditional expectation $E_B$ for any ball $B$.  This forms a standard net of subfactors in the sense of \cite{Longo:1994xe}, see the discussion on transportability in Sec \ref{SecII}. Acting on $\Omega$, the algebras ${\cal A}(B)$ are cyclic in a smaller space ${\cal H}_{\cal A}\subset {\cal H}_{\cal B}$. The projector on ${\cal H}_{\cal A}$ is a common Jones projection $e$ for all balls.
In the finite index case we are considering,  we can then associate a relative index $I$ to such subfactor, that by transportability, does not depend on the choice of ball. In this representation of net, ${\cal A}$ automatically satisfies Haag duality for balls in its Hilbert space ${\cal H}_{\cal A}$. This is because the commutant of ${\cal B}(B)\vee e$ is ${\cal A}(B')$ and then ${\cal A}(B')'\cap {\cal B}({\cal H}_A)= {\cal A}(B)$.

 In proposition $24$ of Ref. \cite{Kawahigashi:1999jz} it was shown for conformal nets in $d=1$ that the global indices (two-ball index) $\mu_{\mathcal{A}}$ and $\mu_{\mathcal{B}}$ of each net and the relative index $I$ satisfy
\be 
\mu_{\mathcal{A}}=I^2 \mu_{\mathcal{B}}\;.
\ee
The global index $\mu_\mathcal{A}$ is the Jones index of $\mathcal{A}(R)\subset\hat{\mathcal{A}}(R)$ within ${\cal H}_{\cal A}$, where $R=B_1\cup B_2$ is a two separated ball region, which in $d=1$ is a region composed of two finite segments. The first step in the proof of this result is to notice that
 within the Hilbert space  $\mathcal{H}_{\mathcal{B}}$ the index of $\mathcal{A}(R)\subset \hat{\cal A}(R)$   (where in ${\cal H}_{\cal B}$ we write $\hat{\cal A}(R)={\mathcal{A}}(R')^\triangledown$) is $I^2 \mu_{\mathcal{A}}$.  The defining representation of $\mathcal{B}$ is unitarily equivalent to the defining representation of $\mathcal{A}$ composed with the dual canonical endomorphism \cite{Longo:1994xe}.
 Then the inclusion $\mathcal{A}(R)\subset \hat{\mathcal{A}}(R)$  within $\mathcal{H}_{\mathcal{B}}$ is unitarily equivalent to the inclusion $\rho(\mathcal{A}(R))\subset\rho(\hat{\mathcal{A}}(R))$ within $\mathcal{H}_{\mathcal{A}}$, where $\rho$ is the dual canonical endomorphism associated to the net of subfactors $\mathcal{A}\subset \mathcal{B}$, that can be taken as an extension of the one corresponding to a ball $B\subset B_1$.

 But we also have
\be
\mathcal{A}(R)\subset \mathcal{B}(R)\subset \hat{\mathcal{B}}(R)\subset \hat{\mathcal{A}}(R)\;.
\ee
The index is multiplicative under chains of inclusions such as this one. But $\mathcal{A}(R)\subset \mathcal{B}(R)$ has index $I^2$ due to the split property, $\mathcal{B}(R)\subset \hat{\mathcal{B}}(R)$ has index $\mu_{\mathcal{B}}$ by definition, and $\hat{\mathcal{B}}(R)\subset \hat{\mathcal{A}}(R)$ has index $I^2$. The reason for the latter is that by taking commutants we get again $\mathcal{A}(R)\subset \mathcal{B}(R)$ since we are in $d=1$ and the complement of two segments are two segments. This completes the proof.

The next proposition is the modification of this result for $d>1$
\begin{prop}\label{Itomu}
    Let $\mathcal{A}\subset \mathcal{B}$ be a finite-index inclusion of nets of factors. Assuming $\mathcal{A}$ has finite global index $\mu_{\mathcal{A}}$ and $d>1$ we have
\be
\mu_{\mathcal{A}}=I \mu_{\mathcal{B}}\;.
\ee
where $I$ is the relative index of the net inclusion $\mathcal{A}\subset \mathcal{B}$.
\end{prop}
{\sl Proof:} All but one of the steps in the proof of the $d=1$ case in  \cite{Kawahigashi:1999jz} remain valid. The only  change is that $\hat{\mathcal{B}}(R)\subset \hat{\mathcal{A}}(R)$ has index $I$ (instead of $I^2$). The reason is that by taking commutants we get an inclusion ${\cal A}(R')\subseteq {\cal B}(R')$ in a connected domain in $d>1$. This has index equal to that of a ball, i.e. equal to $I$, by transportability. $\square$

The next theorem uses all previous results and, to our knowledge, has not appear in this form elsewhere. 
\begin{theorem}\label{Thdr}
    Let $\mathcal{A}\subseteq {\cal B}$ be a net of subfactors as described above. Let ${\cal B}$  be $\pi_0$-complete, and $d>1$. In this scenario, the relative index $I$ of the net extension equals to
    \be 
I=\sum_r d_r^2\;,
    \ee
    where the sum runs over all irreducible DHR endomorphisms of the net $\mathcal{A}$ and $d_r$ are their dimensions. Furthermore, the dual canonical endomorphism (for a ball) reads
    \be 
\rho\simeq \bigoplus_r d_r\rho_r\;,
    \ee
    i.e. all DHR endomorphisms appear with multiplicities equal to their dimensions. In particular, the dimensions $d_r$ are integer numbers.
\end{theorem}
{\sl Proof:} A $\pi_0$-complete net has $\mu_{\cal B}=1$. Then, proposition \eqref{Itomu} shows that such extensions requires $I=\mu_{\mathcal{A}}$. Then proposition \eqref{mun} proves the first part of the theorem. For the second part we notice that the canonical endomorphism associated with the extension can be reduced on general grounds to
   \be 
\rho\simeq \bigoplus_r N_r\rho_r\;,
    \ee
where $\rho_r$ are DHR endomorphisms due to transportability of the net of subfactors and $N_r$ are some multiplicities. But these multiplicities are bounded by $0\leq N_r \leq d_r$, see e.g. \cite{Longo:1994xe}. Since the index $I=\sum_r d_r^2$, and $I$ equals the dimension of the associated canonical endomorphism, it ought to be the case that $N_r=d_r$ for all DHR sectors. $\square$.

This theorem proves the main aspects of the DHR theorem, namely the fact that symmetry originates from groups in $d>1$ and the spin-statistics connections, as we now show. First, from the general discussion in the previous section, see \cite{Longo:1994xe}, a finite index inclusion such as $\mathcal{A}\subset \mathcal{B}$ that restricts to a subfactor $\mathcal{A}(B)\subset \mathcal{B}(B)$ in any ball $B$, implies the existence of a partial isometry $v\in \mathcal{B}(B)$ intertwining the identity representation with the canonical endomorphism in $\mathcal{B}$, namely
\be 
vb=\gamma(b)v\,,\quad b\in\mathcal{B}(B)\;,
\ee
and satisfying the defining equations of a $Q$-system, i.e. eq. \eqref{Qsystem}. Then, choosing a set of partial isometries $\omega_r^{i}\in \mathcal{A}(B)$, where $r$ runs over representations and $i=1,\cdots ,d_r$, satisfying \eqref{omegpart}, we can explicitly write the canonical endormophism as in \eqref{endoexpl}. Besides we can find the charged intertwiners for all sectors using \eqref{chargephi}. More explicitly we define
\be
\psi_r^i\equiv \sqrt{\frac{I}{d_r}}\,\omega_r^{i\,\dagger}\,v\;,
\ee
which can be verified to be a partial isometry by direct computations using $Q$-system calculus. But in the present case, where the multiplicity $N_r$ of representation $r$ has been shown to be $d_r$, we immediately obtain more. First, as demonstrated in Ref. \cite{Rehren:1993yu}, when $N_r=d_r$ we have that
\be 
\sum\limits_{i=1}^{d_r}\,\psi_r^i\psi_r^{i\dagger}=\mathds{1}\;,
\ee
implying that the intertwining relation
\be 
\psi_r^i\,a=\rho_r(a)\,\psi_r^i\,,\quad a\in\mathcal{A}(B)\;,
\ee
can be resolved to provide an explicit implementation of endomorphisms
\be 
\rho_r(a)=\sum_{i}\, \psi_r^i\,a\,\psi_r^{i\dagger}\,,\quad a\in\mathcal{A}(B)\;.
\ee
Equivalently, this shows there exists a Cuntz algebra \cite{Cuntz:1977ut} associated with each irreducible endomorphism. Second, in Ref. \cite{Longo:1994zza} it was shown that in these scenarios (finite index with $N_r=d_r$) the local subfactors are the fixed-point subfactors under the action of a Hopf $C^*$ algebra. Equivalently, in $\mathcal{A}(B)\subset \mathcal{B}(B)$, the subalgebra $\mathcal{A}(B)$ is the invariant part of $\mathcal{B}(B)$ under the action of a Hopf $C^*$ algebra. 

A group algebra is an example of a Hopf $C^*$ algebra, but there are further examples. The difference between a group and a more general Hopf $C^*$ algebra, such as a quantum group, is whether the comultiplication is symmetric. In terms of its category of representations, the difference lies on the way the Hopf $C^*$ algebra acts in the tensor product of two representations. While for groups this action is symmetric, namely for two irreps $r$ and $r'$, the tensor product is simply $\pi(g)=\pi_r(g)\otimes\pi_{r'}(g)$, for more general Hopf $C^*$ algebras this action is more complicated. Equivalently, from a physical point of view, a group algebra requires charged particles to have permutation statistics while a quantum group requires charged particles to have more general braided statistics.

 But in the present context locality constraints the statistics to be symmetric. This is one of the key inputs of the DHR theorem as we described above. But having shown that $d_r$ are integers, that $N_r=d_r$, and the fact we are in the complete local net, we can now provide an easier proof that the braiding is indeed a permutation symmetry. 

 \begin{theorem}\label{thgroup}
    Let $\mathcal{A}\subset \mathcal{B}$ be a net of subfactors in $d>1$. Assume the global index $\mu_{\mathcal{B}}=1$, the relative index $I$ to be finite, and the $\pi_0$-complete net $\mathcal{B}$ to be local or graded local, then the net $\mathcal{A}$ is the fixed-point of the net $\mathcal{B}$ under the action of a finite group $G$. 
\end{theorem}
{\sl Proof:} The previous theorem showed that $\mathcal{A}$ is the fixed-point under the action of a $C^*$ Hopf algebra. To constraint this algebra to be a group algebra we need to show that the category of representations is symmetric.\footnote{From a physical standpoint, in $d>1$ there is no geometric way to say what is the ``order'' in a tensor product of two charged states. Therefore, one cannot define the action of a quantum group in a local QFT in $d>1$ without explicitly breaking Lorentz covariance.} Within this scenario, this was shown e.g. in Ref. \cite{Rehren:1993yu}. First, by transportability we have unitary intertwiners $U^r$ in the net $\mathcal{A}$ moving the localization support of the charged intertwiners $\psi_r^i$. Then by (graded)-locality of $\mathcal{B}$
\be 
U^r\psi_r^i\,\psi_r^j=\pm\, \psi_r^j\,U^r\psi_r^i=\pm\,\rho_r(U^r)\,\psi_r^j\psi_r^i\;.
\ee
This can be rewritten as
\be
\rho_r(U^r)^*\,U^r\psi_r^i\,\psi_r^j=\pm\,\psi_r^j\psi_r^i\;.
\ee
But $\rho_r(U^r)^*\,U^r$ is the statistics operator $\varepsilon_r$ of the sector $r$ \cite{Doplicher:1971wk},\footnote{Sometimes this is written as $\varepsilon=U_2^*\rho (U_1^*)U_1\rho (U_2)$ for two charged transporters $U_1$,$U_2$. Here we are fixing $U_2$ to be the identity.} so graded-locality of the $\pi_0$-complete net $\mathcal{B}$ is the statement that
\be
\varepsilon_r\psi_r^i\,\psi_r^j=\pm\,\psi_r^j\psi_r^i\;.
\ee
But using the previously derived completeness properties of the partial isometries $\psi_r^i$, this equation can be solved for the statistics operators, obtaining
\be
\varepsilon_r=\pm\,\sum\limits_{ij}\psi_r^i\,\psi_r^j\,\psi_r^{i\dagger}\,\psi_r^{j\dagger}\;,
\ee
where we notice that the sign is meaningless for an isometric intertwiner (if $V$ is an isometry, so is $-V$). This relation implies $\varepsilon_r^2=1$, i.e. permutation group statistics. $\square$

Theorems \eqref{Thdr} and \eqref{thgroup} are naturally seen as a converse of the DHR reconstruction theorem \cite{Doplicher:1971wk,Doplicher:1973at,doplicher1990there} in the finite index scenario. The DHR theorem shows any Haag dual net can be $\pi_0$-completed to a graded local net for $d>1$ where symmetries are group-like,  and sectors are bosonic or fermionic, corresponding to local or graded-local charged operators. Here our starting point was a graded-local $\pi_0$-complete net, and the theorems presented constrain the possible subnets to be the fixed points under the action of a symmetry group, and the charged operators to have permutation statistics. This complementary approach extends to the case where the original subnet ${\cal A}$ is not Haag dual for  balls.  We note that the uniqueness of the extension in the DHR reconstruction allows to reach the same conclusions, see  \cite{carpi2005classification}.

\subsection{Relation between ${\cal A}$, ${\cal A}^d$ and ${\cal B}$ in spontaneously broken symmetry scenarios}

 The previous discussion shows that under the present assumptions, if we have an inclusion of nets ${\cal A}\subset {\cal A}^d\subset {\cal B}$, where ${\cal B}$ is a $\pi_0$-complete net, then ${\cal A}$ is the fixed point of ${\cal B}$ under the action of a group $G$, and ${\cal A}^d$ is the fixed point of ${\cal B}$ under the action of a group $H$. The action of these groups leave the algebras on balls globally stable. Clearly $H$ is a subgroup of $G$. 
 It was essential for this result that ${\cal B}$ is $\pi_0$-complete, having $\mu_B=1$. Therefore, it is not expected in general that ${\cal A}$ is the fixed point of ${\cal A}^d$ under the action of a group since  the latter net is not necessarily $\pi_0$-complete. In this sense, ${\cal A}^d$ might display symmetry structures that are not group-like. However, this will be still the case if the coset $G/H$ is itself a group, i.e. $H$ is a normal subgroup of $G$. Also, it is clear that all symmetry structures and selecection rules just arise as possible quotients of the larger symmetry group $G$.
 
 In summary we have 
\bea
&& E_G:{\cal B}(B)\rightarrow {\cal A}(B)\,,\\
&& E_H: {\cal B(B)}\rightarrow {\cal A}^d(B)\,,\\
&& E_{G/H}:  {\cal A}^d(B)\rightarrow {\cal A}(B)\,, 
\eea
where we have written $E_{G/H}=E_G|_{{\cal A}^d}$. The indices are $|G|$, $|H|$, and $|G|/|H|$ respectively.  We can write 
\bea
&& {\cal B}(B)= {\cal A}(B)\vee \{\psi_r^i\}\,,\\ 
&& {\cal B}(B)={\cal A}^d(B)\vee \{\tilde{\psi}_{\tilde{r}}^{\tilde{i}}\}\,,\\
&& {\cal A}^d(B)= {\cal A}(B)\vee \{E_H (\psi_r^i) \}\,,
\eea
 where $\psi_r^i$ and ${\tilde{\psi}_{\tilde{r}}^{\tilde{i}}}$ are isometries corresponding to  the irreducible representations of $G$ and $H$ respectively. 

This result is deduced using a faithful representation of the algebras of the nets in a Hilbert space generated by a cyclic and separating vector $|\Omega\rangle$ for ${\cal B}(B)$ that is invariant under the conditional expectation to any of the subalgebras. So the subalgebras act cyclically in smaller spaces ${\cal H}_{{\cal A}^d}$ and ${\cal H}_{\cal A}$ containing this vector. 

This representation coincides with the DHR reconstruction of ${\cal B}$ from a net ${\cal A}^d$. This latter is Haag dual for single balls but not in general for more than one disjoint ball. This inclusion describes the standard unbroken global symmetry scenario, where there is a state, usually taken as the vacuum and here $|\Omega\rangle$,  which is cyclic and separating for ${\cal B}(B)$ and cyclic and separating for the neutral subalgebra  ${\cal B}/H={\cal A}^d$ in the subspace of neutral states. The symmetry group $H$ can be implemented globally in this case, leaving ${\cal H}_{{\cal A}^d}$ pointwise invariant.  

However, while the algebraic relations for the algebras of balls $B$  remain correct, 
in this construction if ${\cal A}\subset {\cal A}^d$ are different, ${\cal H}_{\cal A}\subset {\cal H}_{{\cal A}^d}$, and then ${\cal A}$ is not irreducible in the Hilbert space ${\cal H}_{{\cal A}^d}$. This is not the case of the original space where these nets where defined, where ${\cal H}={\cal H}_{\cal A}={\cal H}_{{\cal A}^d}$. The difference lies in that in this original Hilbert space the operators corresponding to the elements $g\in G, g\notin H$ cannot be implemented globally. Otherwise ${\cal A}$ would not be irreducible in ${\cal H}$, since this net would commute with $E_H(g)$. However, the symmetry can always be implemented locally due to the split property. 
The essential difference is that this Hilbert space is constructed from a cyclic and separating vector for both ${\cal A}(B)$ and ${\cal A}^d(B)$, and not invariant under the conditional expectation $E_{G/H}$ (while still invariant under the conditional expectation $E_H$ in ${\cal H}_{\cal B}$).      Because of that ${\cal A}(B)$ is not Haag dual in ${\cal H}$, since charged operators exist in ${\cal H}$ and are generated in $({\cal A}(B'))'$.

Then this case corresponds to the case of a spontaneous symmetry breaking, where global symmetry operators cannot be implemented, and the vacuum state is ``charged''. The full symmetry group is $G$ while the unbroken symmetry group is $H$. These generic features  were described in \cite{doplicher1990there, buchholz1992new}, 
 but a complete account in the general case was not given because the situation for infinite index is more involved. Here we restricted ourselves to the finite index case. 

We remark that not all of the DHR endomorphisms of $\mathcal{A}$ correspond to superselection sectors in the net $\mathcal{A}$. By construction, all superselection sectors in $\mathcal{A}^d$ correspond to endomorphisms of $\mathcal{A}^d$. These descend to DHR endomorphisms of $\mathcal{A}$ by restriction in subfactor theory \cite{Longo:1994xe}. But there are further endomorphisms of $\mathcal{A}$ coming from the inclusion $\mathcal{A}^d\supset\mathcal{A}$. By construction, these are DHR endormorphisms (localizable and transportable) due to transportability in the net of subfactors. But also by construction they do not generate superselection sectors because the isometries generating $\mathcal{A}^d$ from $\mathcal{A}$ do not exit the vacuum Hilbert space, since the Hilbert space generated from $\mathcal{A}^d$ is the same as the one generated from $\mathcal{A}$. Equivalently, the inclusion structure provides a set of DHR endomorphisms for $\mathcal{A}$ coming from superselection sectors of the net, and a further set of DHR endomorphisms coming from the inclusion with the dual net. This structure is nicely encoded in the canonical endomorphism associated with two balls since charged intertwiners for both types of sectors will appear as HDV isometries in the two-ball region.

However, as emphasized in \cite{doplicher1990there}, the set of all DHR endomorphisms of $\mathcal{A}$ still forms a symmetric $C^*$ category with conjugates and subobjects. This full category of DHR endormorphisms of the invariant net $\mathcal{A}$ appears in the HDV of a region composed of disjoint balls, since charge transporters and related isometries associated to both sets of DHR will violate Haag duality in the region. The algebraic relations of the inclusions of nets for bounded regions are the same in any faithful representation, and the set of DHR endomorphism is also intrinsic to the net.  
The commutants may change however, and thus the subfactor ${\cal A}(R)\subset \hat{\cal A}(R)$, in particular its associated canonical endomorphism.  
As a concrete example, the dual canonical endomorphism of, e.g. the $2$-ball region $R=B_1\cup R_2$, associated with the inclusion $
\mathcal{A}(R)\subset\hat{\mathcal{A}}(R)$ 
will be of the form $\bigoplus \rho_a\bar{\rho}_a$ if there is Haag duality for single balls, as previously described. However, if there is no Haag duality for balls, and ${\cal A}^d$ is $\pi_0$-complete, by the split property, the canonical endomorphism will be of the form  $\bigoplus N_r N_{r'}\rho_r \rho_{r'}$, where the $\rho_r$ are the irreducible DHR of $\mathcal{A}$.
  This neatly shows that completeness in a region composed by two balls implies completeness for a single ball, but not the other way around. In the general case the global index of ${\cal A}\subset \hat{\cal A}$ is $|G|^2/|H|$.
 This arises as the product of the index ${\cal A}(R)\subset {\cal A}^d(R)$, that is $(|G|/|H|)^2$ by the split property and the index of a single ball inclusion, and the index ${\cal A}^d(R)\subset \hat{\cal A}^d(R)=\hat{\cal A}(R)$, which is $|H|$.

After these general comments, we summarize with the following proposition
\begin{prop} \label{SSB}
    Let $\mathcal{A}$ be a net satisfying assumptions $1-9$, and ${\cal A}^d$ the dual net, $\mathcal{A}\subseteq\mathcal{A}^d$, which in addition satisfies assumption $8'$ (Haag duality for balls). Then there is a unique $\pi_0$-complete graded-local extension $\mathcal{B}$, i.e. a net with global index $\mu_{\mathcal{B}}=1$, and assumptions $1,2',3-6,7',8',9$, such that
\be 
\mathcal{B}\supseteq \mathcal{A}^d\supseteq\mathcal{A}\;,
\ee
is a net of subfactors, the $\mathcal{A}^d$ is the invariant part of $\mathcal{B}$ under the action of a group $H$, $\mathcal{A}$ is the invariant part of $\mathcal{B}$ under the action of a group $G$. In particular, if $\mathcal{A}^d$ is  $\pi_0$-complete, $\mathcal{A}$ is the invariant part of $\mathcal{A}^d$ under a group $G$.
\end{prop}

\section{HDV in $k$-dimensional loops: abelianity of higher form symmetries \label{SecIV}}

We now consider the case of $d\ge 3$ (four or more space-time dimensions) and regions with the topology of ``$k$-dimensional loops'',  $S^{k}\sim \mathbb{S}^{k}\times \tilde{B}$, where $\tilde{B}$ is a $d-k$ dimensional ball. We take simple loops, unknotted in the ambient space. We have already dealt with the case $k=0$ and $k=d-1$. These cases will be assumed to have no HDV, that is, we assume ${\cal A}$ is a $\pi_0$-complete theory. As we have seen this entails no loss of generality.  So we restrict attention to $1\le k \le d-2$. Note that as the full space can be thought as $E=\mathbb{S}^d$, the HDV for a loop $S^k$ are dual to the HDV for a loop $S^{d-k-1}$. Then, both are trivial or non trivial together. The prototypical example, probably the most important, is the case of $k=1$, associated with pure gauge theories \cite{Casini:2020rgj}.   

The main idea is to drive the problem to that of single ball HDV scenario in an appropriate reduced dimensionality. This will imply that such sectors are classified as representations of groups. If this happens for a region and its complement simultaneously, we end up with abelian groups. The only exception to this lore are standard internal symmetries, i.e. ball or $S^{0}$ sectors (for $\pi_0$ incomplete theories), where the complementary regions have non-trivial $S^{d-1}$ sectors, whose ``dimensional reduction'' gives rise to sectors in $d=1$, which can obey more general fusion rules.

Hence consider a  net ${\cal A}$ in $\mathbb{R}^d$. We construct a new dimensionally reduced theory $\tilde{\cal A}$ in the $d-k$ dimensional semi-space $\tilde{E}=\{\tilde{x}=(x^1,\cdots, x^{d-k}); x^1,x^2,\cdots, x^{d-k}>0\}$. Notice only $x^{d-k}>0$. Define the unit $k$- dimensional sphere in $k+1$ coordinates $C=\{y=(y^1,\cdots,y^{k+1}), y^2=1\}$.   
For any region $\tilde{R}\subset \tilde{E}$ we take a $d$-dimensional region that is rotationally symmetric in $k$ directions:
\be
R(\tilde{R})=\{x=(x^1,\cdots,x^{d-k}. (y^1,\cdots, y^{k+1})); \tilde{x}\in\tilde{R}, y\in C\}\,.
\ee
It follows that $R(\tilde{R}')=(R(\tilde{R}))'$. 
We define the net $\tilde{\cal A}$ in $\tilde{E}$ as
\be
\tilde{\cal A}(\tilde{R})={\cal A}(R(\tilde{R}))\,,\quad \tilde{R}\subset \tilde{E}\,.
\ee
For any region $\tilde{R}\subset \tilde{E}$ the maximal algebra $\hat{\tilde{{\cal A}}}(\tilde{R})=({\cal A}(R(\tilde{R}')))'=({\cal A}(R(\tilde{R})'))'=\hat{\cal A}(R(\tilde{R}))$. The proof of the following proposition is immediate.

\begin{prop}
For a net ${\cal A}$ satisfying assumptions $1-9$ the net $\tilde{A}$ satisfies assumptions $1-9$. 
\end{prop}

However, assumption $8'$ (Haag duality for balls) in ${\cal A}$ does not imply assumption $8'$ for $\tilde{\cal A}$. If the original net is $\pi_0$-complete, the dimensionally reduced one may have one-ball sectors, but not additional two ball sectors:

\begin{prop}
For a $\pi_0$-complete net ${\cal A}$, the dual net of the dimensionally reduced net $\tilde{A}^d$ is $\pi_0$-complete. 
\end{prop}
\noindent {\sl Proof:}
For any ball $\tilde{B}\subset \tilde{E}$ the dual net algebra is the maximal algebra $\tilde{{\cal A}}^d(\tilde{B})=\hat{\tilde{{\cal A}}}(\tilde{B})=\hat{\cal A}(R(\tilde{B}))$, and we have $\hat{\tilde{{\cal A}}}^d(\tilde{B})={\tilde{\cal A}}^d(\tilde{B})$, as must be the case for the dual net.
Taking now two separated balls $\tilde{B}_1,\tilde{B}_2\subset \tilde{E}$ we have $\hat{\tilde{{\cal A}}}^d(\tilde{B}_1\cup \tilde{B}_2)=\hat{\tilde{{\cal A}}}(\tilde{B}_1\cup \tilde{B}_2)=\hat{\cal A}(R(\tilde{B}_1\cup \tilde{B}_2))=\hat{\cal A}(R(\tilde{B}_1))\vee \hat{\cal A}(R( \tilde{B}_2))={\tilde{\cal A}}^d(\tilde{B}_1)\vee {\tilde{\cal A}}^d(\tilde{B}_1)$, were we have used the split property and $\pi_0$-completeness (see proposition (\ref{propsplit}) below for more details) for the net ${\cal A}$.  $\square$

As $\tilde{{\cal A}}^d$ is $\pi_0$-complete, according to proposition (\ref{SSB}), the category of endomorphisms of $\tilde{{\cal A}}(\tilde{B})$ induced by the subfactor $\hat{\tilde{{\cal A}}}(\tilde{B})/\tilde{{\cal A}}(\tilde{B})$ is the one of representations of a (finite) group. This subfactor is in fact the one ${\hat{{\cal A}}(R(\tilde{B}))/{\cal A}(R(\tilde{B}))}$ expressed in terms of the original, non dimensionally reduced theory. By transportability, this gives a classification of the possible non local HDV sectors of any simple $k$ dimensional loop in $d$ dimension. Both $k$-dimensional loops and $d-k-1$-dimensional loops have non local sectors corresponding to group representations. As they come from dual subfactors, the only possibility is that the dual sectors correspond to dual abelian groups.       

\begin{theorem}
For a $\pi_0$-complete theory the subfactor $\hat{\cal A}(S^k)/{\cal A}(S^k)$, corresponding to simple loops of dimension $k$, $k\in \{1\,\cdots\,  d-2\}$, has sectors $\rho_r$ with finite abelian group $G^*$ fusion rules. The fusion of the endomorphisms of the dual loop $S^{d-k-1}$ corresponds to the dual group $G$.  
We can expand a generic element  $x\in \hat{\cal A}(S^k)$ as
\be
x=\sum_{r\in G^*} x_r\, a_r\,.
\ee
The $a_r$, $r\in G^*$, can be chosen unitary non local operators that obey the group multiplication law, $a_{r_1} a_{r_2}=a_{r_1r_2}$, and $x_r\in {\cal A}(S^k)$ are additive operators. We can write $\rho_r(z)=a_r \, z\, a_r^*$ for $z\in {\cal A}(S^k)$.   Write the corresponding expansion for the dual loop, $y\in \hat{\cal A}(S^{d-k-1})$ as
\be
y=\sum_{g\in G} y_g\, b_g\,,
\ee
where $y_g\in {\cal A}(S^{d-k-1})$, and the unitary non local operators satisfy $y_{g_1}y_{g_2}=y_{g_1 g_2}$. For dual simple dual loops we have
\be
a_r \, b_g=\chi(r,g)\,b_g\, a_r\,,\label{45}
\ee
with $\chi(r,g)$ the character of the representation $r$ evaluated on the group element $g$. 
\end{theorem}
 \noindent {\sl Proof:} 
 The fusion of the representations of a group can form a group only for abelian groups, in which case these fusion rules correspond to  the dual of the group. The sectors have dimension $1$ and the charged operators can be taken unitary. Eq. (\ref{45}) is the action of the dual group symmetry on the charged intertwiners, as in eq. (\ref{tyty}), and follows because the charged isometries implement the endomorphisms. See \cite{Hollands:2022gab} Appendix A for a description of the action of non local operators in a Q-system in the case of a finite group. The charged operators corresponding to $S^k$ are unitaries that implement a group of automorphisms on the maximal algebra  of the complement $\hat{\cal A}(S^{d-k-1})$. This automorphism group can be implemented in a standard way using the standard cone construction  in such a way that the charged operators obey the group multiplication rules (see \cite{haag2012local}, section V.2.2). $\square$  
 
 Some remarks follow. First, a slightly different argument for the abelianity of higher dimensional form-symmetries for $\pi_0$-complete theories was previously given in  \cite{Casini:2020rgj}. Results in the next two sections will put such an argument in a rigorous footing, see below. Second, although the previous theorem classifies sectors for $k$-dimensional loops in $\pi_0$-complete theories, notice that for non $\pi_0$-complete theories the structure of sectors is still very simple since it comes from an abelian group quotiented by another group. Finally, in \cite{Gaiotto:2014kfa} it was argued that if the sectors form a group it must be abelian. This is because the fusion itself must be abelian since two endomorphisms can be localized to act with spatially separated support.

\section{Further consequences of $\pi_0$ completeness}\label{SecV}

The DHR theorem allows us to complete two-ball sectors by uplifting the theory to the theory ${\cal B}$. Hence, in the following we will assume that this step has been taken and we have a theory without two-ball sectors.  We assume such a $\pi_0$-complete theory, satisfying assumptions $1,2',3-6,7',8',9$, and for simplicity we still call it ${\cal A}$ . As we have seen, this considerably simplifies the classification of sectors for other regions. It is the purpose of this section to establish some further basic consequences of the absence of two ball sectors. Note the theory also does not have sectors for shells, that are dual to two balls. Hence $\pi_0$ completeness is the same as $\pi_{d-1}$ completeness. In this section,  for the first time, we will make essential use of the weak modularity assumption.

\subsection{Topologically separated regions}

We assume a $\pi_0$-complete theory. Then the additive algebra of two disjoint topological balls $B_1,B_2$ separated by a non zero distance is also the maximal algebra. Because we can insert a split between the two balls, this algebra is isomorphic to the tensor product of the algebras of each ball.  Consider two regions $R_1\subset B_1$ and $R_2\subset B_2$. We say that in this situation $R_1,R_2$ are {\sl topologically separated}. 

\begin{prop} \label{propsplit} For two topologically separated regions $R_1, R_2$ in a $\pi_0$-complete theory with the split property we have $\hat{\cal A}(R_1\cup R_2)=\hat{{\cal A}}(R_1)\vee \hat{\cal A}(R_2)\simeq \hat{{\cal A}}(R_1)\otimes \hat{\cal A}(R_2)$. \end{prop}

\noindent {\sl Proof:}
We have from monotonicity of maximal algebras and the split property 
$
\hat{\cal A}(R_1\cup R_2)\subset {\cal A}(B_1)\vee {\cal A}(B_2)\simeq {\cal A}(B_1)\otimes {\cal A}(B_2)$. 
Using the theorem of distributivity of intersections of factors in a tensor product \cite{ge1996tensor} it follows that
$ \hat{\cal A}(R_1\cup R_2)=\hat{\cal A}(R_1\cup R_2)\cap ( {\cal A}(B_1)\vee {\cal A}(B_2))\simeq (\hat{\cal A}(R_1\cup R_2)\cap  {\cal A}(B_1))\otimes ( \hat{\cal A}(R_1\cup R_2)\cap {\cal A}(B_2))
 =  \hat{{\cal A}}(R_1)\otimes \hat{\cal A}(R_2)$. $\square$
 
We also have ${\cal A}(R_1\cup R_2) \simeq  {\cal A}(R_1)\otimes {\cal A}(R_2)$. Therefore, the theory of sectors for topologically separated regions is just the tensor product of the ones for the two regions $R_1,R_2$. Any irreducible endomorphism coming from the subfactor for $R_1\cup R_2$ is equivalent to $\rho_1 \,\rho_2$, with $\rho_1$, $\rho_2$, irreducible endomorphisms for $R_1$ and $R_2$, and viceversa. Clearly, the same happens for a larger number of topologically separated regions, allowing to partially reduce the problem of classification of possible sectors to regions that are not topologically separated.  

 The following is an immediate consequence

\begin{corollary}
In a $\pi_0$ complete theory the following pairs of regions are transportable: $R$ and $R\cup B$ for a ball $B$ separated from $R$, and, dually, $R-B$ and $R$ for $B\subset R$ a ball separated from the boundary of $R$.

\end{corollary}

\subsection{Connected sums}

In this section we prove other transportability properties  for $\pi_0$-complete theories. In particular, this is the case for direct sums of regions. This property will allow us to understand the structure of sectors for general regions in $d=4$ in complete generality. 

\noindent {\bf Definition:} Given two topologically separated regions $R_1\subseteq B_1$ and $R_2\subseteq B_2$, where $B_1,B_2$ are separated topological balls,  and a topological ball $C$ with $C\cap B_1=\emptyset$, $C\cap B_2=\emptyset$, we say that $R_1\oplus R_2=R_1\cup R_2\cup C$ is a {\sl connected sum} of $R_1$ and $R_2$ if $R_1\cup C\cup B_2$ is a topological deformation of $R_1$ and $R_2\cup C\cup B_1 $ a topological deformation of $R_2$. 

We will call a shell to the topology of a region $S=B_1-B_2$ where $B_2\Subset B_1$ are two balls in $\mathbb{S}^d$.

\begin{prop} For a $\pi_0$ complete theory,  for two topologically separated regions $R_1,R_2$ the union $R_1\cup R_2$ is transportable to a connected sum $R_1\oplus R_2$.\end{prop}

\noindent {\sl Proof:} 
Let $R_1$, $R_2$, $B_1$, $B_2$ and $C$ as above. We call $R_1\cup R_2=R$, $R\cup C=\tilde{R}$.  We have to prove $\hat{{\cal A}}(R)\vee {\cal A}(C)=\hat{{\cal A}}(\tilde{R})$, and $\hat{{\cal A}}(\tilde{R}')\vee {\cal A}(C)=\hat{{\cal A}}(R')$. 
Both assertions follow quite directly from weak modularity.  
For the second equation we can use the split property instead. We first prove this second equation without using weak modularity. The dual, using Haag duality for balls, is, ${\cal A}(\tilde{R})\cap {\cal A}(C')={\cal A}(R)$. We have ${\cal A}(B_1\cup C\cup B_2)\cap {\cal A}(C')={\cal A}(B_1\cup B_2)$. This is because, under the assumption of of $\pi_0$-completeness, all the involved algebras are maximal. Then it follows that the algebra ${\cal A}(\tilde{R})\cap {\cal A}(C')\subset {\cal A}(B_1\cup B_2)={\cal A}(B_1)\vee {\cal A}(B_2)\simeq {\cal A}(B_1)\otimes {\cal A}(B_2)$, where the last relation follows from the split property. Distributing the intersection in the tensor product, we have to compute  ${\cal A}(\tilde{R})\cap {\cal A}(C')\cap {\cal A}(B_1)={\cal A}(\tilde{R})\cap {\cal A}(B_1)$, and analogously for $B_2$. We have, using strong additivity,  ${\cal A}(R_1)\subseteq {\cal A}(\tilde{R})\cap {\cal A}(B_1)\subseteq  ({\cal A}(R_1)\cup {\cal A}(C\cup B_2))\cap {\cal A}(C\cup B_2)'={\cal A}(R_1)$. This last equation follows from  the topological transportability between $R_1$ and $R_1\cup C\cup B_2$. Therefore we get ${\cal A}(\tilde{R})\cap {\cal A}(C')={\cal A}(R_1)\vee {\cal A}(R_2)={\cal A}(R)$. 

The first equation,  $\hat{{\cal A}}(R)\vee {\cal A}(C)=\hat{{\cal A}}(\tilde{R})$, is valid in general, even if DHR sectors are not eliminated. It is equivalent to ${\cal A}(R')\cap {\cal A}(C')={\cal A}(\tilde{R}')$.   Consider another topological ball $C_1$, $C\Subset C_1$, and call its intersection with $\tilde{R}'\cap C_1=V$ . Then we apply weak modularity (splitting $R'=(V\cup C)\cup (\tilde{R}'-V)$, where this last region is interior to $C'$, to get $ {\cal A}(C')\cap {\cal A}(R')={\cal A}(\tilde{R}'-V)\vee ({\cal A}(C')\cap {\cal A}(V\cup C))$. Now ${\cal A}(C')\cap {\cal A}(V\cup C)= \hat{{\cal A}}(V)$ because it is an intersection of maximal algebras. The geometry is such there is a shell $S\subset R'$, $C\cup V\subset S$, $S-C=\tilde{R}'\cap S$. So $S-C$ is a topological ball contained in $\tilde{R}'$, and 
we can replace ${\cal A}(\tilde{R}'-V)\vee \hat{\cal A}(V)$ by ${\cal A}(\tilde{R}'-V)\vee {\cal A}(V)={\cal A}(\tilde{R}')$ by transportability of $S-C$ and $S-C-V$. $\square$

Notice that the transportability under connected sums is equivalent to its dual, the transportability under taking a region which contains a non trivial shell $S$, and piercing a hole (only one) through $S$.

\subsection{Transportability under the action of a conditional expectation}

Now we extend the previous result about connected sums to the case in which the region is not the union of topologically separated regions. In fact we are now gluing with a topological ball two patches of the boundary  of a region, independently of whether the two boundaries belong to the same component or to separated components of the region. Of course, in this case it is in general not true the region is transportable to the glued one, but we will prove that there is transportability under the action of a conditional expectation. 

To be more precise, let us recall the definition of transportability, 
where two regions $R_1, R_2$, $R_1\subset R_2$,  are said transportable to each other when we have  
 $\hat{{\cal A}}(R_1)\vee {\cal A}(R)=\hat{\cal A}(R_2)$ and  $\hat{{\cal A}}(R_2')\vee {\cal A}(R)=\hat{\cal A}(R_1')$, where we have called $R=R_2-R_1=R_1'-R_2'$. This leads to proposition \ref{prop1} that states the equivalence between the subfactors ${\cal A}(R_1)\subset\hat{\cal A}(R_1)$ and ${\cal A}(R_2)\subset\hat{\cal A}(R_2)$. In the proof of this equivalence,  only three points about this definition are relevant. First, the algebras ${\cal A}$ are additive, which in the present case only requires  ${\cal A}(R_1)\vee {\cal A}(R)={\cal A}(R_2)$ and ${\cal A}(R_2')\vee {\cal A}(R)={\cal A}(R_1')$. Second, the definition of $\hat{\cal A}$ for a region is the commutant of the additive algebra of the complement, and then is determined only by the additive algebra. Third, we need that what plays the role of the additive algebras are still causal for disjoint regions. Under this more abstract understanding, the idea of transportability and proposition 1 can be greatly generalized.

 Here we are interested in the situation described above. We have a region $R_1$ and two disjoint (topological) disks $D_1,D_2,$ in the boundary of $R_1$. Let $R$ be a topological ball, $R\cap R_1=\emptyset$, such that its boundary intersects $R_1$ in $D_1\cup D_2$. We call $R_2=R_1\cup R$ a gluing of boundaries in $R_1$, and $R$ the gluing tube. In general there is no ordinary transportability between $R_1$ and $R_2$ in this case. In particular, in the complement $R_2'$ there is a simple loop region $S$ surrounding $R$ (whose boundary covers the boundary of $R$ except for $D_1,D_2$) that is in general not contractible inside $R_2'$ and then can have non trivial non-local operators for $R_2'$. These, however, are local in $R_1'$ since $S$ is contractible in $S\cup R\subset R_1'$. 
 
 In this scenario, we are going to prove transportability where the role of additive algebras are played by ${\cal A}(R_1)$, $ {\cal A}(R_2)={\cal A}(R_1)\vee {\cal A}(R)$, and, for the complementary regions, $\tilde{{\cal A}}(R_2')={\cal A}(R_2')\vee \hat{\cal A}(S)$, and $\tilde{{\cal A}}(R_1')=\tilde{{\cal A}}(R_2')\vee {\cal A}(R)={\cal A}(R_1')$.  In terms of the maximal algebras, the transportability reads $\hat{\cal A}(R_1)\vee {\cal A}(R)=E_S(\hat{\cal A}(R_2))$, $\hat{\cal A}(R_2')\vee {\cal A}(R)=\hat{\cal A}(R_1')$. 
 We have written $E_S$ to the conditional expectation induced by the non local operators of $S$, $E_S(x)=|G|^{-1}\, \sum_{g\in G} g\, x\, g^{-1}$. The algebra $E_S(\hat{\cal A}(R_2))$ only contains operators that commute with the non local operators of $S$ and coincides with $(\tilde{{\cal A}}(R_2'))'$. We call this condition  ``transportability under the action of a conditional expectation''.

 \begin{prop} \label{ppp}
 A region is transportable to a gluing of its boundaries under the conditional expectation generated by a loop  surrounding the gluing tube (see previous paragraph for notation).
 \end{prop}

\noindent{\sl Proof:}
The statement follows directly from weak modularity in an analogous way as the proof of transportability under direct sums. Namely, the proof of $({\cal A}(R))'\cap {\cal A}(R_2)={\cal A}(R_1)$ follows from weak modularity by separating from $R_2-R$ two topological balls $B_1,B_2$ whose boundaries cover $D_1,D_2$, and the fact that for the two balls there is Haag duality. The complementary relation $({\cal A}(R))'\cap {\cal A}(R_1')=\tilde{{\cal A}}(R_2')$ follows from weak additivity by separating the loop $S$ whose maximal algebra joined with ${\cal A}(R_2'-S)$ gives $\tilde{{\cal A}}(R_2')$.
$\square$

\section{Links and knots in $d=3$}
\label{SecVI}

 The previous analysis has simplified the classification problem in any dimensions in several aspects. First, we can restrict ourselves to $\pi_0$-complete theories. This trivializes the cases of regions containing shells or, dually, topologically separated regions. The case of topological sums can also be decomposed into the summands. In particular, for $d=2$ there are no further sectors for $\pi_0$-completed theories. 

In this section we classify the theory of sectors associated to arbitrary connected compact regions and their complements in $d=3$. A connected compact submanifold of $\mathcal{S}^3$ is the complement of a union of handlebodies, that in general can be embedded in a linked and knotted way \cite{fox1948imbedding}. Unlinked and unknotted handlebodies embedded in $\mathcal{S}^3$ can be already understood from the previous results: the sectors form the abelian group $G^n$ where $n$ is the total number of handles. Thus, we essentially need to understand the case of linked, knotted, and glued loops. 

We first start by understanding the role of the loop orientation and uncover a new constraint that reduces the possibilities for the group of ring sectors. 

\subsection{Orientation and a restriction on the possible abelian groups}

Let us fix a single unknotted ring-like region $R_0$ with topology $\mathbb{S}^1\times D^2$. Any other unknotted ring $R$ can be obtained by  smoothly deforming $R_0$ in two inequivalent ways labeled by the orientation of the mapping. 
Therefore, in labeling the irreducible endomorphisms (or equivalently the loop non local operators) corresponding to $R_0$ in terms of the elements of the abstract group $G$, it is convenient to fix an orientation to $R_0$ for this association. For any other oriented ring $R$ we associate the irreducible endomorphism $\sigma_g$ in $R$ to the one corresponding to $g$ in $R_0$ transported by preserving the orientation. This gives us for each ring two possible labelings of the endomorphisms, say $\sigma_g^\pm$ according to the two orientations.

Now we take two rings $R_1,R_2$ that share an interval $I$, such that $R_1-I, R_2-I$, are topological balls, and $(R_1\cup R_2)-I$ is a new simple ring $R_{12}$. The theorem on direct sums shows the sectors of $R_{12}$ decompose into products of the ones of $R_1,R_2$. Let us take the two rings $R_1,R_2$ with the same orientation dictated by a fixed orientation of $R_{12}$. This orientation is opposite along $I$ for the two rings. It follows by testing with dual loops simply linked with any of the rings that $\sigma_g(R_{12})\sim \sigma_g(R_1)\sigma_g(R_2)$. That is, the product of the same group elements $g$ for two rings that share some segment, having opposite orientations along the segment, is transportable to the same $g$ for the composed ring with the compatible orientation. 

Then, taking a simple ring $R_3$ linked once with $R_1$ and $R_2$ in the above construction it follows that $\sigma_g(R_1)\sigma_g(R_2)$, being equivalent to the sector $\sigma_g(R_{12})$ inside $R_1\cup R_2$, that is unlinked to $R_3$, is equivalent to the identity in $R_3'$. But in $R_3'$ the product $\sigma_g(R_1)\sigma_g(R_2)$ is equivalent to the product of $\sigma_g^+$ and $\sigma_g^-$ for sectors of a single ring. Therefore, we conclude the labeling for different orientations corresponds to conjugate sectors:
\be
\sigma_g^-=\sigma_{g^{-1}}^+\,.
\ee
In other words, by rotating a round ring an angle $\pi$ along an axis on the plane of the ring, we take sectors to their conjugate ones. We note that this observation, together with the results on direct sums in the previous section, puts the original heuristic argument of Ref. \cite{Casini:2020rgj} concerning the abelianity of ring-like sectors on a rigorous footing.  

For $d=3$ the complement of a simple ring is again a simple ring. 
We take a simple oriented ring and transport it to its complement with the orientation determined by having a linking number $1$ (or using the right hand rule to fix this orientation). We call this the standard relative orientation for the linked loops. In this way we obtain a fixed isomorphism  $\phi:G\rightarrow G^*$ between the group and its dual.  It follows that the standard orientation is symmetric under interchanging the order of the loops. The transport to the other orientation gives the same isomorphism $\phi$ composed with conjugation. We write the commutation relation for the loop operators for these simply linked simple loops as
\be
\psi^1_{g_1}\, \psi_{g_2}^2= \chi_{g_1,g_2} \, \psi_{g_2}^2\,\psi^1_{g_1}\,,\quad\textrm{(standard relative orientation)}\,,\label{eqq}
\ee
where we have identified $\phi(g_2)\rightarrow  g_2$. This defines the character table with a particular ordering of the rows (or columns), that obeys $\chi_{g_1,h}\,\chi_{g_2,h}=\chi_{g_1 g_2, h}$, $\chi_{h,g_1}\,\chi_{h,g_2}=\chi_{h,g_1 g_2}$.  
Interchanging the role of the two oriented regions we get
\be
\psi^2_{g_2}\, \psi_{g_1}^1= \chi_{g_2,g_1} \, \psi_{g_1}^1\,\psi^2_{g_2}\,.
\ee
From this and (\ref{eqq}) we get that the character matrix is Hermitian
\be
\chi_{g_1,g_2}=\chi_{g_2,g_1}^*\,.
\ee
This significantly reduces the space of possible abelian groups. In particular, diagonal elements of the character table have to be $\pm 1$. A loop operator either commutes or anticommutes when linked to itself. Note that a Hermitian character table automatically ensures there is an isomorphism between the group of columns and rows given by the assignation of column $j$ to row $j$.

\begin{prop} \label{qqq}
 The group $G$ of loop sectors in $d=3$ is either of the form $H\times H$ or $\mathbb{Z}_2\times H\times H$. 
 \end{prop}

\noindent {\sl Proof:} 
We start from a finite abelian group $G$ with an isomorphism $G\stackrel{*}{\leftrightarrow} G^*$, such that the character table $\chi_{g_1,g_2}$ is Hermitian. We are identifying the elements of $G^*$ with the ones of $G$ using the isomorphism. The group $G$ is divided into $G^+$ formed by elements $g$ with $\chi_{g,g}=1$, and $G^-$, formed by elements with $\chi_{g,g}=-1$. Because of the Hermiticity of the character table, and that all entries are phases, it follows that $\chi_{g_1,g_1}\chi_{g_2,g_2}=\chi_{g_1,g_1}
\chi_{g_1,g_2}\chi_{g_2,g_1}\chi_{g_2,g_2}=\chi_{g_1 g_2,g_1 g_2}$.  Therefore, $G^+$ is a subgroup of $G$, and for $g_1,g_2\in G^-$, the product $g_1 g_2\in G^+$. Then, if $G^-$ is non empty, taking the quotient $G/G^+$ it follows that there is only one class generated by the elements of $G^-$, since the product of any two elements gives the identity class. Therefore, if $G^-$ is non empty it follows that $G\simeq\mathbb{Z}_2\times G^+$, where $G^+$ has a Hermitian character table with entries $1$ in the diagonal. 

Then consider the case in which $G=G^+$. 
For any subset $H\subseteq G$ call $H'\subseteq G$ to the set of $g$ such that $\chi_{H,g}=1$. It follows $H'$ is a subgroup. It follows that $H\subseteq H''$ is the smallest subgroup containing $H$, and $H''=H$ for a subgroup $H$. It also follows that for a subgroup $H\subset G$, $H'\subseteq G$ is isomorphic to $G^*/H^*$ and hence to $G/H$. Let us consider subgroups $H$ such that $H\subset H'$. We seek to find a maximal subgroup with this property. These types of groups always exist, 
 an example is the group generated by any element $g$. Then, parting from a group $H\subset G$ such that $H\subset H'$, if there is a $g\in H', g\notin H$, we can form the group $[g,H]$ generated by $g$ and $H$. This also satisfies $[g,H]\subseteq [g,H]'$. In this way we enlarge the group $H$ until we get $H=H'$. But $H'\simeq G/H$, therefore $G\simeq H\times H$. 
$\square$

In the case where $G$ is a group square, the proof shows that $H$ can be chosen such that $H=H'$, that is, the non local operators of $H$ commute with themselves and form a maximal subgroup of such operators. This implies a Haag-Dirac (HD) net can be constructed \cite{Casini:2020rgj}. A HD net is an intermediate  net
\be
{\cal A}(R)\subseteq {\cal A}_{HD}(R)\subseteq ´\hat{\cal A}(R)\,,
\ee
that is causal and Haag dual at the same time, though if there are non local operators, it is not additive.  
In the present case it is given by choosing the algebras ${\cal A}_{HD}(R)={\cal A}(R)\vee H$ for any loop region. This choice always exist in this case, and in general is non unique. 
On the contrary, in the case in which there is an additional factor $\mathbb{Z}_2$ it is not possible to enlarge the additive algebras of loops with non local operators in the same way for all regions such that the net is Haag dual. This is because if $a^2=1, \chi_{a,a}=-1$ is the non trivial element in the $\mathbb{Z}_2$ factor, we have $\chi_{a, h_1 h_2}=1$. Then we would have to add $a$ for obtaining Haag duality, at the expense of loosing locality.

Examples of these type of sectors in $d=3$, without the extra $\mathbb{Z}_2$ factor, are given by Lagrangian models based on a Lie algebra $\mathbb{L}$ for non abelian gauge groups. For pure gauge theories (no matter fields), the group of loops for ring-like regions is given by $G=H\times H^*$, where $H$ is the center of the universal cover of $\mathbb{L}$. Here $H^*$ labels non additive Wilson loops, and $H$ non additive 't Hooft loops. $G$ labels generic ``dyonic'' loops.  Any finite abelian group $H$ can be attained in this way, for example, by multiplying theories based on $SU(n)$ gauge groups, which have $H=\mathbb{Z}_n$.  The character table for this duplicate group $G$ is simply\footnote{See \cite{Casini:2020rgj,Casini:2021zgr} for a review in line with the setup of the present article.}
\be
\chi^{(G)}_{(h_1,h_1^*),(h_2,h_2^*)}=\chi^{(H)}_{h_1,h_2^*}\,(\chi^{(H)}_{h_2,h_1^*})^*\,.
\ee
This is automatically Hermitian for any $H$.

Mater fields of different gauge group representations can be incorporated to the model. Let the algebra generated by the matter fields have some specific charges $d=(h,h^*)$ with respect to $G$. The algebra of all these charged fields generates a subgroup $D\subset G$. For any $d\in D$ the corresponding loop operator is not any more a non local operator in the ring, but rather an additive operator. This is because it now can be decomposed into Wilson lines ended by the charged operators. As these dyon loop operators  are additive for complementary rings, it must be the case that they commute to their representatives in the linked loops:
\be
 D\subset D'\,.
\ee
This is a generalized Dirac quantization condition for the physical gauge charges. The operators that can be still considered loop operators (and not surface operators) must also commute with $D$, and hence be part of $D'\simeq G/D$. But two such operators that can be transformed to each other by an additive operator in $D$ belong to the same non local class. Then the group of non local opertors is $\tilde{G}=D'/D$ or equivalently $\tilde{G}=(G/D)\cap (D)'$. In particular, the dimension of the groups generated in this way is always a square $|G|^2/|D|^2$. More concretely, it is given by elements $(h,h^*)D$ such that
\be
\chi_{(h,h^*)D,D}=1\,,
\ee
and the new character matrix is 
\be
\chi_{(h_1,h_1^*)D,(h_2,h_2^*)D}=\chi_{(h_1,h_1^*),(h_2,h_2^*)}\,,
\ee
that is again Hermitian. 

It seems there are no known examples or candidates for models where $G$ has an additional factor of $\mathbb{Z}_2$.

\subsection{Self-dual case in other dimensions}

This restriction of the group does not hold for loops that are not topologically equivalent to their duals. Since the loop of dimension $k$ is dual to a loop of dimension $d-1-k$, this may happen only for odd $d$ and the loop dimension $k=(d-1)/2$. Let us analyze then the general case of odd $d$ and $k=(d-1)/2$. To understand the generalization of the geometry of the above construction in more dimensions we can think we have a $(d-1)/2$ volume form $\omega$ on the loop, giving its orientation. Let $e$ be a unit orthogonal vector, separating the loop from the dual loop. The form $\omega'=(\omega\wedge e)^*$ gives the analogous ``right hand rule'' for the volume form for the linked loop. Here the symbol $*$ is the Hodge operation on forms. The question is: start now from this new volume form and construct again the dual in this way, do we get the original orientation for the original loop? In doing so we have to replace $e\rightarrow -e$ for the relative distance vector. That is, we have to compute
\be
((\omega\wedge e)^* \wedge -e)^*= (-1)^{(d+1)/2}\, \omega\,.
\ee
The change of sign according to the dimension comes from the square of the Hodge operation.  Therefore when $(d+1)/2$ is even  (i.e. $d=4 n -1$, as for $d=3$), by the same calculation in the last subsection  we get hermitian character matrices, and the same possible structures for the group $G$. When $(d+1)/2$ is odd (i.e. $d=4 n+1$, as for $d=5$) the orientation has changed after two operations of taking a loop to its dual, and following the above calculation we only get a symmetric character matrix 
\be
\chi_{g_1,g_2}=\chi_{g_2,g_1}.
\ee
This can always be achieved when ordering the character table with an isomorphism to the dual group. To see this, notice that all finite abelian groups are products of cyclic groups. For a cyclic group of order $n$ the elements can be labeled by an integer $q=0,1,\cdots,n-1$, and the dual group by integers $g=0,1,\cdots,n-1$, such that the character table reads
\be
\chi_{q,g}=e^{2\pi i\, \frac{q\,g}{n}}\,.
\ee
This gives a symmetric matrix identifying $\phi(q)=q$.
Therefore, no restriction on the group appears for these dimensions. 

Simple, relativistic examples with $k=(d-1)/2$ and non Hermitian character table, that however have sectors with group $\mathbb{R}$, and hence infinite index, can be constructed with free fields. These are the self dual free massless one form field (a chiral scalar) for $D=2$, a self dual free massless three form field (i.e. a gauge two-form field) for $D=6$, and analogous examples for $D=2+4 n$. For interacting models see \cite{sen2020self}. Non local operators are given by exponentials of the (smeared) conserved fluxes of these form fields over surfaces $\Sigma$, that are non local on the thickened boundary of $\Sigma$. The analogous for $D=4$ would be a self dual Maxwell two-form field:
\be
F^*=\pm i\, F\,.\label{68}
\ee
The $i$ in this expression is necessary for self-consistency because the square of the Hodge operation (in Minkowski signature) gives $F^{**}=-F$ in $D=4$. These fields describe fotons with only one helicity $\pm$ \cite{bialynicki1981note}. However, because of (\ref{68}), the field cannot be self-adjoint. Therefore, it does not have observables or local von Neumann algebras. Incorporating the adjoint of the field recovers the full Maxwell field that has a Hermitian character table.   

Under the impression of this argument, a putative example of group of loop non local operators $G=\mathbb{Z}_2$ in $d=3$ would resemble a ``chiral SU(2)'' gauge theory. In any case, it is to be noted that this type of model, if it exists, could not be a weakly coupled theory. According to the entropic certainty relation \cite{Casini:2020rgj} it could not be in a spontaneous symmetry breaking phase or a confining phase either. These scenarios dynamically choose a HD net, that does not exist for this group. If such a model exists, it is probably stuck in a conformal phase. 

\subsection{Discrete space-time symmetries}

In the general setup of this paper we have studied topological properties of the nets without invoking space-time symmetries. In this subsection  we make a short depart from this line and study the transformation properties of loop operators under discrete space-time symmetries.  

For a relativistic theory with Lorentz covariance, the CRT antiunitary symmetry acts geometrically as the transformation $(x^0,x^1, x^2,\cdots,x^{d})\rightarrow (-x^0,-x^1, x^2,\cdots,x^{d})$. This transformation is implemented by the Tomita-Takesaki reflection $J$ corresponding to the algebra of the Rindler wedge and the vacuum state. It is interesting to understand how the sectors transform under $J$. 
 Let us call $\bar{R}$ to the CRT transformed region, that we take with the orientation that results from the orientation of $R$ when composed with the CRT operation. In the $x^0=0$ description of a region, rings that are  symmetric for rotations around the $z$ axis have a dimensionally reduced description in the $xz$ plane, $x>0$, according to the analysis of section \ref{SecIV}. In this description the loop sector is seen as a DHR sector of the dimensionally reduced $d=2$ theory.
 Taking reflections on the $xy$ plane, the action of $J$ on the particular rotationally symmetric loops is equivalent in the original and dimensionally reduced theory.  $J$ takes a DHR sector to a conjugate one   \cite{guido1992relativistic}. Therefore we get 
  for these loops 
 \be
J\, \sigma^{(R)}_g\, J\simeq  \bar{\sigma}^{(\bar{R})}_g=\sigma_{g^{-1}}^{(\bar{R})}\,.
\ee
 In this case, the loop $\bar{R}$ is parallel to $R$ and has the orientation obtained by translation. It is not difficult to see that composing with deformations, the same formula holds for all other loops.  In particular, if we take a loop orthogonal to the plane of reflection,  $J$ will change its orientation at the same time that conjugates the sector. This means it will keep the sector invariant. For linked loops the operation changes relative standard orientation by the opposite one and conjugate both sectors.  
 This is necessary for consistency  because the antiunitary $J$ conjugates the character in the commutation relation     $\chi_{g_1,g_2}^*=\chi_{g_1^*,g_2}=\chi_{g_1,g_2^*}$.

Suppose now the theory has a time reflection symmetry implemented by the antiunitary operator $T$. This keeps the regions we are considering, the additive algebras, and the maximal ones invariant. Then it has to map irreducible sectors into irreducible sectors. This induces a group isomorphism $t:G\rightarrow G$, with $t^2=1$. We keep the orientation invariant in this definition of $t$. Applying time reflection to the relation (\ref{eqq}) we get 
\be
\chi_{t(g_1),t(g_2)}= \chi_{g_1,g_2}^*=\chi_{g_1^*,g_2}=\chi_{g_1,g_2^*}\,.
\ee
 When there are complex characters, this relation roughly tells that time reflection conjugates half of the sectors, while leaving intact the other half. 

Finally, a reflection is unitary and takes $R$ to $\bar{R}$ that we again take with the reflected orientation. For linked loops it must not change the character but changes the standard relative orientation to the opposite one. Then
\be
\chi_{r(g_1),r(g_2)}=\chi_{g_1,g_2}^*\,.
\ee

\subsection{Abelianity of sectors for linked and knotted loops}

We now finally analyze the sectors for arbitrary linked and knotted handlebodies immersed in $\mathbb{S}^3$. A description of these general links, with possible joinings, is simplified when we think about them thickened one dimensional graphs. In turn, these can be described by their projection on a two dimensional plane as a bounded graph where the vertices are the meeting of lines. The vertices can be of three types. One is a join of lines, corresponding to a glue of tubes in the handlebody. The other possibilities describe an upper or a lower crossing of lines, corresponding to the projection of an upper or lower crossing in three dimensions. 

\begin{prop}\label{www}
The sectors of a knotted and linked multicomponent handlebody embedded in $\mathbb{S}^3$ form an abelian group $G^n$, with $n=\sum n_i$, and $n_i$ the genus of the different connected components.
\end{prop}

\noindent {\sl Proof:} We convert all upper and lower crossings of lines into gluings of four lines with corresponding conditional expectations, using the result of proposition \ref{ppp}.
Then, the resulting theory of sectors is the one of a graph where all vertices are gluings, subject to several conditional expectations generated by loops wrapped around some of the vertices of four lines. These loops cross above and below the subsequent edges (four in total) that appear as the loop turns around the vertex. The only difference between the original upper or lower crossing is that any given edge is under-crossed or upper-crossed by the loop generating the conditional expectation. By the theorem on direct sums, the sector theory of the graph with all gluings is just the abelian group $G^{n_L}$, with $n_L$ being the sum of the total number of loops (with empty interior) on each connected component of the graph. A generic non-local operator can be chosen of the form  $\psi_{g_1} \cdots \psi_{g_{n_L}}$, where each $\psi_{g_k}$ is a non local operator corresponding to the loop $k$, and all non local operators commute with each other. If $\psi_{g_1}, \psi_{g_2}, \psi_{g_3}, \psi_{g_4}$  are non local operators corresponding to four loops that meet in a crossing, and these are labeled with the same orientation,  the conditional expectations corresponding to this vertex gives the relation $\psi_{g_1} \psi_{g_2}^{*} \psi_{g_3} \psi_{g_4}^{*}=1$.  If only three loops meet at a crossing because the vertex is on the boundary,  we get the relation $\psi_{g_1} \psi_{g_2}^{*} \psi_{g_3}=1$.  
Therefore the final result is again an abelian group, of the form $G^q$, for some integer $q$. We only need to determine how many of the constraints are independent. To see this note that the constraints are the same for the upper or lower crossings. Then, for each loop of the handle-body we can change the crossings such as to make it unknotted and unlinked. The result follows.  
$\square$

 Therefore, knotting and linking does not change the group describing the fusion of the sectors.  The proof also teaches us how the non local operators for knots are actually constructible from simple loop operators. This is done by first including the knot into a set of loops in direct sum, and using the simple loop operators, local additive operators, and projections to some particular sectors by using conditional expectations generated by dual loop operators.  

The sector theory for the complement $R'$ of the knotted and linked multi-component handlebody $R$, which is an arbitrary single component region, is again the Pontriagyn dual of the group $G^n$, that for finite groups is isomorphic to $G^n$. It is evident that the non local operators for these more general regions can be constructed using loop operators for loops in $R'$ non trivially linked with $R$.  For general multi-component regions in $d=3$ the same argument gives an abelian group sector theory that is a  power of $G$ with an exponent that is the sum of the ones of each single component. This follows again using the proposition \ref{ppp} to convert these multi-component regions to direct sums under some conditional expectations.  

For an abelian group theory of sectors, the non local operators $\psi$ can always be chosen such that they furnish a faithful representation of the group. This is done e.g. by taking the standard unitary implementation of the automorphism group that the non local operators induce on the maximal algebra of the complement.  Therefore, they can always be chosen to commute with each other. In some situations this is not the most natural choice. For example, for a region formed by two linked simple loops, a natural choice  consists of products of loop operators for each of the two loops, and these operators do not commute with each other.

We turn now to analyze the commutation relations of loop operators more generally. Consider a knot $K$. The associated group of sectors is $G$. According to the proof of proposition \ref{www}, these sectors can be constructed from the ones of the graph $M$ of a projection of $K$ on a plane. In this graph all crossings are gluings of lines. The general sector of $K$ is a product of simple loop sectors, and the sectors of $K$ are obtained by projecting with particular conditional expectations at the crossings. It follows that, taking all loops in $M$ oriented in the same way in the plane, and considering the constraints of the conditional expectation at a vertex, the total charge carried by a line (the product of the elements of the loops shared by the line) with orientation towards a vertex is the same as the one carried by the continuation of the line from the vertex with the same orientation. Therefore, we can assign a consistent orientation to $K$, and associated to this orientation there is a definite labeling of non local operators.         

Consider now two oriented knots $K_1, K_2$ that are linked to each other. We want to compute the commutation relation between the corresponding no local operators $\psi_{g_1}^{(1)}, \psi_{g_2}^{(2)}$.

\begin{prop}\label{yyy}
The non local operators of two oriented knots $K_1,K_2$ satisfy
\be
\psi_{g_1}^{(1)} \,\psi_{g_2}^{(2)}=(\chi_{g_1,g_2})^{l(K_1,K_2)}\, \psi_{g_2}^{(2)}\,\psi_{g_1}^{(1)} \,, \label{ooo}
\ee
where $l(K_1,K_2)$ is the linking number of the oriented knots.
\end{prop}

\noindent {\sl Proof:} 
 Again we look at a projection of the two knots to a plane. There are a number of intersection points of the graphs of the two knots with upper or lower crossings. Suppose $K_2$ upper crosses $K_1$ at a certain vertex $v$. Take two points $x_1,x_2\in K_2$ at each side of $v$ and near enough of $v$. This determines an interval $J\subset K_2$ that contains $v$. We also join $x_1,x_2$ by an arc $I$ outside $K_2$ that is under-crossing $K_1$, such that $J\cup I$ forms a simple loop that is simply linked to $K_1$. By the theorem of direct sums, the non local operators of $K_2$ can be decomposed as products of commuting ones of $J\cup I$ and $I\cup (K_2-J)$. We take these regions with the orientation inherited from $K_2$ along $J$, and the orientation inherited by $K_2-J$, respectively. The element $\psi_{g_2}^{(2)}\simeq \psi_{g_2}^{J\cup I}\, \psi_{g_2}^{I\cup (K_2-J)}$, where these operators are transformed to each other by unitary additive operators in the union, that commute with non local operators of $K_1$. Therefore 
 \bea
&& (\psi_{g_1}^{(1)})^{-1} \,(\psi_{g_2}^{(2)})^{-1}\,\psi_{g_1}^{(1)} \,\psi_{g_2}^{(2)}=(\psi_{g_1}^{(1)})^{-1} \,(\psi_{g_2}^{I\cup (K_2-J)})^{-1}\,(\psi_{g_2}^{J\cup I})^{-1}\, \psi_{g_1}^{(1)} \,\psi_{g_2}^{J\cup I}\, \psi_{g_2}^{I\cup (K_2-J)}\nonumber\\
&&=\chi_{g_1,g_2}^{s}\,(\psi_{g_1}^{(1)})^{-1} \,(\psi_{g_2}^{I\cup (K_2-J)})^{-1}\, \psi_{g_1}^{(1)} \, \psi_{g_2}^{I\cup (K_2-J)}\,,
\eea
where $s=1$ when the relative orientation between $J\cup I$ and $K_1$ is standard, and $-1$ in the opposite scenario. In this way we have changed the problem of computing the phase between $K_1$ and $K_2$ to the one of $K_1$ and $(K_2-J)\cup I$, changing one upper crossing to an under-crossing. It follows that this relation is compatible with (\ref{ooo}), because the linking number is given by $(n_+-n_-)/2$, where $n_+$ and $n_-$ are the number of crossings that respect the right-hand rule or have the opposite orientation, respectively. The result follows  by taking the knots to unlinked ones.  
 $\square$

\section{Discussion}\label{SecVII}

In this work we have constrained the possible categorical structures associated with confinement order/disorder parameters and further generalized symmetries. The starting idea has been to formalize the DHR theorem as the classification of certain HDV, that are the essential features of QFT determining symmetries and order parameters.  In this light, the power of DHR is not only that it demonstrates the group-like origin of $0$-form internal symmetries, but that it allows to eliminate the HDV simplifying the problem for further investigation. Using this new perspective we have been able to classify HDV for quite general regions, in particular for all regions in $D=4$. The results in the important case of $D=4$ give abelian groups with Hermitian character tables and commutators determined by linking numbers. Apart from the intriguing scenario of a potential extra $Z_2$ factor in the group decomposition, this demonstrates the Yang-Mills theory phenomenology from a first principles perspective.

We end now by describing various interesting directions for future progress.

\noindent \textbf{On assumptions:} In section \ref{SecII} we have described a large set of assumptions. Although most of those are natural from a certain algebraic point of view, it is also plain that a more economic starting point of view is desirable. Indeed, this task is important not only for the results of this article but from a more foundational standpoint. In this vein, we believe it should be possible to derive some of the more abstract assumptions, such as the split property, transportability and weak modularity from the existence of a Wightman field stress tensor. Also, transportability and weak modularity may well follow from the split property and space-time symmetries. From a different perspective, we expect some of the assumptions used above to be superfluous, i.e. not really required for the results to follow. In this vein, we expect one can relax the finite index assumption, leading to potential proof of the abelianity of ``QED-like'' sectors. Also, we expect one can relax the assumption that the algebra acts irreducibly in the Hilbert space, namely assumption \ref{as:1}. A completely local perspective would be to start from the type III factor of a topological ball in spacetime, and then derive the theorem just from the net properties inside the ball. This would proceed by defining endomorphisms, transportability, strong additivity, relative commutants, etc and generalizing the methods to such scenario.

\noindent \textbf{On anomalies:} Above we have classified certain aspects of HDV and/or generalized DHR and/or generalized symmetries. But not all. In particular, symmetries can display ABJ and 't Hooft anomalies \cite{PhysRev.177.2426,Bell:1969ts,hooft1980naturalness,Gaiotto:2014kfa}. The first problem here is that the local imprints of these anomalies is not understood in full detail. Typically, anomalies are analyzed by coupling the QFT with further degrees of freedom, or by studying partition functions in topologically non-trivial manifolds. A local understanding of the ABJ anomaly in the case of standard $0$-form symmetries was put forward in \cite{Benedetti:2023owa} as global symmetries that interchange the sectors of other generalized symmetries. The case of 't Hooft anomalies lacks a throughout understanding. Such an understanding would potentially lead to a proof of the famous conjectural classification \cite{Chen:2011pg,chen2013symmetryprotectedtopologicalorders}.

\noindent \textbf{On sectors for infinite regions and theories with reduced transportability:} In this work we have focused on topologically non-trivial, but bounded regions or their complements. It is interesting to extend this HDV analysis to infinite regions. This will naturally connect with Buchholz-Fredehangen (BF) selection criteria \cite{Buchholz:1981fj}. A relaxation of the transportability criterion may potentially open a venue to use the present tools for more disparate systems such as those exhibiting fractonic behavior \cite{Chamon:2004lew,Haah:2011drr,PhysRevB.92.235136,Vijay:2016phm}.

\noindent \textbf{On the classification of topological order:} Our theorem seems to be related with the classification of topological field theories in one dimension higher. In particular, bosonic topological field theories appear to have a similar classification, see the recent \cite{Johnson-Freyd:2020usu} and references therein. The fact that only bosonic are needed might follow from the fact derived above, stating that in local QFT, all generalized order parameters can be taken bosonic since in the $\pi_0$-completion one can neutralize a putative fermionic non-local operator with a local fermion insertion. It would be interesting to investigate the connection between the classification of HDV and topological order in one dimension higher, most probably along the lines of \cite{freed2014relativequantumfieldtheory}.

\section*{Acknowledgements} 
We are indebted to Roberto Longo for insightful discussions. We wish to thank discussions with Philip C. Argyres, Valentin Benedetti, Sebastiano Carpi, Davide Gaiotto, Luca Giorgetti, Jaume Gomis, Stefan Hollands, Marina Huerta, Yasuyuki Kawahigashi, Leandro Martinek, Vincenzo Morinelli, Diego Pontello, Gonzalo Torroba and Feng Xu. We  thank hospitality at the Perimeter Institute of Theoretical Physics and at the Mathematics department at University Tor Vergata in Rome, where part of this work was developed. The work of H.C is partially supported by CONICET, CNEA and Universidad Nacional de Cuyo, Argentina. The work of J.M is supported by a Ramón y Cajal fellowship from the Spanish Ministry of Science. In the first stages of this project, the work of J.M was supported by Conicet, Argentina.

\bibliographystyle{utphys}
\bibliography{EE}

\providecommand{\href}[2]{#2}\begingroup\raggedright\begin{thebibliography}{10}

\bibitem{PhysRevD.10.2445}
K.~G. Wilson, ``Confinement of quarks,''
  \href{http://dx.doi.org/10.1103/PhysRevD.10.2445}{{\em Phys. Rev. D}
  {\bfseries 10} (Oct, 1974) 2445--2459}.
  \url{https://link.aps.org/doi/10.1103/PhysRevD.10.2445}.

\bibitem{tHooft:1977nqb}
G.~'t~Hooft, ``{On the Phase Transition Towards Permanent Quark Confinement},''
  \href{http://dx.doi.org/10.1016/0550-3213(78)90153-0}{{\em Nucl. Phys. B}
  {\bfseries 138} (1978) 1--25}.

\bibitem{Doplicher:1971wk}
S.~Doplicher, R.~Haag, and J.~E. Roberts, ``{Local observables and particle
  statistics. 1},'' \href{http://dx.doi.org/10.1007/BF01877742}{{\em Commun.
  Math. Phys.} {\bfseries 23} (1971) 199--230}.

\bibitem{Doplicher:1973at}
S.~Doplicher, R.~Haag, and J.~E. Roberts, ``{Local observables and particle
  statistics. 2},''
\href{http://dx.doi.org/10.1007/BF01646454}{{\em Commun. Math. Phys.}
  {\bfseries 35} (1974) 49--85}.

\bibitem{doplicher1990there}
S.~Doplicher and J.~E. Roberts, ``Why there is a field algebra with a compact
  gauge group describing the superselection structure in particle physics,''
  {\em Communications in Mathematical Physics} {\bfseries 131} no.~1, (1990)
  51--107.

\bibitem{Deligne2007}
P.~Deligne, {\em Cat{\'e}gories tannakiennes},
  \href{http://dx.doi.org/10.1007/978-0-8176-4575-5_3}{pp.~111--195}.
\newblock Birkh{\"a}user Boston, Boston, MA, 2007.
\newblock \url{https://doi.org/10.1007/978-0-8176-4575-5_3}.

\bibitem{Doplicher1989}
R.~J.~E. Doplicher, Sergio, ``A new duality theory for compact groups.'' {\em
  Inventiones mathematicae} {\bfseries 98} no.~1, (1989) 157--218.
  \url{http://eudml.org/doc/143725}.

\bibitem{roberts1974spontaneously}
J.~E. Roberts, ``Spontaneously broken gauge symmetries and superselection
  rules,'' tech. rep., Centre National de la Recherche Scientifique, 1974.

\bibitem{Gaiotto:2014kfa}
D.~Gaiotto, A.~Kapustin, N.~Seiberg, and B.~Willett, ``{Generalized Global
  Symmetries},'' \href{http://dx.doi.org/10.1007/JHEP02(2015)172}{{\em JHEP}
  {\bfseries 02} (2015) 172}, \href{http://arxiv.org/abs/1412.5148}{{\ttfamily
  arXiv:1412.5148 [hep-th]}}.

\bibitem{McGreevy:2022oyu}
J.~McGreevy, ``{Generalized Symmetries in Condensed Matter},''
  \href{http://dx.doi.org/10.1146/annurev-conmatphys-040721-021029}{{\em Ann.
  Rev. Condensed Matter Phys.} {\bfseries 14} (2023) 57--82},
  \href{http://arxiv.org/abs/2204.03045}{{\ttfamily arXiv:2204.03045
  [cond-mat.str-el]}}.

\bibitem{Schafer-Nameki:2023jdn}
S.~Schafer-Nameki, ``{ICTP lectures on (non-)invertible generalized
  symmetries},'' \href{http://dx.doi.org/10.1016/j.physrep.2024.01.007}{{\em
  Phys. Rept.} {\bfseries 1063} (2024) 1--55},
  \href{http://arxiv.org/abs/2305.18296}{{\ttfamily arXiv:2305.18296
  [hep-th]}}.

\bibitem{Brennan:2023mmt}
T.~D. Brennan and S.~Hong, ``{Introduction to Generalized Global Symmetries in
  QFT and Particle Physics},''
  \href{http://arxiv.org/abs/2306.00912}{{\ttfamily arXiv:2306.00912
  [hep-ph]}}.

\bibitem{Bhardwaj:2023kri}
L.~Bhardwaj, L.~E. Bottini, L.~Fraser-Taliente, L.~Gladden, D.~S.~W. Gould,
  A.~Platschorre, and H.~Tillim, ``{Lectures on generalized symmetries},''
  \href{http://dx.doi.org/10.1016/j.physrep.2023.11.002}{{\em Phys. Rept.}
  {\bfseries 1051} (2024) 1--87},
  \href{http://arxiv.org/abs/2307.07547}{{\ttfamily arXiv:2307.07547
  [hep-th]}}.

\bibitem{Shao:2023gho}
S.-H. Shao, ``{What's Done Cannot Be Undone: TASI Lectures on Non-Invertible
  Symmetries},'' in {\em {Theoretical Advanced Study Institute in Elementary
  Particle Physics 2023}: {Aspects of Symmetry}}.
\newblock 8, 2023.
\newblock \href{http://arxiv.org/abs/2308.00747}{{\ttfamily arXiv:2308.00747
  [hep-th]}}.

\bibitem{Casini:2020rgj}
H.~Casini, M.~Huerta, J.~M. Magan, and D.~Pontello, ``{Entropic order
  parameters for the phases of QFT},''
  \href{http://dx.doi.org/10.1007/JHEP04(2021)277}{{\em JHEP} {\bfseries 04}
  (2021) 277}, \href{http://arxiv.org/abs/2008.11748}{{\ttfamily
  arXiv:2008.11748 [hep-th]}}.

\bibitem{Casini:2021zgr}
H.~Casini and J.~M. Magan, ``{On completeness and generalized symmetries in
  quantum field theory},''
  \href{http://dx.doi.org/10.1142/S0217732321300251}{{\em Mod. Phys. Lett. A}
  {\bfseries 36} no.~36, (2021) 2130025},
  \href{http://arxiv.org/abs/2110.11358}{{\ttfamily arXiv:2110.11358
  [hep-th]}}.

\bibitem{Casini:2019kex}
H.~Casini, M.~Huerta, J.~M. Magán, and D.~Pontello, ``{Entanglement entropy
  and superselection sectors. Part I. Global symmetries},''
  \href{http://dx.doi.org/10.1007/JHEP02(2020)014}{{\em JHEP} {\bfseries 02}
  (2020) 014}, \href{http://arxiv.org/abs/1905.10487}{{\ttfamily
  arXiv:1905.10487 [hep-th]}}.

\bibitem{Magan:2020ake}
J.~M. Magan and D.~Pontello, ``{Quantum Complementarity through Entropic
  Certainty Principles},''
  \href{http://dx.doi.org/10.1103/PhysRevA.103.012211}{{\em Phys. Rev. A}
  {\bfseries 103} no.~1, (2021) 012211},
  \href{http://arxiv.org/abs/2005.01760}{{\ttfamily arXiv:2005.01760
  [hep-th]}}.

\bibitem{Pedro}
H.~Casini, J.~M. Magan, and P.~J. Martinez, ``{Entropic order parameters in
  weakly coupled gauge theories},''
  \href{http://dx.doi.org/10.1007/JHEP01(2022)079}{{\em JHEP} {\bfseries 01}
  (2022) 079}, \href{http://arxiv.org/abs/2110.02980}{{\ttfamily
  arXiv:2110.02980 [hep-th]}}.

\bibitem{casini2021generalized}
V.~Benedetti, H.~Casini, and J.~M. Magan, ``{Generalized symmetries of the
  graviton},'' \href{http://dx.doi.org/10.1007/JHEP05(2022)045}{{\em JHEP}
  {\bfseries 05} (2022) 045}, \href{http://arxiv.org/abs/2111.12089}{{\ttfamily
  arXiv:2111.12089 [hep-th]}}.

\bibitem{Magan:2021myk}
J.~M. Magan, ``{Proof of the universal density of charged states in QFT},''
  \href{http://dx.doi.org/10.1007/JHEP12(2021)100}{{\em JHEP} {\bfseries 12}
  (2021) 100}, \href{http://arxiv.org/abs/2111.02418}{{\ttfamily
  arXiv:2111.02418 [hep-th]}}.

\bibitem{Benedetti:2022zbb}
V.~Benedetti, H.~Casini, and J.~M. Magan, ``{Generalized symmetries and
  Noether\textquoteright{}s theorem in QFT},''
  \href{http://dx.doi.org/10.1007/JHEP08(2022)304}{{\em JHEP} {\bfseries 08}
  (2022) 304}, \href{http://arxiv.org/abs/2205.03412}{{\ttfamily
  arXiv:2205.03412 [hep-th]}}.

\bibitem{Casini:2022rlv}
H.~Casini and M.~Huerta, ``{Lectures on entanglement in quantum field
  theory},'' \href{http://dx.doi.org/10.22323/1.403.0002}{{\em PoS} {\bfseries
  TASI2021} (2023) 002}, \href{http://arxiv.org/abs/2201.13310}{{\ttfamily
  arXiv:2201.13310 [hep-th]}}.

\bibitem{Benedetti:2023owa}
V.~Benedetti, H.~Casini, and J.~M. Magan, ``{ABJ anomaly as a U(1) symmetry and
  Noether's theorem},'' \href{http://arxiv.org/abs/2309.03264}{{\ttfamily
  arXiv:2309.03264 [hep-th]}}.

\bibitem{Benedetti:2023ipt}
V.~Benedetti, P.~Bueno, and J.~M. Magan, ``{Generalized Symmetries for
  Generalized Gravitons},''
  \href{http://dx.doi.org/10.1103/PhysRevLett.131.111603}{{\em Phys. Rev.
  Lett.} {\bfseries 131} no.~11, (2023) 111603},
  \href{http://arxiv.org/abs/2305.13361}{{\ttfamily arXiv:2305.13361
  [hep-th]}}.

\bibitem{benedetti2023charges}
V.~Benedetti, H.~Casini, and J.~M. Mag{\'a}n, ``Charges in the uv completion of
  neutral electrodynamics,'' {\em Journal of High Energy Physics} {\bfseries
  2023} no.~6, (2023) 1--35.

\bibitem{Casini:2023vrb}
H.~Casini and L.~Martinek, ``{Standard translation twists and an
  operator-bounded energy inequality},''
  \href{http://dx.doi.org/10.1103/PhysRevD.109.045001}{{\em Phys. Rev. D}
  {\bfseries 109} no.~4, (2024) 045001},
  \href{http://arxiv.org/abs/2310.06961}{{\ttfamily arXiv:2310.06961
  [hep-th]}}.

\bibitem{Benedetti:2024dku}
V.~Benedetti, H.~Casini, Y.~Kawahigashi, R.~Longo, and J.~M. Magan, ``{Modular
  invariance as completeness},''
  \href{http://dx.doi.org/10.1103/PhysRevD.110.125004}{{\em Phys. Rev. D}
  {\bfseries 110} no.~12, (2024) 125004},
  \href{http://arxiv.org/abs/2408.04011}{{\ttfamily arXiv:2408.04011
  [hep-th]}}.

\bibitem{Benedetti:2024utz}
V.~Benedetti, H.~Casini, and J.~M. Magan, ``{Selection rules for RG flows of
  minimal models},'' \href{http://arxiv.org/abs/2412.16587}{{\ttfamily
  arXiv:2412.16587 [hep-th]}}.

\bibitem{Shao:2025mfj}
S.-H. Shao, J.~Sorce, and M.~Srivastava, ``{Additivity, Haag duality, and
  non-invertible symmetries},''
  \href{http://dx.doi.org/10.1007/JHEP08(2025)009}{{\em JHEP} {\bfseries 08}
  (2025) 009}, \href{http://arxiv.org/abs/2503.20863}{{\ttfamily
  arXiv:2503.20863 [hep-th]}}.

\bibitem{Jia:2025bui}
Q.~Jia and J.~Tian, ``{Symmetry, Symmetry Topological Field Theory and von
  Neumann Algebra},'' \href{http://arxiv.org/abs/2507.17103}{{\ttfamily
  arXiv:2507.17103 [hep-th]}}.

\bibitem{Martinek:2025xik}
L.~Martinek, ``{Optimal symmetry operators},''
  \href{http://arxiv.org/abs/2509.09670}{{\ttfamily arXiv:2509.09670
  [hep-th]}}.

\bibitem{Evans:2025msy}
D.~E. Evans and C.~Jones, ``{An operator algebraic approach to fusion category
  symmetry on the lattice},'' \href{http://arxiv.org/abs/2507.05185}{{\ttfamily
  arXiv:2507.05185 [math-ph]}}.

\bibitem{Abate:2025ywp}
N.~Abate, H.~Casini, M.~Huerta, and L.~Martinek, ``{Exact Mutual Information
  Difference: Scalar vs. Maxwell Fields},''
  \href{http://arxiv.org/abs/2511.04742}{{\ttfamily arXiv:2511.04742
  [hep-th]}}.

\bibitem{inproceedings}
K.~Fredenhagen, ``Structural aspects of gauge theories in the algebraic
  framework of quantum field theory,''
\newblock 11, 1982.

\bibitem{Longo:1994xe}
R.~Longo and K.-H. Rehren, ``{Nets of subfactors},''
  \href{http://dx.doi.org/10.1142/S0129055X95000232}{{\em Rev. Math. Phys.}
  {\bfseries 7} (1995) 567--598},
\href{http://arxiv.org/abs/hep-th/9411077}{{\ttfamily arXiv:hep-th/9411077
  [hep-th]}}.

\bibitem{Xu1997JONESWASSERMANNSF}
F.~Xu, ``Jones-wassermann subfactors for disconnected intervals,'' {\em
  Communications in Contemporary Mathematics} {\bfseries 02} (1997) 307--347.
  \url{https://api.semanticscholar.org/CorpusID:9411193}.

\bibitem{Kawahigashi:1999jz}
Y.~Kawahigashi, R.~Longo, and M.~Muger, ``{Multiinterval subfactors and
  modularity of representations in conformal field theory},''
  \href{http://dx.doi.org/10.1007/PL00005565}{{\em Commun. Math. Phys.}
  {\bfseries 219} (2001) 631--669},
\href{http://arxiv.org/abs/math/9903104}{{\ttfamily arXiv:math/9903104
  [math-oa]}}.

\bibitem{Kawahigashi:2002px}
Y.~Kawahigashi and R.~Longo, ``{Classification of local conformal nets: Case c
  \ensuremath{<} 1},'' {\em Annals Math.} {\bfseries 160} (2004) 493--522,
  \href{http://arxiv.org/abs/math-ph/0201015}{{\ttfamily
  arXiv:math-ph/0201015}}.

\bibitem{Kawahigashi:2003gi}
Y.~Kawahigashi and R.~Longo, ``{Classification of two-dimensional local
  conformal nets with c less than 1 and 2 cohomology vanishing for tensor
  categories},'' \href{http://dx.doi.org/10.1007/s00220-003-0979-1}{{\em
  Commun. Math. Phys.} {\bfseries 244} (2004) 63--97},
  \href{http://arxiv.org/abs/math-ph/0304022}{{\ttfamily
  arXiv:math-ph/0304022}}.

\bibitem{BELAVIN1984333}
A.~Belavin, A.~Polyakov, and A.~Zamolodchikov, ``Infinite conformal symmetry in
  two-dimensional quantum field theory,''
  \href{http://dx.doi.org/https://doi.org/10.1016/0550-3213(84)90052-X}{{\em
  Nuclear Physics B} {\bfseries 241} no.~2, (1984) 333--380}.
  \url{https://www.sciencedirect.com/science/article/pii/055032138490052X}.

\bibitem{francesco2012conformal}
P.~Francesco, P.~Mathieu, and D.~S{\'e}n{\'e}chal, {\em Conformal field
  theory}.
\newblock Springer Science \& Business Media, 2012.

\bibitem{Choi:2022jqy}
Y.~Choi, H.~T. Lam, and S.-H. Shao, ``{Noninvertible Global Symmetries in the
  Standard Model},''
  \href{http://dx.doi.org/10.1103/PhysRevLett.129.161601}{{\em Phys. Rev.
  Lett.} {\bfseries 129} no.~16, (2022) 161601},
  \href{http://arxiv.org/abs/2205.05086}{{\ttfamily arXiv:2205.05086
  [hep-th]}}.

\bibitem{Cordova:2022ieu}
C.~Cordova and K.~Ohmori, ``{Noninvertible Chiral Symmetry and Exponential
  Hierarchies},'' \href{http://dx.doi.org/10.1103/PhysRevX.13.011034}{{\em
  Phys. Rev. X} {\bfseries 13} no.~1, (2023) 011034},
  \href{http://arxiv.org/abs/2205.06243}{{\ttfamily arXiv:2205.06243
  [hep-th]}}.

\bibitem{Witten:1988}
E.~Witten, ``{Quantum Field Theory and the Jones Polynomial},''
  \href{http://dx.doi.org/10.1007/BF01217730}{{\em Commun. Math. Phys.}
  {\bfseries 121} (1989) 351--399}.

\bibitem{haag1962postulates}
R.~Haag and B.~Schroer, ``Postulates of quantum field theory,'' {\em Journal of
  Mathematical Physics} {\bfseries 3} no.~2, (1962) 248--256.

\bibitem{araki1964neumann}
H.~Araki, ``Von neumann algebras of local observables for free scalar field,''
  {\em Journal of Mathematical Physics} {\bfseries 5} no.~1, (1964) 1--13.

\bibitem{araki1963lattice}
H.~Araki, ``A lattice of von neumann algebras associated with the quantum
  theory of a free bose field,'' {\em Journal of Mathematical Physics}
  {\bfseries 4} no.~11, (1963) 1343--1362.

\bibitem{buchholz1987universal}
D.~Buchholz, C.~D'Antoni, and K.~Fredenhagen, ``The universal structure of
  local algebras,'' {\em Communications in Mathematical Physics} {\bfseries
  111} no.~1, (1987) 123--135.

\bibitem{gabbiani1993operator}
F.~Gabbiani and J.~Fr{\"o}hlich, ``Operator algebras and conformal field
  theory,'' {\em Communications in mathematical physics} {\bfseries 155} no.~3,
  (1993) 569--640.

\bibitem{brunetti1993}
R.~Brunetti, D.~Guido, and R.~Longo, ``Modular structure and duality in
  conformal quantum field theory,'' {\em Comm. Math. Phys.} {\bfseries 156}
  no.~1, (1993) 201--219.
  \url{https://projecteuclid.org:443/euclid.cmp/1104253522}.

\bibitem{Witten:2018lha}
E.~Witten, ``{APS Medal for Exceptional Achievement in Research: Invited
  article on entanglement properties of quantum field theory},''
  \href{http://dx.doi.org/10.1103/RevModPhys.90.045003}{{\em Rev. Mod. Phys.}
  {\bfseries 90} no.~4, (2018) 045003},
\href{http://arxiv.org/abs/1803.04993}{{\ttfamily arXiv:1803.04993 [hep-th]}}.

\bibitem{Longo:1979dw}
R.~Longo, ``{NOTES ON ALGEBRAIC INVARIANTS FOR NONCOMMUTATIVE DYNAMICAL
  SYSTEMS},'' \href{http://dx.doi.org/10.1007/BF01197443}{{\em Commun. Math.
  Phys.} {\bfseries 69} (1979) 195--207}.

\bibitem{cmp/1103839939}
H.-J. Borchers, ``{A remark on a theorem of B. Misra},'' {\em Communications in
  Mathematical Physics} {\bfseries 4} no.~5, (1967) 315 -- 323.

\bibitem{Jones1983}
V.~Jones, ``Index for subfactors.'' {\em Inventiones mathematicae} {\bfseries
  72} (1983) 1--26. \url{http://eudml.org/doc/143011}.

\bibitem{KOSAKI1986123}
H.~Kosaki, ``Extension of jones' theory on index to arbitrary factors,''
  \href{http://dx.doi.org/https://doi.org/10.1016/0022-1236(86)90085-6}{{\em
  Journal of Functional Analysis} {\bfseries 66} no.~1, (1986) 123 -- 140}.
  \url{http://www.sciencedirect.com/science/article/pii/0022123686900856}.

\bibitem{Longo:1989tt}
R.~Longo, ``{Index of subfactors and statistics of quantum fields. I},''
\href{http://dx.doi.org/10.1007/BF02125124}{{\em Commun. Math. Phys.}
  {\bfseries 126} (1989) 217--247}.

\bibitem{CONNES1980153}
A.~Connes, ``On the spatial theory of von neumann algebras,''
  \href{http://dx.doi.org/https://doi.org/10.1016/0022-1236(80)90002-6}{{\em
  Journal of Functional Analysis} {\bfseries 35} no.~2, (1980) 153--164}.
  \url{https://www.sciencedirect.com/science/article/pii/0022123680900026}.

\bibitem{bischoff2015tensor}
M.~Bischoff, Y.~Kawahigashi, R.~Longo, and K.-H. Rehren, {\em Tensor categories
  and endomorphisms of von neumann algebras: with applications to quantum field
  theory}, vol.~3 of {\em Springer Series in Mathematical Physics}.
\newblock Springer, 2015.
\newblock \url{https://www.springer.com/gp/book/9783540566236}.

\bibitem{Rehren:1993yu}
K.-H. Rehren, ``{Subfactors and coset models},'' in {\em {3rd EGS4 Users'
  Meeting in Japan}}, pp.~338--356.
\newblock 7, 1993.
\newblock \href{http://arxiv.org/abs/hep-th/9308145}{{\ttfamily
  arXiv:hep-th/9308145}}.

\bibitem{Doplicher:1984zz}
S.~Doplicher and R.~Longo, ``{Standard and split inclusions of von Neumann
  algebras},''
\href{http://dx.doi.org/10.1007/BF01388641}{{\em Invent. Math.} {\bfseries 75}
  (1984) 493--536}.

\bibitem{buchholz1986noether}
D.~Buchholz, S.~Doplicher, and R.~Longo, ``On noether's theorem in quantum
  field theory,'' {\em Annals of Physics} {\bfseries 170} no.~1, (1986) 1--17.

\bibitem{horuzhy2012introduction}
S.~S. Horuzhy, {\em Introduction to algebraic quantum field theory}, vol.~19.
\newblock Springer Science \& Business Media, 2012.

\bibitem{buchholz1986causal}
D.~Buchholz and E.~H. Wichmann, ``Causal independence and the energy-level
  density of states in local quantum field theory,'' {\em Communications in
  mathematical physics} {\bfseries 106} no.~2, (1986) 321--344.

\bibitem{buchholz1990nuclear}
D.~Buchholz, C.~D'Antoni, and R.~Longo, ``Nuclear maps and modular structures
  ii: Applications to quantum field theory,'' {\em Communications in
  mathematical physics} {\bfseries 129} (1990) 115--138.

\bibitem{ge1996tensor}
L.~Ge and R.~Kadison, ``On tensor products of von neumann algebras,'' {\em
  Inventiones mathematicae} {\bfseries 123} no.~3, (1996) 453--466.

\bibitem{zacharias2001splitting}
J.~Zacharias, ``Splitting for subalgebras of tensor products,'' {\em
  Proceedings of the American Mathematical Society} {\bfseries 129} no.~2,
  (2001) 407--413.

\bibitem{Bisognano:1975ih}
J.~J. Bisognano and E.~H. Wichmann, ``{On the Duality Condition for a Hermitian
  Scalar Field},''
\href{http://dx.doi.org/10.1063/1.522605}{{\em J. Math. Phys.} {\bfseries 16}
  (1975) 985--1007}.

\bibitem{brunetti1993modular}
R.~Brunetti, D.~Guido, and R.~Longo, ``Modular structure and duality in
  conformal quantum field theory,'' {\em Communications in Mathematical
  Physics} {\bfseries 156} (1993) 201--219.

\bibitem{Harlow:2025cqc}
D.~Harlow, S.-H. Shao, J.~Sorce, and M.~Srivastava, ``{Disjoint additivity and
  local quantum physics},'' \href{http://arxiv.org/abs/2509.03589}{{\ttfamily
  arXiv:2509.03589 [hep-th]}}.

\bibitem{buchholz1992new}
D.~Buchholz, S.~Doplicher, R.~Longo, and J.~E. Roberts, ``A new look at
  goldstone’s theorem,'' {\em Reviews in Mathematical Physics} {\bfseries 4}
  no.~spec01, (1992) 49--83.

\bibitem{loi1992theory}
P.~H. Loi, ``On the theory of index for type iii factors,'' {\em Journal of
  Operator Theory} (1992) 251--265.

\bibitem{longo2003conformal}
R.~Longo, ``Conformal subnets and intermediate subfactors,'' {\em
  Communications in mathematical physics} {\bfseries 237} no.~1, (2003) 7--30.

\bibitem{d1983interpolation}
C.~D'Antoni and R.~Longo, ``Interpolation by type i factors and the flip
  automorphism,'' {\em Journal of Functional Analysis} {\bfseries 51} no.~3,
  (1983) 361--371.

\bibitem{Buchholz:1981fj}
D.~Buchholz and K.~Fredenhagen, ``{Locality and the Structure of Particle
  States},''
\href{http://dx.doi.org/10.1007/BF01208370}{{\em Commun. Math. Phys.}
  {\bfseries 84} (1982) 1}.

\bibitem{haag2012local}
R.~Haag, {\em Local quantum physics: Fields, particles, algebras}.
\newblock Springer Science \& Business Media, 2012.

\bibitem{guido1992relativistic}
D.~Guido and R.~Longo, ``Relativistic invariance and charge conjugation in
  quantum field theory,'' {\em Communications in Mathematical Physics}
  {\bfseries 148} no.~3, (1992) 521--551.

\bibitem{Cuntz:1977ut}
J.~Cuntz, ``{Simple C* Algebras Generated by Isometries},''
  \href{http://dx.doi.org/10.1007/BF01625776}{{\em Commun. Math. Phys.}
  {\bfseries 57} (1977) 173--185}.

\bibitem{kastler1990algebraic}
D.~Kastler, {\em Algebraic Theory Of Superselection Sectors, The: Introduction
  And Recent Results-Proceedings Of The Covegno Internazionale" Algebraic
  Theory Of Superselection Sectors And Field Theory"}.
\newblock World Scientific, 1990.

\bibitem{Halvorson:2006wj}
H.~Halvorson and M.~Muger,
  \href{http://dx.doi.org/10.1016/B978-044451560-5/50011-7}{``{Algebraic
  quantum field theory},''} in {\em Philosophy of physics}, J.~Butterfield and
  J.~Earman, eds., pp.~731--864.
\newblock 2007.
\newblock
\href{http://arxiv.org/abs/math-ph/0602036}{{\ttfamily arXiv:math-ph/0602036
  [math-ph]}}.
\newblock

\bibitem{guido1995algebraic}
D.~Guido and R.~Longo, ``An algebraic spin and statistics theorem,'' {\em
  Communications in Mathematical Physics} {\bfseries 172} no.~3, (1995)
  517--533.

\bibitem{carpi2005classification}
S.~Carpi and R.~Conti, ``Classification of subsystems for graded-local nets
  with trivial superselection structure,'' {\em Communications in mathematical
  physics} {\bfseries 253} no.~2, (2005) 423--449.

\bibitem{cmp/1104200513}
R.~Longo, ``{Index of subfactors and statistics of quantum fields. II.
  Correspondences, braid group statistics and Jones polynomial},''
  \href{http://dx.doi.org/cmp/1104200513}{{\em Communications in Mathematical
  Physics} {\bfseries 130} no.~2, (1990) 285 -- 309}. \url{https://doi.org/}.

\bibitem{Longo:1994zza}
R.~Longo, ``{A duality for Hopf algebras and for subfactors. 1.},''
  \href{http://dx.doi.org/10.1007/BF02100488}{{\em Commun. Math. Phys.}
  {\bfseries 159} (1994) 133--150}.

\bibitem{Hollands:2022gab}
S.~Hollands, ``{Anyonic Chains {\textendash} $\alpha $-Induction {\textendash}
  CFT {\textendash} Defects {\textendash} Subfactors},''
  \href{http://dx.doi.org/10.1007/s00220-022-04581-w}{{\em Commun. Math. Phys.}
  {\bfseries 399} no.~3, (2023) 1549--1621},
  \href{http://arxiv.org/abs/2205.15243}{{\ttfamily arXiv:2205.15243
  [cond-mat.str-el]}}.

\bibitem{fox1948imbedding}
R.~H. Fox, ``On the imbedding of polyhedra in 3-space,'' {\em Annals of
  Mathematics} {\bfseries 49} no.~2, (1948) 462--470.

\bibitem{sen2020self}
A.~Sen, ``Self-dual forms: action, hamiltonian and compactification,'' {\em
  Journal of Physics A: Mathematical and Theoretical} {\bfseries 53} no.~8,
  (2020) 084002.

\bibitem{bialynicki1981note}
I.~Bialynicki-Birula, E.~Newman, J.~Porter, J.~Winicour, B.~Lukacs, Z.~Perjes,
  and A.~Sebestyen, ``A note on helicity,'' {\em Journal of Mathematical
  Physics} {\bfseries 22} no.~11, (1981) 2530--2532.

\bibitem{PhysRev.177.2426}
S.~L. Adler, ``Axial-vector vertex in spinor electrodynamics,''
  \href{http://dx.doi.org/10.1103/PhysRev.177.2426}{{\em Phys. Rev.} {\bfseries
  177} (Jan, 1969) 2426--2438}.
  \url{https://link.aps.org/doi/10.1103/PhysRev.177.2426}.

\bibitem{Bell:1969ts}
J.~S. Bell and R.~Jackiw, ``{A PCAC puzzle: $\pi^0 \to \gamma \gamma$ in the
  $\sigma$ model},'' \href{http://dx.doi.org/10.1007/BF02823296}{{\em Nuovo
  Cim. A} {\bfseries 60} (1969) 47--61}.

\bibitem{hooft1980naturalness}
G.~Hooft, ``Naturalness, chiral symmetry, and spontaneous chiral symmetry
  breaking,'' in {\em Recent developments in gauge theories}, pp.~135--157.
\newblock Springer, 1980.

\bibitem{Chen:2011pg}
X.~Chen, Z.-C. Gu, Z.-X. Liu, and X.-G. Wen, ``{Symmetry protected topological
  orders and the group cohomology of their symmetry group},''
  \href{http://dx.doi.org/10.1103/PhysRevB.87.155114}{{\em Phys. Rev. B}
  {\bfseries 87} no.~15, (2013) 155114},
  \href{http://arxiv.org/abs/1106.4772}{{\ttfamily arXiv:1106.4772
  [cond-mat.str-el]}}.

\bibitem{chen2013symmetryprotectedtopologicalorders}
X.~Chen, Z.-C. Gu, Z.-X. Liu, and X.-G. Wen, ``Symmetry protected topological
  orders in interacting bosonic systems,'' 2013.
\newblock \url{https://arxiv.org/abs/1301.0861}.

\bibitem{Chamon:2004lew}
C.~Chamon, ``{Quantum Glassiness},''
  \href{http://dx.doi.org/10.1103/physrevlett.94.040402}{{\em Phys. Rev. Lett.}
  {\bfseries 94} no.~4, (2005) 040402},
  \href{http://arxiv.org/abs/cond-mat/0404182}{{\ttfamily
  arXiv:cond-mat/0404182}}.

\bibitem{Haah:2011drr}
J.~Haah, ``{Local stabilizer codes in three dimensions without string logical
  operators},'' \href{http://dx.doi.org/10.1103/physreva.83.042330}{{\em Phys.
  Rev. A} {\bfseries 83} no.~4, (2011) 042330},
  \href{http://arxiv.org/abs/1101.1962}{{\ttfamily arXiv:1101.1962
  [quant-ph]}}.

\bibitem{PhysRevB.92.235136}
S.~Vijay, J.~Haah, and L.~Fu, ``A new kind of topological quantum order: A
  dimensional hierarchy of quasiparticles built from stationary excitations,''
  \href{http://dx.doi.org/10.1103/PhysRevB.92.235136}{{\em Phys. Rev. B}
  {\bfseries 92} (Dec, 2015) 235136}.
  \url{https://link.aps.org/doi/10.1103/PhysRevB.92.235136}.

\bibitem{Vijay:2016phm}
S.~Vijay, J.~Haah, and L.~Fu, ``{Fracton Topological Order, Generalized Lattice
  Gauge Theory and Duality},''
  \href{http://dx.doi.org/10.1103/PhysRevB.94.235157}{{\em Phys. Rev. B}
  {\bfseries 94} no.~23, (2016) 235157},
  \href{http://arxiv.org/abs/1603.04442}{{\ttfamily arXiv:1603.04442
  [cond-mat.str-el]}}.

\bibitem{Johnson-Freyd:2020usu}
T.~Johnson-Freyd, ``{On the Classification of Topological Orders},''
  \href{http://dx.doi.org/10.1007/s00220-022-04380-3}{{\em Commun. Math. Phys.}
  {\bfseries 393} no.~2, (2022) 989--1033},
  \href{http://arxiv.org/abs/2003.06663}{{\ttfamily arXiv:2003.06663
  [math.CT]}}.

\bibitem{freed2014relativequantumfieldtheory}
D.~S. Freed and C.~Teleman, ``Relative quantum field theory,'' 2014.
\newblock \url{https://arxiv.org/abs/1212.1692}.

\end{thebibliography}\endgroup

\end{document}